\pdfoutput=1
\documentclass{aastex701}

\usepackage{amsmath}
\usepackage{makecell}
\usepackage{multirow}
\usepackage{threeparttable}
\usepackage{float}
\usepackage{microtype}

\begin{document}

\title{Classical Be Stars and Classical Be Star Binaries from LAMOST DR12} 

\correspondingauthor{Wei-Min Gu}
\email{guwm@xmu.edu.cn}

\author[0009-0009-3791-0642]{Qian-Yu An}\email{anqianyu@stu.xmu.edu.cn}
\affiliation{Department of Astronomy, Xiamen University, Xiamen, Fujian 361005, People's Republic of China} 

\author[0000-0003-3137-1851]{Wei-Min Gu}\email{guwm@xmu.edu.cn}
\affiliation{Department of Astronomy, Xiamen University, Xiamen, Fujian 361005, People's Republic of China}

\author[0000-0002-2419-6875]{Zhi-Xiang Zhang}\email{zzx@qztc.edu.cn}
\affiliation{College of Physics and Information Engineering, Quanzhou Normal University, Quanzhou 362000,
People’s Republic of China}

\begin{abstract}
Classical Be (CBe) stars are rapidly rotating B-type stars with Balmer emission lines that originate from the decretion disks surrounding them in their spectra. Accounting for $\sim$20\% of all B-type stars, most CBe stars are thought to form through mass and angular momentum transfer from their companions. It follows that in most close CBe star binaries, the companions are expected to be post-main-sequence stars rather than main-sequence (MS) stars. Hitherto, $\sim$100 CBe star binaries have been identified, the majority of which are Be/X-ray binaries. As expected, none of the others have indeed been confirmed as CBe+MS binary stars. To further study and verify the origin of CBe stars, identifying additional CBe star binaries is indispensable. In this study, we report 504 CBe stars identified using data from Data Release 12 of the Large sky Area Multi-Object fiber Spectroscopic Telescope. Among these, 141 are newly identified and 14 exhibiting radial velocity variations are identified as CBe star binaries. Besides, 60 CBe stars with high normalized unit weight error (\texttt{RUWE}) but not confirmed by dynamics are proposed as potential CBe star binaries. We also find that 34 CBe stars are potential cluster members. By calculating peculiar velocities, 37 runaway stars are identified with peculiar velocities ranging from $\sim$40 km s$^{-1}$ to $\sim$101 km s$^{-1}$.  

\end{abstract}


\section{Introduction} 

Classical Be (CBe) stars are rapidly rotating B-type stars with Balmer emission lines that originate from the decretion disks surrounding them \citep[e.g.,][]{2003PASP..115.1153P, 2013A&ARv..21...69R}. The rotational velocities of CBe stars can be at least 0.7 times their critical rotation velocities \citep[e.g.,][]{2013A&ARv..21...69R, 2016A&A...595A.132Z}, and various channels have been proposed to explain why they rotate rapidly. \par \citet{1995ARA&A..33..199B} advocated that certain B-type stars are endowed with a substantial amount of angular momentum inherited
from their parent molecular clouds, causing them to be fast rotators right from their formation, ultimately becoming CBe stars. However, some studies on open clusters suggested that Be stars are not only young stars, but in all
ages \citep{1979ApJ...230..485A, 1982A&A...109...48M, 1985ApJS...59..769S}. Moreover, it has been shown that most young B stars have rotational velocities that are well below the limit for Be star formation \citep{2004ApJ...601..979W, 2010ApJ...722..605H}.
\par Another single-star evolution channel that forms CBe stars is angular momentum transfer from the core to the envelope to accelerate B stars \citep[e.g.,][]{2008A&A...478..467E, 2013A&A...553A..25G, 2020A&A...633A.165H}. On the other hand, \citet{2020A&A...633A.165H} also pointed out that CBe stars formed through such a channel should have much larger surface nitrogen enhancement, while only a small portion of CBe stars exhibit the model prediction of nitrogen enrichment \citep{2011A&A...536A..65D}.
\par The single-star evolution channel does not seem to dominate the formation of CBe stars, implying the existence of alternative formation channels. The binary interaction channel proposes that a star can be spun up through mass and angular momentum transfer from an evolved companion \citep[e.g.,][]{1975BAICz..26...65K, 1991A&A...241..419P}. Supporting this, \citet{2005ApJS..161..118M} conducted a photometric survey of 55 southern open clusters and concluded that up to 73\% of Be stars may have been spun up via binary mass transfer. Later, \citet{2014ApJ...796...37S} provided theoretical support using the BSE code \citep{2002MNRAS.329..897H}, demonstrating that most Be stars likely originate from binary interactions. Consequently, if a companion of a CBe star survives from the binary interactions, it should be a post-main-sequence object, such as an O/B-type hot subdwarf (sdO/B), a stripped helium star, or a compact object, rather than a main-sequence star.
\par There is no doubt that searching for more CBe stars and CBe star binaries is necessary to further study and verify the origin of CBe stars. Since the first identification of CBe star, $\gamma$ Cas, also the first star found to show emission lines in the spectrum \citep{1866AN.....68...63S}, thousands of CBe stars have been identified by numerous photometric or spectroscopic measurements \citep[e.g.,][]{2015AJ....149....7C, 2015RAA....15.1325L, 2016RAA....16..138H}, and compilation of the CBe star catalogs has also been carried out accordingly \citep[e.g.,][]{1982IAUS...98..261J, 2011AJ....142..149N}. Among them, the Be Star Spectra database established by \citet{2011AJ....142..149N} is not only a catalog but also a collection of 323,030 spectra for 2455 Be stars as of June 18, 2025. Additionally, optical counterparts of several X-ray sources have been identified as CBe stars \citep[e.g.,][]{1999MNRAS.306..100R, 2011ApJ...728...86T}, expanding both the known population of CBe stars and Be/X-ray binaries. In Be/X-ray binaries, all the confirmed companions in the Milky Way are neutron stars, while several cases for white dwarf companions have been found in the Large Magellanic Cloud and Small Magellanic Cloud \citep[e.g.,][]{2006A&A...458..285K, 2021MNRAS.508..781K, 2025ApJ...980L..36M}. Meanwhile, progress has also been made in identifying non-X-ray source CBe binaries based on the CBe star samples, with most confirmed companions in these systems being sdO/Bs \citep[e.g.,][]{1983PASP...95..311P, 1991A&A...250..437W, 2018ApJ...853..156W, 2021AJ....161..248W, 2025A&A...699L...1N}. Rarely, several CBe stars were found or suspected to reside in triple systems, e.g., V1371 Tau \citep{2025A&A...694A.172R, 2026ApJ...996...61R}, $\nu$ Gem \citep{2021ApJ...916...24K}, $\delta$ Sco \citep{2013ApJ...766..119M} and 60 Cyg \citep{2022ApJ...926..213K}. Evidence or hints for multiplicity in these systems have primarily been obtained through orbital analysis, long-term RV monitoring, as well as high-angular-resolution techniques such as speckle observations and interferometry observations. 
\par In this work, we report the identification of {504} CBe stars ({141} newly identified) from the Data Release 12 (DR12) of the Large Sky Area Multi-Object Fiber Spectroscopy Telescope (LAMOST), of which 14 exhibit radial velocity (RV) variations. The structure of this paper is organized as follows. We introduce the spectra collected in Section \ref{sec2}. In Section \ref{sec3}, we describe the process of selecting CBe stars. In Section \ref{sec4}, we present 14 CBe star binaries identified in this work and attach 60 potential CBe star binaries. In Section \ref{sec5}, we study the spatial distribution of these CBe stars and identify potential cluster members and runaway stars among them. We summarize our results and make a discussion in Section \ref{sec6}.

\section{LAMOST spectral observation}\label{sec2}
LAMOST, also known as the GuoShouJing Telescope, is a special 4-meter aperture reflecting Schmidt telescope with an installation of 4000 fibers within the 5 deg field of view \citep{2012RAA....12.1197C}, thus possessing extremely high spectral acquisition efficiency. It launched the first 5-year survey with low-resolution spectrographs ($R=1800$) since 2012 \citep{2015RAA....15.1095L}, later expanding to include medium-resolution spectrographs ($R=7500$) in 2017 \citep{2020arXiv200507210L}. The low-resolution spectra (LRS) cover a broad wavelength range of 3700–9000 $\mathrm{\AA}$, while the medium-resolution spectra (MRS) cover wavelength ranges from 4950–5350 $\mathrm{\AA}$ and 6300–6800 $\mathrm{\AA}$ for the blue arms and the red arms, respectively. For stars with spectral type late A, F, G, K, or M, stellar parameters are derived using the LAMOST Stellar Parameter Pipeline \citep{2015RAA....15.1095L, 2018SPIE10707E..2BL, 2021RAA....21..202D}. However, for hotter stars, only spectral classifications are available in LRS, while neither stellar parameters nor spectral classifications are available for hot stars in MRS. 
\par As of DR12, LAMOST has collected an impressive total of 12,602,390 LRS and 15,475,985 MRS\footnote{\url{https://www.lamost.org/dr12/}}, representing one of the largest stellar spectroscopic databases available. Among them, 12,110,344 MRS belong to time-domain survey \citep{2021RAA....21..292W}; therefore, numerous sources have multiepoch spectra, which can be used for RV monitoring and search for binaries.
\par Our initial sample contains 14,981 sources classified as O/B-type stars (\texttt{subclass}:B, B6, B9, O or OB) in at least one LRS observation. After crossmatching these sources with the medium-resolution database, we finally obtain a total of 34,558 LRS and 25,809 MRS corresponding to these sources. 

\section{sample selecting}\label{sec3}
\subsection{Preliminary selection of CBe stars by checking spectra}\label{sec3.1}
We inspect the obtained spectra of the initial sample by eye checking and apply the following two criteria to select CBe stars preliminarily:
\begin{enumerate}
\item The H$\alpha$ exhibits a strong emission line (with a full width $>$ 5 $\mathrm{\AA}$ in MRS or $>$ 10 $\mathrm{\AA}$ in LRS, respectively) in at least one spectrum,
\item There are no forbidden emission lines such as [O III] $\lambda$4959, $\lambda$5007, [N II] $\lambda$6583, and [S II] $\lambda$6717, $\lambda$6731 in the spectra.
\end{enumerate}
The second criterion is applied to exclude Herbig Ae (HAe) or Be (HBe) stars and sources located in the H II regions. While these sources exhibit H$\alpha$ emission similar to CBe stars, they can be distinguished by their additional forbidden line emission—a characteristic feature of low-density environments. In contrast, CBe stars possess high-density circumstellar disks that preclude forbidden line emission. Applying these two criteria yields a preliminarily sample of 538 CBe star candidates.

\subsection{Crossmatching with known CBe star catalogs}\label{sec:3.2}
Building upon the preliminary sample selection described in Section \ref{sec3.1}, we crossmatch it with the available CBe star catalogs compiled by \citet{1982IAUS...98..261J, 2011AJ....142..149N, 2015AJ....149....7C, 2015RAA....15.1325L, 2016MNRAS.463.1162C, 2016RAA....16..138H, Vioque2020, 2021RAA....21..288S} and \cite{2022ApJS..260...35W}. This crossmatching serves two purposes: (i) to quantify the number of newly identified CBe stars in our sample, and (ii) to eliminate misclassified or controversial sources in our sample. Besides, we also retrieve the \texttt{SIMBAD} database \citep{2000A&AS..143....9W} to obtain the main type and other types, and record whether they contain ``Be'' to supplement the missing sources in the above catalogs. Following this crossmatching procedure, we identify 24 sources requiring exclusion: 12 show conflicting classifications as a CBe star and other types (evolved, [H II], {HBe} star or B[e]) in different sources, while another 12 are never classified as CBe stars in any available catalog. After removing these sources, 514 sources are reserved, including 147 newly identified sources. 

\subsection{IR test}\label{sec3.3}
To further remove {HAe/HBe} stars from the sample and strengthen the accuracy of the sample as much as possible, we perform the IR test, suggested by the fact that the IR excess in {HAe/HBe} stars is stronger than that of CBe stars \citep{1984A&AS...55..109F}. We identified {HAe/HBe} stars by applying a modified, more stringent version of the criterion established by \citet{2016RAA....16..138H}: $H-K>0.4$ or $K-W_1>0.8$, where $H$, $K$ and $W_1$ are the $H$ band of Two Micron All Sky Survey \citep[2MASS;][]{2006AJ....131.1163S}, the $K_{\rm s}$ band of 2MASS, and the $W_1$ band of Wide-field Infrared Survey Explorer \citep[WISE;][]{2010AJ....140.1868W}, respectively. We retrieve photometric data of 2MASS and WISE, respectively, from II/246/out \citep{2003yCat.2246....0C} and II/328/allwise \citep{2014yCat.2328....0C} of \texttt{Vizier} catalogs \citep{2000A&AS..143...23O}. To correct for interstellar extinction, we obtain extinction values from \texttt{GALExtin}\footnote{\url{http://galextin.org/}} \citep{2021MNRAS.508.1788A}, using the model constructed by \citet{2019ApJ...887...93G}. The distances required for this process are taken from \citet{2021AJ....161..147B}. Following the reddening law of \citet{1989ApJ...345..245C} with $R_{\rm V} = 3.1$, we convert the extinction values into the color excess of $E(H-K)$ and $E(K-W_{1})$ using the wavelength-dependent extinction ratios from \texttt{VOSA}\footnote{\url{http://svo2.cab.inta-csic.es/theory/fps/}} \citep{2008A&A...492..277B}. After applying these corrections, we calculate dereddened $H-K$ and $K-W_1$, identifying 505 CBe stars, including 142 newly identified sources.

\subsection{Stellar properties}\label{sec3.4}
To investigate the stellar properties of these CBe stars, we plot them on the Hertzsprung–Russell (H-R) diagram and crossmatch their LAMOST/LRS spectra with high signal-to-noise ratio (SNR) with observed spectra of other B stars with spectral classification.
\par
\begin{figure*}[ht!]
\plotone{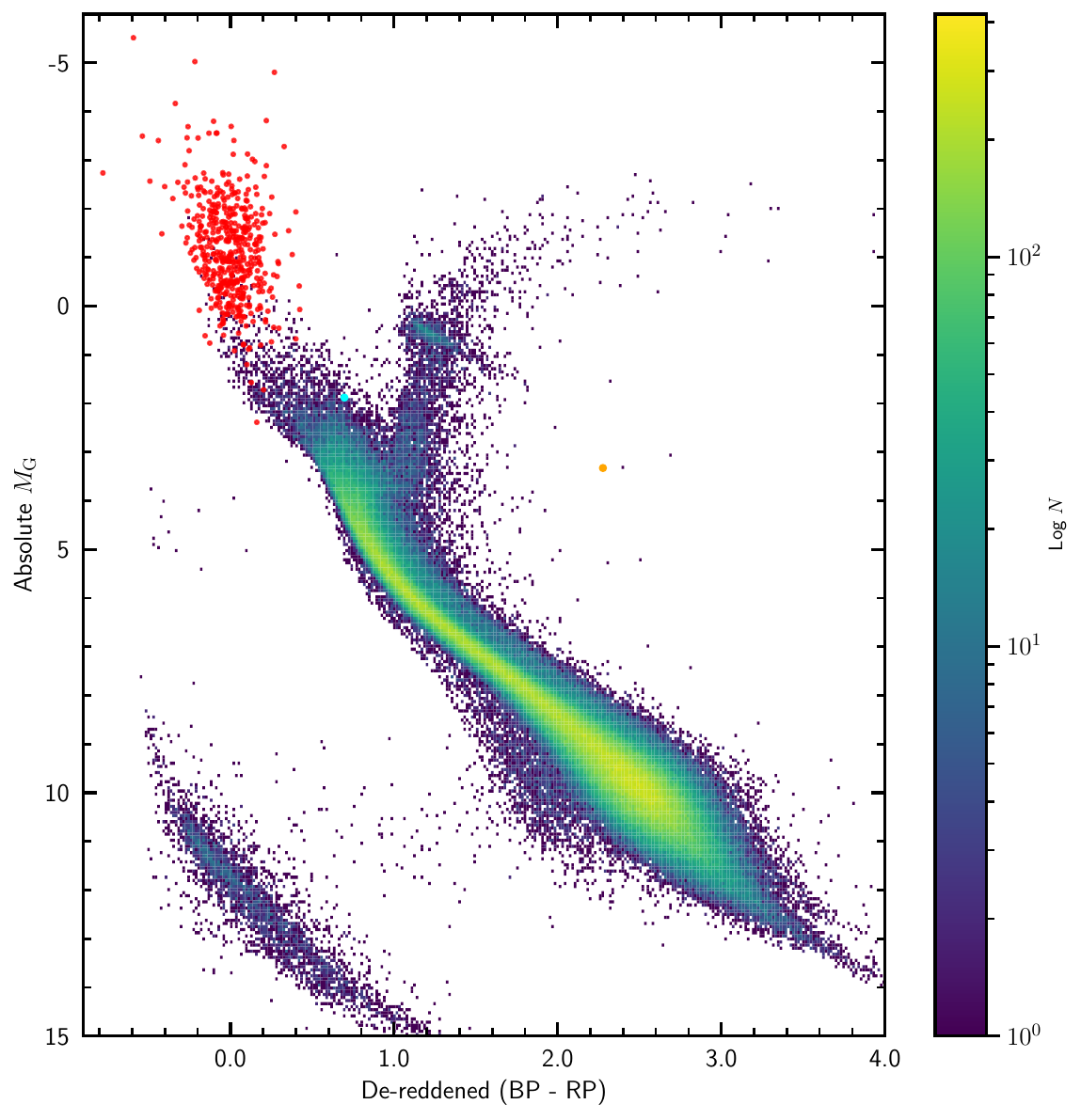}
\caption{The distribution of CBe stars in our sample on the H-R diagram. Most sources in our sample are marked by red dots, while two outliers, J033630.81+480913.2 and J063809.24+034454.6, are marked by an orange dot and an aquamarine dot, respectively.
\label{fig:HR_distribution}}
\end{figure*}

 We take 400,000 stars from \emph{Gaia} DR3 as the background of the H-R diagram, selected by the following criteria:
\begin{enumerate}
\item \texttt{Parallax $>$ 2 mas} and {parallax\_over\_err $>$ 10},
\item $\lvert \mathrm{Galactic~latitude} \rvert$ $>$ 30,
\item \texttt{RUWE $<$ 1.4},
\item \texttt{phot\_g\_mean\_flux\_over\_error $>$ 50},
\item \texttt{phot\_rp\_mean\_flux\_over\_error $>$ 20},
\item \texttt{phot\_bp\_mean\_flux\_over\_error $>$ 20},
\item \texttt{1.0 + 0.015 * power(phot\_bp\_mean\_mag - phot\_rp\_mean\_mag,2) $<$ phot\_bp\_rp\_excess\_factor $<$ 1.3 + 0.060 * power(phot\_bp\_mean\_mag - phot\_rp\_mean\_mag,2)},
\item \texttt{visibility\_periods\_used $>$ 8}.
\end{enumerate}
We then select 475 sources with relative distance uncertainty not exceeding 20\% to calculate their dereddened absolute $G_\mathrm{mag}$ and BP – RP, following the similar method as described in Section \ref{sec3.3}, and plot them on the H-R diagram (see Figure \ref{fig:HR_distribution}). Most sources in our sample lie on or near
the MS, and none occupies the subdwarf region. Two objects—J063809.24+034454.6 (marked by an aquamarine dot in Figure \ref{fig:HR_distribution}) and J033630.81+480913.2 (marked by an orange dot in Figure \ref{fig:HR_distribution})—appear as outliers with significantly redder BP – RP colors. Inspection of their spectra reveals strong Ca II triplet absorption line but no He I absorption lines in J033630.81+480913.2, indicating an F-type (or later-type) star. Although there is only one spectrum with low SNR ($<$ 30 across most wavelength ranges) for J033630.81+480913.2, preventing an in-depth analysis of it, we exclude J033630.81+480913.2 from our CBe star sample to ensure a purer sample. In contrast, J063809.24+034454.6 is confirmed to be a B-type star based on the presence of Balmer series, He I lines, and Paschen series in its spectrum. However, its blue-arm flux is substantially lower than that of the medium arm, implying unusually strong interstellar extinction. This extinction is likely underestimated by the universal model, leading to a severely overestimated dereddened BP – RP color. Ultimately, we identify 504 CBe stars, including 141 newly identified sources. In {Tables \ref{tab:final_sample} and \ref{tab:exluded}}, we list these 504 identified CBe stars {and 33 sources excluded by Section \ref{sec:3.2} and \ref{sec3.3}, respectively.}
\par We also derive the spectral types and luminosity classes of 405 sources in our sample by matching their optical spectra with the spectral library of B-type MK standard
stars. These 405 sources satisfy the requirement of having at least one LAMOST/LRS spectrum with a SNR in the $g$ band of $\geq$ 40. For sources with multiple spectra meeting this criterion, we consistently adopt the spectrum with the highest SNR. The spectra of B-type MK standard stars are obtained from the IACOB project, which has developed a new grid of northern standards for the spectral classification of more than 150 B-type stars and four A-type stars \citep{2024A&A...690A.176N}. This grid covers the
classical classification spectral range from 3900 to 5100 $\mathrm{\AA}$ with spectral resolution of $\sim$4000. For consistency, we degrade the spectra of B-type MK standard stars to 1800 to match the spectral resolution of LAMOST/LRS spectra. Furthermore, to minimize the effects caused by the differences of RVs and projected rotational velocities between the standard stars and our sample, we apply a systematic grid of RVs and $V$sin$i$ to the LAMOST/LRS spectra of our sample and the spectra of standard stars, respectively, in the matching process. The grids explore RVs and $V$sin$i$s ranging from -100 to 100 km s$^{-1}$ and from 50 to 410 km s$^{-1}$, with steps of 10 and 20 km s$^{-1}$, respectively.
\par We identify the match with minimal \( \chi^2 \)/$N$ as the best match for each source. The \( \chi^2 \)/$N$ is defined as:
\[
\chi^2/N = \frac{1}{N}\sum_{i} \left( \frac{f_{\rm obs}^i - f_{\rm lib}^i}{f_{\rm err}^{i}} \right)^2\text{,}
\]
where $f_{\rm obs}^i$ refers to the flux of the LAMOST/LRS spectrum of a given source at wavelength $i$, $f_{\rm lib}^i$ refers to the flux of a spectrum of standard star at wavelength $i$, $f_{\rm err}^i$ refers to the observed uncertainty of flux at wavelength $i$, and $N$ is the number of wavelength points. During the calculation of \( \chi^2 \)/$N$, we select the spectral range from 4,000 $\mathrm{\AA}$ to 4,800 $\mathrm{\AA}$, masking the Balmer series and the diffuse interstellar bands. There are no O-type standard stars, so potential Oe stars in our sample will be misclassified. Considering the rarity of O-type stars, it will not cause a significant impact. In Table \ref{tab:spectral_match}, we show the best matches for sources with \( \chi^2 \)/$N$ $\leq$ 9, and $\sim$90\% of them are with luminosity classes of III, IV or V.

\begin{deluxetable*}{cccccc}
\tablewidth{0pt}
\tablecaption{CBe stars identified from LAMOST DR12 \label{tab:final_sample}}
\tablehead{
\colhead{LAMOST}  & \colhead{Simbad} & \colhead{\emph{Gaia} DR3} & \colhead{R.A.}  &  \colhead{Decl.}  & \colhead{New} \\
\colhead{designation}  & \colhead{name} &  \colhead{ID} & \colhead{(deg)} & \colhead{(deg)} & \colhead{CBe star?}}
\startdata
J000003.86+635429.6 & EM* VES 976
 & 431630851621447552 & 0.016093 & 63.908246 & No \\
J000305.76+632002.1 & --- & 431579342069149184 & 0.774036 & 63.33392 & Yes \\
J000336.71+635514.3 & --- & 431622845802749696 & 0.902974 & 63.92066 & Yes \\
J000352.84+573550.0 & --- & 422602452401685120 & 0.970206 & 57.597233 & No \\
J000429.45+644140.6 & LS I +64 14 & 432136141624375808 & 1.122739 & 64.694625 & Yes \\
... & ... & ... & ... & ... & ... \\
\enddata
     \begin{tablenotes}
        \item[1] {(This table is available in its entirety in machine-readable form in the \href{https://doi.org/10.3847/1538-4365/ae450d}{online article}.)} 
     \end{tablenotes}
\end{deluxetable*}

\begin{deluxetable*}{cccccccc}
\setlength{\tabcolsep}{3pt}
\tablewidth{0pt}
\tablecaption{{33 sources excluded by cross-matching with known CBe star catalogs and IR test}  \label{tab:exluded}}
\tablehead{
\colhead{{LAMOST}} & \colhead{{\emph{Gaia} DR3}} & \colhead{{R.A.}}  &  \colhead{{Decl.}}  & \colhead{{Excluded by}} & \colhead{{Other type(s)}} & \colhead{{H - K}} & \colhead{{K - W$_{1}$}} \\
\colhead{{designation}} & \colhead{{ID}} & \colhead{{(deg)}} & \colhead{{(deg)}} & \colhead{} & \colhead{{[Ref.]}} & \colhead{{(mag)}} & \colhead{{(mag)}}}
\startdata
{J015650.77+574037.6} & {505371935395794432} & {29.211578} & {57.677124} & {Cross-matching} & {Evolved [3]} & {0.14} & {0.23} \\
{J034158.92+575733.6} & {449123153896296704} & {55.4955} & {57.959361} & {IR test} & & {-0.08} & {1.37} \\
{J035952.32+481339.9} & {246810265606108544} & {59.968015} & {48.227753} & {Cross-matching} & {HBe [3]} & {-0.03} & {1.18} \\
{J041805.77+533707.2} & {275377398760016000} & {64.524051} & {53.618671} & {Cross-matching} & {HBe [3]} & {-0.05} & {0.97} \\
{J042118.65+451430.8} & {232751566334445568} & {65.32773} & {45.241901} & {Cross-matching} & {A[e] [3]} & {-0.01} & {0.14} \\
{J043000.64+422259.6} & {252274593578484992} & {67.502696} & {42.383247} & {Cross-matching} & {HBe [3]} & {-0.05} & {1.25} \\
{J043638.32+545049.5} & {274087843419934208} & {69.159697} & {54.847097} & {Cross-matching} & {HII [1]} & {0.04} & {0.07} \\
{J044531.85+544316.3} & {274177694136042496} & {71.382746} & {54.721206} & {IR test} & & {-0.06} & {0.88} \\
{J050104.85+262444.2} & {3420337681942539520} & {75.270218} & {26.412281} & {IR test} & & {-0.10} & {0.89} \\
{J051838.65+350814.2} & {183624252932329728} & {79.661056} & {35.137287} & {Cross-matching} & {HII [1]} & {-0.02} & {0.01} \\
{J052019.20+320817.0} & {180790154334589696} & {80.08001} & {32.138056} & {Cross-matching} & {HII [1]; HBe [3]} & {-0.01} & {0.98} \\
{J052958.80+330642.2} & {3449029781465926400} & {82.495015} & {33.111744} & {Cross-matching} & {HII [1]} & {0.00} & {0.08} \\
{J053123.44+383157.2} & {190415420860586624} & {82.847667} & {38.532583} & {Cross-matching} & {PMSc [2]} & {0.23} & {0.76} \\
{J053141.10+091327.9} & {3337843940146649984} & {82.921254} & {9.22443} & {Cross-matching} & {HII [1]} & {0.19} & {-0.05} \\
{J053201.66+363519.2} & {183489867698309888} & {83.006942} & {36.588679} & {IR test} & & {-0.04} & {1.02} \\
{J053304.40+300057.1} & {3446048597423458432} & {83.268348} & {30.015879} & {Cross-matching} & {B[e] [3]} & {0.10} & {0.08} \\
{J053308.45+261924.4} & {3441911547485643136} & {83.285249} & {26.323461} & {IR test} & & {-0.05} & {1.13} \\
{J053309.04+291103.0} & {3445858794228638592} & {83.287708} & {29.184194} & {IR test} & & {0.03} & {0.81} \\
{J054443.01+263002.0} & {3441355396462179456} & {86.179232} & {26.500569} & {Cross-matching} & {B[e] [3]} & {0.07} & {-0.10} \\
{J054848.25+283547.7} & {3443142313313938816} & {87.201083} & {28.596611} & {Cross-matching} & {B[e] [3]} & {0.13} & {0.12} \\
{J055554.65+284706.3} & {3431561569657351936} & {88.977746} & {28.785103} & {IR test} & & {-0.01} & {2.03} \\
{J055838.98+201108.5} & {3422833371476299264} & {89.662417} & {20.185722} & {Cross-matching} & {HBe [1,3]; PMSc [2]} & {0.76} & {1.07} \\
{J060147.79+404022.4} & {3458471704427619456} & {90.44915} & {40.67291} & {Cross-matching} & {B[e] [3]} & {0.08} & {-0.01} \\
{J060414.76+240402.4} & {3426211861471165952} & {91.061522} & {24.067347} & {Cross-matching} & {HBe [3]} & {0.00} & {1.02} \\
{J062029.71+040604.6} & {3125446193770750976} & {95.123809} & {4.101283} & {Cross-matching} & {HBe [3]} & {-0.02} & {0.91} \\
{J062531.59+200550.4} & {3372586102405306496} & {96.381646} & {20.097346} & {Cross-matching} & {Evolved [3]} & {0.15} & {0.54} \\
{J064056.90+115646.4} & {3352022383108018176} & {100.23712} & {11.946228} & {Cross-matching} & {HII [1]} & {0.14} & {0.22} \\
{J064409.32+112626.8} & {3351184898843187072} & {101.03887} & {11.440784} & {Cross-matching} & {HII [1]} & {0.03} & {-0.03} \\
{J065910.35-003708.4} & {3112643823976115072} & {104.793129} & {-0.619007} & {IR test} & & {0.23} & {3.74} \\
{J074409.19+020131.1} & {3088632444230894592} & {116.03831} & {2.025315} & {IR test} & & {0.05} & {2.36} \\
{J213341.54+415208.5} & {1967510469468468096} & {323.4231} & {41.86905} & {Cross-matching} & {Evolved [3]} & {0.18} & {0.37} \\
{J214356.11+390537.4} & {1954023035205245312} & {325.983794} & {39.093748} & {Cross-matching} & {HII [1]} & {--} & {--} \\
{J224515.96+563739.7} & {2006972762119220480} & {341.316504} & {56.627719} & {Cross-matching} & {HBe [3]} & {0.05} & {1.01} \\
\enddata
     \begin{tablenotes}
        \item[1] {\textbf{Notes.} HII: spectra contaminated by H II regions. PMSc: pre-main-sequence candidates.} 
        \item[2] {\textbf{Reference:} [1] \citet{2016RAA....16..138H}; [2] \citet{Vioque2020}; [3] \citet{2021RAA....21..288S}.}   
        \item[3] {(This table is available in machine-readable form in the \href{https://doi.org/10.3847/1538-4365/ae450d}{online article}.)} 
     \end{tablenotes}
\end{deluxetable*}

\begin{deluxetable*}{ccccccc}
\tablewidth{0pt}
\tablecaption{Spectral match results of sources with spectra of high SNR \label{tab:spectral_match}}
\tablehead{
\colhead{LAMOST} &
\colhead{\emph{Gaia} DR3} & \colhead{SNR} & \colhead{Best-matched} & \colhead{Spectral}  & \colhead{Luminosity} & \colhead{\( \chi^2 \)/$N$} \\
\colhead{designation} & \colhead{ID} & \colhead{($g$ band)} & \colhead{standard star} & \colhead{type} & \colhead{class} & \colhead{}}
\startdata
J000003.86+635429.6 & 431630851621447552 & 207.76 & HD 144470 & B1 & V & 7.53\\
J000305.76+632002.1 & 431579342069149184 & 80.78 & HD 35299 & B1.5 & V & 7.95\\
J000336.71+635514.3 & 431622845802749696 & 161.67 & HD 34503 & B6 & III & 5.04\\
J000352.84+573550.0 & 422602452401685120 & 165.31 & HD 207330 & B2.5 & III & 1.61\\
J000429.45+644140.6 & 432136141624375808 & 156.31 & HD 21483 & B3 & III & 6.21\\
... & ... & ... & ... & ... & ... \\
\enddata
     \begin{tablenotes}
        \item[1] {(This table is available in its entirety in machine-readable form in the \href{https://doi.org/10.3847/1538-4365/ae450d}{online article}.)} 
    \end{tablenotes}
\end{deluxetable*}

\section{CBe star binaries}\label{sec4}
\subsection{Identified CBe star binaries}\label{sec:4.1}
\begin{figure*}[ht!]
\plotone{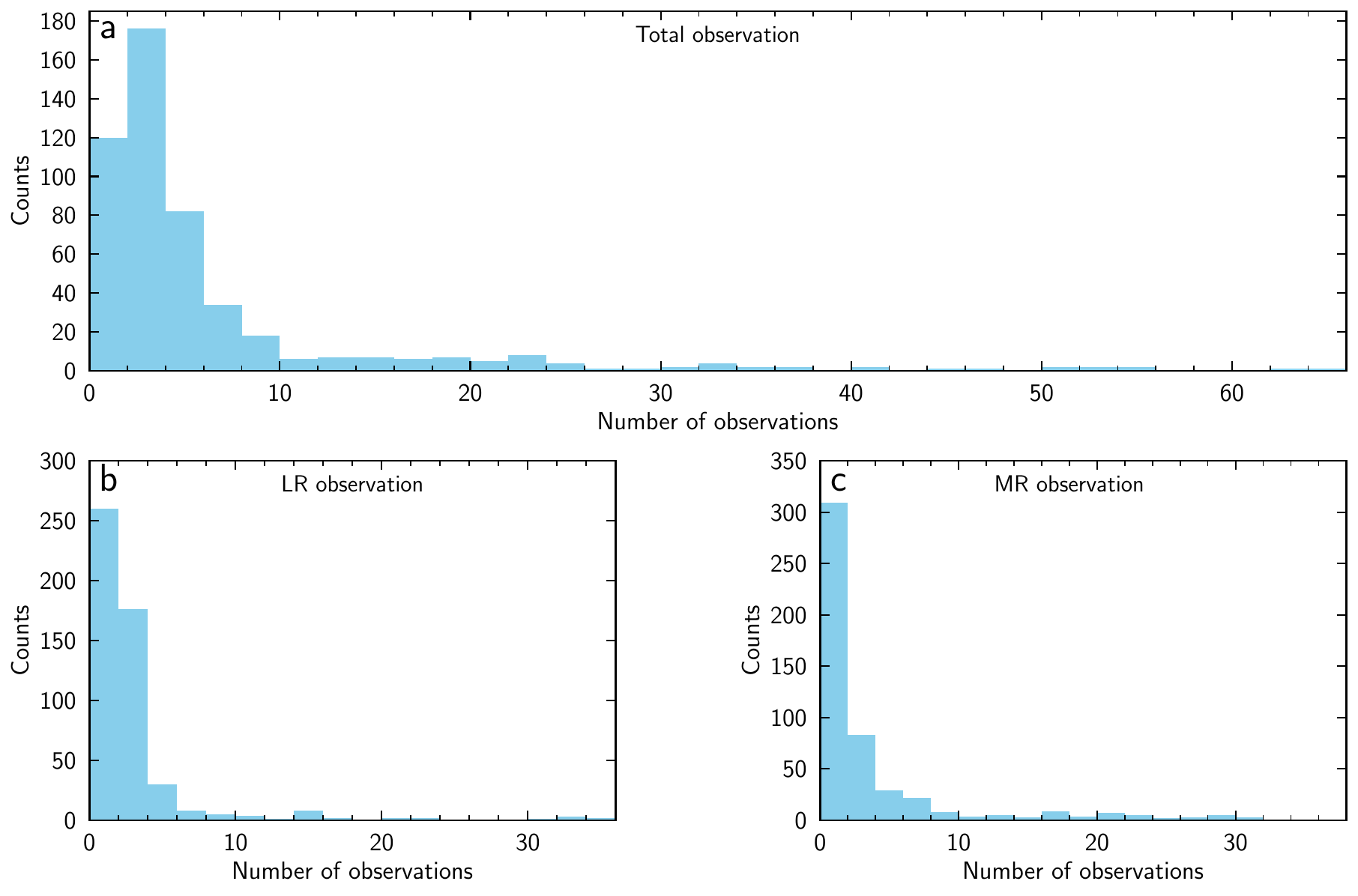}
\caption{(a) Distribution of the total number of observations of the 504 CBe stars in LAMOST. (b) Distribution of the number of observations of the 504 CBe stars
in the low-resolution survey of LAMOST. (c) Distribution of the number of observations of the 504 CBe stars in the medium-resolution survey of LAMOST.
\label{fig:obs_counts}}
\end{figure*}

Each of the {504} CBe stars in our sample has been observed by LAMOST an average of 6.2 times, enabling monitoring of their RV variations. Figure \ref{fig:obs_counts} shows the distribution of the number of observations for these sources. To identify CBe stars with significant RV variations, we preselect 15 sources exhibiting visible spectral line shifts through visual inspection. We then apply the criterion from \citet{2013A&A...550A.107S} to solidify the result—a source is classified as RV variable if at least one pair of its RV measurements simultaneously satisfies the following equations:
\begin{equation}\label{eq1}
\sigma_\mathrm{detect} = \frac{|v_i - v_j|}{\sqrt{\sigma_i^2+\sigma_j^2}}  >  4.0  \hspace*{3mm}\mathrm{and}\hspace*{3mm}|v_i - v_j|> C, 
\end{equation} 
where $v_i$ and $\sigma_i$ are the RVs and their 1 $\sigma$ errors at epoch $i$, and $C$ is the minimum amplitude threshold. \citet{2013A&A...550A.107S} adopted $C = 20$ km s$^{-1}$. These two criteria have been carried forward by many following studies on massive-star multiplicity \citep[e.g.,][]{2021A&A...652A..70B, 2022A&A...658A..69B, 2025arXiv251108675V}. All of these preselected 15 sources satisfy the second criterion in Eq. \ref{eq1}. We measure RVs of these preselected sources mainly using He I 6678 absorption line. For those with a He I 6678 absorption line unavailable (may be due to low SNR or high variation), the He I 4922 absorption line or the wing of the H$\alpha$ emission line (the part below half of the maximum height of the H$\alpha$ emission line) could be a substitute. We employ Gauss-like model to fit the profiles of selected spectral lines to determine their centers and convert the shifts of the centers from the wavelength of the specific lines under a rest frame into RVs by the following equations:
\begin{equation}
F_{\rm \lambda} = h \pm \frac{A}{\sqrt{2 \pi} \sigma}\exp^\frac{-{(\lambda - {\lambda}_{\rm c})} ^ 2}{2 \sigma ^ 2}~(\mathrm{``+"~for~ H\alpha~emission~line~and~``-"~for~ absorption~line}) \hspace*{3mm}\mathrm{and}\hspace*{3mm}RV = \frac{\lambda_\mathrm{c} - \lambda_\mathrm{rest}}{\lambda_\mathrm{rest}}c,   
\end{equation}
where $F_{\rm \lambda}$ denotes the normalized flux; $h$ is approximately 1, representing the local continuum; $\lambda_{\rm c}$ is the central wavelength; $A$ is a factor to adjust the contour longitudinally; $\sigma$ is the standard deviation; $\lambda_{\rm rest}$ is the wavelength of the specific absorption lines under a rest frame; and $c$ is the speed of light in vacuum. We execute \texttt{emcee} \citep{2013PASP..125..306F} with 24 walkers, taking 10,000 steps per walker to sample the parameter space. The initial 5000 steps of each walker are regarded as the burn-in and consequently discarded. For each parameter, we take the 50th percentile, 15.87th percentile and 84.13th percentile of the posterior distribution as the best-fit value, the $1\sigma$ lower limit, and the $1\sigma$ upper limit, respectively. Particularly, LAMOST J045300.92+291405.2 displays a distinctive spectral feature: prominent Fe and Mg absorption lines where emission lines typically appear for CBe stars, particularly among early-type ones, and we measure its RVs by the cross-correlation function technique \citep{1979AJ.....84.1511T} using the blue arms of LAMOST/MRS spectra.
\par
\begin{figure*}[ht!]
\plotone{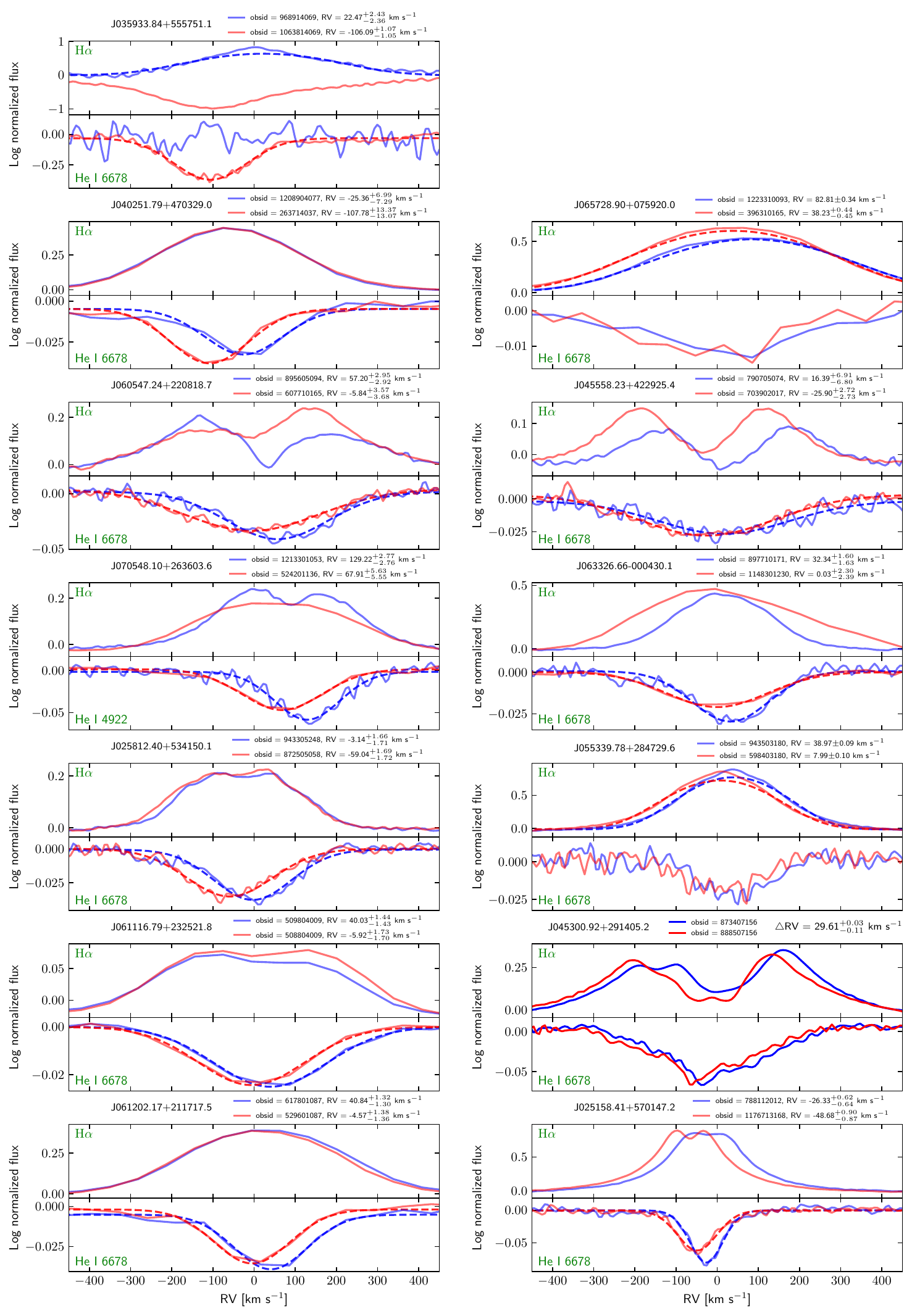}
\caption{The two spectra with maximum RV separation for each 13 newly identified CBe star binaries. Spectra are plotted by light solid lines, while model lines are plotted by dark dashed lines. To strengthen the visual contrast, we plot the logarithm of all the fluxes and the spectrum with obsid of 1063814069 (belongs to J035933.84+555751.1) is amplified by a factor of 7.
\label{fig:show_rv_diff}}
\end{figure*}

 Among the preselected 15 sources, one source, LAMOST J010906.43+562643.1, fails to satisfy the first criterion in Eq. \ref{eq1} and is therefore excluded from the final sample. Of the identified 14 RV-variable systems: Five are newly identified CBe stars in this study, and one, ALS\,8814, was previously reported as a CBe star + black hole (BH) binary candidate \citep{2025arXiv250523151A}, but \citet{2025OJAp....8E.128E} argued that the companion in ALS\,8814 is a visible star by their spectral disentangling. For each of the 13 newly identified CBe star binaries, we list them in Table \ref{tab:CBe star binaries}. Figure \ref{fig:show_rv_diff} displays two spectra corresponding to the epochs of maximum RV separation for each newly identified CBe star binary.

\begin{deluxetable*}{cccccccccc}
\renewcommand{\arraystretch}{1.1}
\tabletypesize{\footnotesize}
\setlength{\tabcolsep}{3pt}
\tablewidth{0pt}
\tablecaption{13 newly identified CBe star binaries from LAMOST DR12 \label{tab:CBe star binaries}}
\tablehead{
\colhead{LAMOST}  &  \colhead{\emph{Gaia} DR3} & \colhead{R.A.}  &  \colhead{Decl.}  & \colhead{N} & \colhead{$\triangle$RV$^{\mathrm{\textcolor{blue}{a}}}$} & \colhead{$\sigma_\mathrm{detect}$} & \colhead{RUWE} & \colhead{New} & \colhead{Notes} \\
\colhead{designation}  & \colhead{ID} & \colhead{(deg)} & \colhead{(deg)} & \colhead{} & \colhead{} & \colhead{} & \colhead{} & \colhead{CBe star?} & \colhead{}}
\startdata
J035933.84+555751.1 & 468716416747466752 & 59.891003 & 55.964199 & 12& 128.56 & 49.61 & 10.32 & Yes & \\
J040251.79+470329.0 & 245895639429336064 & 60.715833	& 47.058073 & 8 & 82.42 & 5.41 & 1.23 & No & \makecell[c]{H$\alpha$ emission line\\seems stationary} \\
J060547.24+220818.7 & 3423721944373852672 & 91.446842	& 22.13855 & 47 & 63.04 & 13.67 & 0.97 & No & \makecell[c]{Astrometric binary; \\H$\alpha$ emission line\\seems stationary} \\
J070548.10+263603.6 & 883218724017482880 & 106.450446 & 26.601008 & 5 & 61.31 & 9.78 & 0.96 & No & \\
J025812.40+534150.1 & 441310466646282880 & 44.551677 & 53.697273 & 28 & 55.90 & 23.25 & 0.96 & No & \\
J061116.79+232521.8 & 3425411417005784448 & 92.81998 & 23.422722 & 62 & 45.95 & 20.47 & 1.15 & No & \\
J061202.17+211717.5 & 3375360788715959552 & 93.009069 & 21.288215 & 45 & 45.41 & 23.95 & 10.80 & No & \\
J065728.90+075920.0 & 3157092612318301696 & 104.370421 & 7.988902 & 5 & 44.58 & 80.17 & 0.91 & Yes & \\
J045558.23+422925.4 & 204958970401866624 & 73.99264 & 42.490414 & 4 & 42.29 & 5.77 & 1.21 & No & \\
J063326.66-000430.1 & 3119304940291183744 & 98.361086 & -0.075046 & 3 & 32.21 & 11.43 & 0.98 & Yes & \\
J055339.78+284729.6 & 3443195399110392192 & 88.415758 & 28.791573 & 31 & 30.98 & 230.27 & 1.82 & No & \\
J045300.92+291405.2 & 155287742339368448 & 73.253872 & 29.234795 & 8 & 29.61 & 987.00 & 1.13 & Yes & Astrometric binary \\
J025158.41+570147.2 & 460613256564267008 & 42.993397 & 57.029804 & 11 & 22.35 & 20.24 & 3.39 & No & \\
\enddata
     \begin{tablenotes}
        \item[1] \textbf{Note.}
        \item[2] $^\mathrm{a}$ Maximum RV separation observed in LAMOST spectra (km s$^{-1}$).  
        \item[3] (This table is available in machine-readable form in the \href{https://doi.org/10.3847/1538-4365/ae450d}{online article}.)
     \end{tablenotes}
\end{deluxetable*}

\subsubsection{Two astrometric binaries in \emph{Gaia} DR3}
Within our sample of 13 newly identified CBe star binaries, two systems are classified as astrometric binaries (\texttt{non\_single\_star} != 0) in Gaia DR3 \citep{2023A&A...674A...1G}. One source, LAMOST J045300.92+291405.2, whose \texttt{nss$_\mathtt{solution}$} type is single-line spectroscopic binary (SB1), has an astrometric published orbital solution in the \texttt{gaiadr3.nss\_two\_body\_orbit} catalog, with an orbital period of 67.82 $\pm$ 0.56 days, a RV semiamplitude of 14.67 $\pm$ 2.81 km s$^{-1}$ and an orbital eccentricity of 0.34 $\pm$ 0.13. The RV semiamplitude reported in \texttt{gaiadr3.nss\_two\_body\_orbit} catalog is consistent with that observed in the available LAMOST spectra (see Figure \ref{fig:show_rv_diff}). The other source, LAMOST J060547.24+220818.7, has a \texttt{nss$_\mathtt{solution}$} type of First Degree Trend SB1. While no orbital solution is available, \emph{Gaia} DR3 reports a \texttt{rv\_amplitude\_robust} (total amplitude in the RV time series after outlier removal) of 125.99 km s$^{-1}$, exceeding the maximum RV separation observed in the available LAMOST spectra.
\subsubsection{Other 11 newly identified CBe star binaries}
The other 11 newly identified CBe star binaries show maximum RV separations ranging from $\sim$20 to $\sim$130 km s$^{-1}$ in available LAMOST spectra. 
\par
\begin{figure*}[ht!]
\plotone{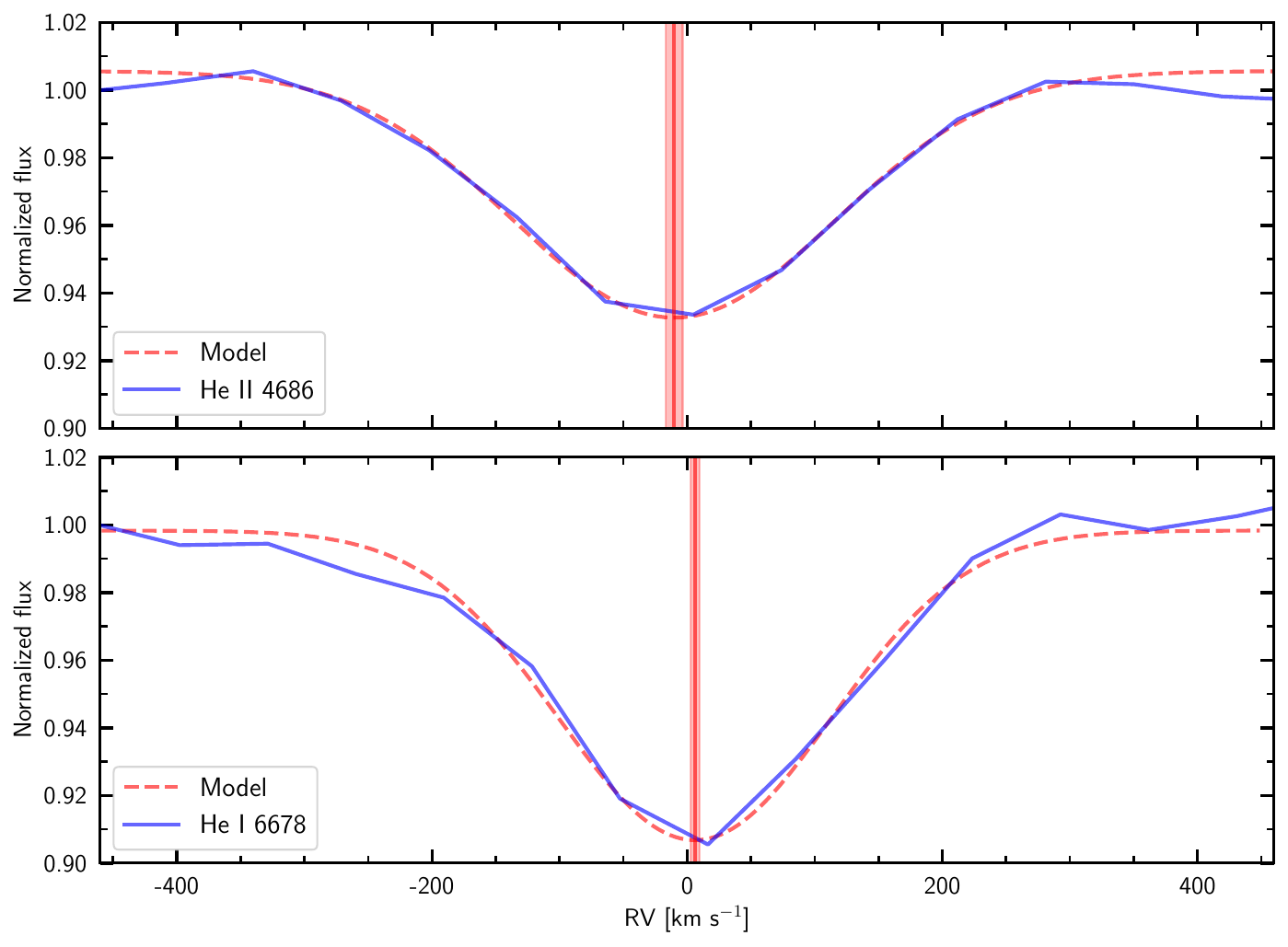}
\caption{The contrast of RVs measured by the He II 4686 line and He I 6678 line of LAMOST 035933.84+555751.1. We convert wavelengths into RV space, and the RV zero points for He II 4686 line and He I 6678 line are 4686.98 $\mathrm{\AA}$ and 6680 $\mathrm{\AA}$, respectively. In both panels, we use solid blue lines and red dashed lines to represent spectra and best-fit model lines, respectively. The vertical red lines show the measured RVs, and the light red shaded areas reflect the 1$\sigma$ uncertainties.  
\label{fig:J0359}}
\end{figure*}

 Among these systems, LAMOST J035933.84+555751.1 stands out with a maximum RV separation of $\sim$130 km s$^{-1}$, higher than that of any known CBe star binaries. Its sole LAMOST/LRS spectrum reveals a He II 4686 absorption line, indicative of a hot component ($T_{\rm eff}>30,000$ K). It is meaningful to discuss whether it originates from the CBe star itself or a potential hot companion. 
We measure the RVs of He II 4686 absorption line and He I 6678 absorption line, and obtain $-10.08^{+6.59}_{-6.65}$ km s$^{-1}$ and $6.08^{+3.71}_{-3.73}$ km s$^{-1}$, respectively (see Figure \ref{fig:J0359}). The $\sigma_\mathrm{detect}$ of this set of measurements is 2.13, which is lower than the threshold for significant differences defined by Eq. \ref{eq1}. However, even if there is no significant RV difference between these two lines, it is not enough to definitively establish a shared stellar origin of these two lines, as the single-epoch spectrum may be obtained (nearly) at the superior conjunction or inferior conjunction. Unfortunately, LAMOST/MRS spectra lack He II 4686 coverage, making it unable to chronically track its movement in available data.
\par
\begin{figure*}[ht!]
\plotone{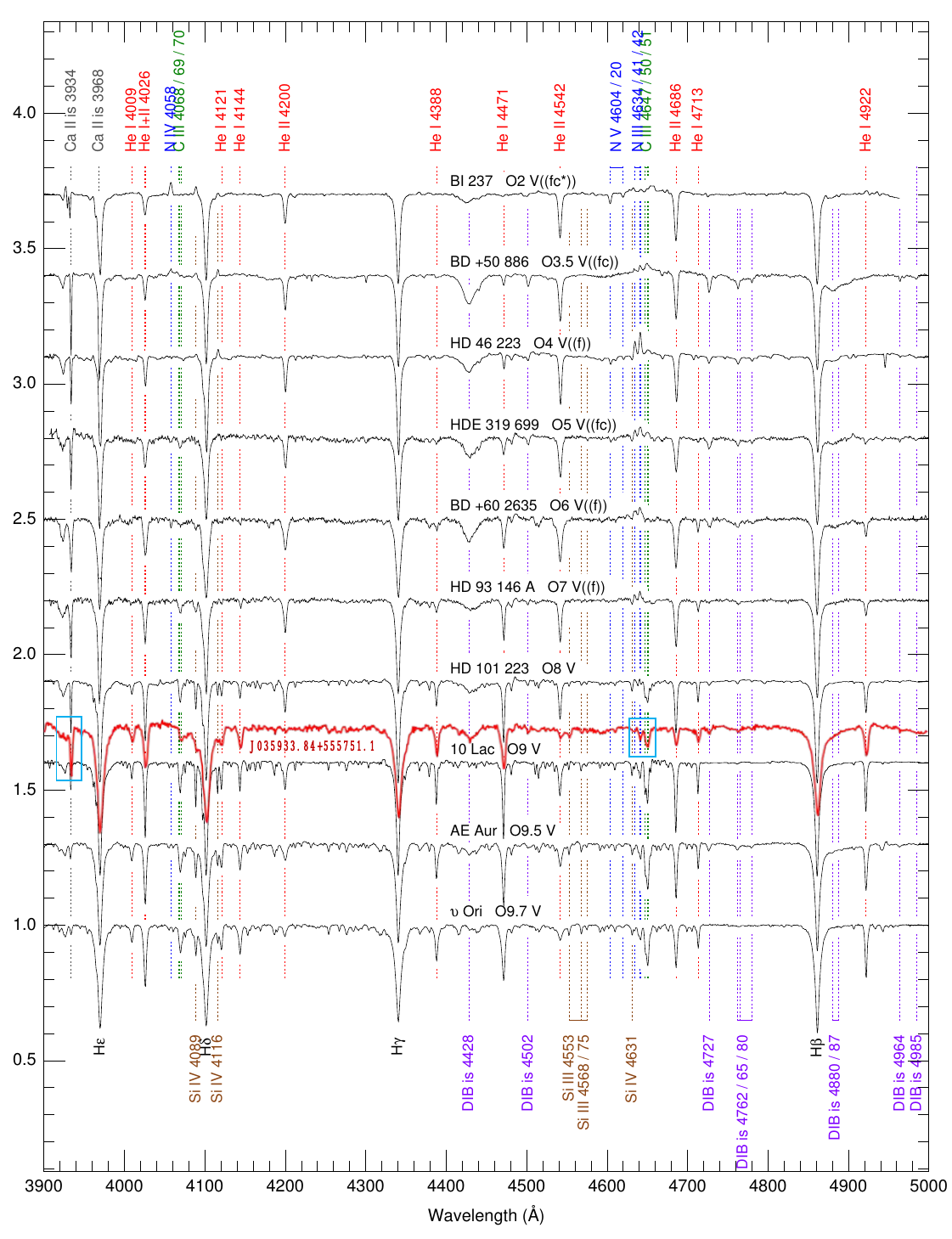}
\caption{The contrast of the spectral type sequence for O stars and LAMOST J035933.84+555751.1. The background figure is taken from \citet{2026enap....2...43M}. The resolutions of the spectra in the background figure are $\sim$2500. The LAMOST/LRS spectrum (resolution: $\sim$1800) of LAMOST J035933.84+555751.1 is plotted by a red line, and the key metal spectral lines to identify the spectral type are highlighted by two blue squares.  
\label{fig:J0359_classfication}}
\end{figure*}

 We also attempt to identify its spectral type by matching spectral characteristics using stellar observational spectra in \citet{2026enap....2...43M}. The strong Ca II 3934 line indicates that the spectral type of LAMOST J035933.84+555751.1 is not later than O9, while the nonweak He I lines and the absence of N III 4634/41/42 emission lines excludes a spectral type of earlier than or equal to O7 (see Figure \ref{fig:J0359_classfication}). Though B-type (super)giants also exhibit the above features, they also exhibit strong Si III 4553/4568/4575 absorption lines, inconsistent with those of LAMOST J035933.84+555751.1. Therefore, we identify LAMOST J035933.84+555751.1 as an O8-O9 star.
\par In summary, we incline to the view that LAMOST J035933.84+555751.1 is actually an Oe star with spectral type of O8-O9, and the He II 4686 absorption line originates from the Oe star itself. Besides, LAMOST J035933.84+555751.1 is a source warranting follow-up observations because of its huge \texttt{RUWE} (see Table \ref{tab:CBe star binaries}), which may hint a wide orbit and consequently a long orbital period. Significant RV variation and long orbital period will yield a high mass function, indicative of a heavy companion (possibly {neutron star} or BH).   

\subsection{Other potential CBe star binaries}
\texttt{RUWE} provided by \emph{Gaia} is an excellent indicator of wide binary systems, particularly when long-term RV monitoring is unavailable. \texttt{RUWE} quantifies the deviation of a source's astrometric observations from the best-fitting single-star model. For sources with well-behaved astrometric solutions, \texttt{RUWE} is expected to be close to $\sim$1 (see \url{https://dms.cosmos.esa.int/COSMOS/doc_fetch.php?id=3757412}). For a star, a common scenario resulting in a \texttt{RUWE} significantly greater than 1 (e.g.,, \texttt{RUWE} $>$ 1.4) is the presence of an unresolved, dim companion orbiting at a wide separation. The orbital motion induced by the companion, which is not accounted for in the single-star model, causes the observed motion to be poorly fitted. \texttt{RUWE} exhibits a specific dependence on the orbital period and the projected semimajor axis. It initially increases with the orbital period and the projected semimajor axis. However, for orbital periods exceeding approximately $\sim$1000 days, \texttt{RUWE} begins to decrease due to the finite 5 year baseline of \emph{Gaia} observations, which is insufficient to resolve very long-period orbits fully. Based on this theory, \citet{2024A&A...686L...2G} identified a binary comprising a G9/K0-type star and a 33 $M_\odot$ BH in an 11.6 year orbit while validating the preliminary \emph{Gaia} astrometric binary solutions. Furthermore, \citet{2025PASP..137i4202N} initiated a large-scale search for dormant BHs in wide binaries by combining spectroscopic follow-up on a sample of low-metallicity stars with acceleration solutions or high \texttt{RUWE}, yielding several promising sources. Thus it can be seen that high \texttt{RUWE} is indeed a wonderful indicator of wide binaries in practice.
\par Of the CBe stars in our sample, 60 exhibit a \texttt{RUWE} exceeding 1.4, and 38 have a \texttt{RUWE} greater than 2.0. Among these, five and four sources, respectively, show RV variations reported in Section \ref{sec:4.1}. In addition, \emph{Gaia} DR3 also reported four RV-variable sources and three RV-variable sources, respectively, for these two \texttt{RUWE}-threshold standards. None of the four sources exhibit RV variations in available LAMOST spectra, but this could also be due to too few observations by LAMOST or false alarms by \emph{Gaia} (RVs measured by \emph{Gaia} are more reliable for cold stars). 
\par However, high \texttt{RUWE} can also be caused by insufficient observation coverage, extreme crowding, background issues (such as nebulosity), magnitude perturbation, among others.
\par As noted in the \emph{Gaia} documentation\footnote{\url{https://gea.esac.esa.int/archive/documentation/GDR3/index.html}}, insufficient observation coverage can lead to spurious solutions, as a too small
degree of freedom may have given a reasonable solution by chance only. To mitigate this effect, \emph{Gaia} recommends requiring \texttt{visibility\_periods\_used $>$ 11}. All 60 sources in our sample satisfy this criterion; therefore, their high \texttt{RUWE} values are unlikely to be caused by insufficient observational coverage.
\par The second and third factors may lead to a regional increase in \texttt{RUWE}. To examine whether the elevated \texttt{RUWE} values of these sources could be attributed to extreme crowding or background-related effects, we compare their \texttt{RUWE} values with those of neighboring sources within an angular distance of $\leq$1$^{\prime}$. The neighboring sources are required to satisfy \texttt{visibility\_periods\_used > 11} and \texttt{astrometric\_params\_solved = 31 or 95} (corresponding to five-parameter and six-parameter astrometric solutions, respectively). We find that the \texttt{RUWE} values of most targets lie at the 100th percentile of the local \texttt{RUWE} distribution, i.e., most of them have the highest \texttt{RUWE} values among their neighbors. In addition, we quantify the local significance of the \texttt{RUWE} excess by defining a \texttt{RUWE} significance parameter:
\begin{equation}
\mathrm{RUWE}\text{\_}\mathrm{sig} = \frac{\mathrm{RUWE}_{*} - \mathrm{median}(\mathrm{RUWE}_\mathrm{neighbors})}{1.4826 \times \mathrm{MAD}(\mathrm{RUWE}_\mathrm{neighbors})},   \end{equation}
where RUWE$_{*}$ is the \texttt{RUWE} value of the target source, RUWE$_\mathrm{neighbors}$ denotes the \texttt{RUWE} values of the neighboring sources, ``median'' means taking the median, and ``MAD'' means taking the median absolute deviation. The factor 1.4826 rescales the \texttt{RUWE} significance parameter to be consistent with the standard deviation for a Gaussian distribution. The RUWE$_\mathrm{sig}$ values span the range 7.5–546.4, with a median of 28.9, indicating that these sources deviate by many robust standard deviations from the local \texttt{RUWE distribution}. This demonstrates that their elevated \texttt{RUWE} values are not mild statistical fluctuations but instead represent genuine local outliers. Generally, high \texttt{RUWE} values of these sources are unlikely to be caused by extreme crowding or background issues.
\par Magnitude perturbations can displace the centroid of the point‐spread function and thus increase \texttt{RUWE}. For CBe stars, this effect may be enhanced by their large, variable circumstellar disks. Although these disks can extend to many stellar radii, the resulting photocenter excursions still remain much smaller than the spatial scale sampled by \emph{Gaia}’s image–parameter–determination (IPD) process. In other words, the perturbations are extended relative to the stellar photosphere but remain effectively unresolved at the IPD level. Consequently, such magnitude perturbations generally do not generate multiple peaks in the IPD profile and therefore are not expected to produce high values of \texttt{ipd\_frac\_multi\_peak}. By contrast, significantly elevated \texttt{ipd\_frac\_multi\_peak} values most plausibly reflect genuine image blending, arising either from relatively wide companions at separations of several to tens of milliarcseconds, or coincidentally from unrelated close neighbors along the line of sight. Sources with high \texttt{ipd\_frac\_multi\_peak} are thus plausible candidates for sources with distant companions, whereas for sources with low \texttt{ipd\_frac\_multi\_peak}, we cannot distinguish whether the elevated \texttt{RUWE} is caused by magnitude perturbations or by unresolved close companions.
\par In summary, high \texttt{RUWE} values of these sources are unlikely caused by insufficient observation coverage, extreme crowding or background issues. We recommend to preferably attend sources with high \texttt{ipd\_frac\_multi\_peak} values when attempting to search for companions among these sources. These potential CBe star binaries are listed in Table \ref{tab:potential CBe star binaries}.

\startlongtable
\begin{deluxetable*}{cccccccccc}
\renewcommand{\arraystretch}{1.1}
\tabletypesize{\footnotesize}
\setlength{\tabcolsep}{4pt}
\tablewidth{0pt}
\tablecaption{Potential CBe star binaries from LAMOST DR12 \label{tab:potential CBe star binaries}}
\tablehead{
\colhead{LAMOST}  &  \colhead{\emph{Gaia} DR3} & \colhead{N}  &  \colhead{\texttt{visibility}}  & \colhead{RUWE} & \colhead{Percentage$^{\mathrm{\textcolor{blue}{a}}}$} & \colhead{RUWE} & \colhead{\texttt{ipd\_frac}} & \colhead{$\triangle$RV$^{\mathrm{\textcolor{blue}{b}}}$} & \colhead{New} \\
\colhead{designation}  & \colhead{ID} & \colhead{} & \colhead{\texttt{\_periods\_used}} & \colhead{} & \colhead{(\%)} & \colhead{\_sig} & \colhead{\texttt{\_multi\_peak} (\%)} & \colhead{} & \colhead{CBe star?}}
\startdata
J040636.03+525022.5 & 275098088451178240 & 2 & 21 & 23.58 & 100.0 & 546.4 & 19 & & No \\
J054415.00+322732.6 & 3448185944950896384 & 2 & 17 & 20.80 & 100.0 & 439.6 & 57 & & Yes \\
J061434.56+225434.2 & 3425138463244391680 & 37 & 17 & 18.58 & 100.0 & 270.8 & 0 & & Yes \\
J053551.06+371645.7 & 189495713746035328 & 1 & 15 & 18.04 & 100.0 & 275.1 & 44 & & No \\
J051554.95+425532.8 & 207456785941426048 & 3 & 16 & 16.86 & 100.0 & 322.1 & 32 & & No \\
J055126.83+170829.6 & 3349946127197206016 & 1 & 16 & 16.59 & 100.0 & 220.9 & 2 & & No \\
J055336.52+401846.2 & 191595540436839296 & 4 & 17 & 14.35 & 100.0 & 256.3 & 46 & & No \\
J042105.03+502714.1 & 270720004945832064 & 4 & 19 & 12.84 & 100.0 & 166.8 & 0 & & No \\
J055052.59+354409.8 & 3455191517646066560 & 2 & 18 & 12.45 & 100.0 & 216.1 & 93 & & No \\
J061202.17+211717.5 & 3375360788715959552 & 45 & 15 & 10.80 & 100.0 & 147.5 & 0 & & No \\
J035933.84+555751.1 & 468716416747466752 & 12 & 22 & 10.32 & 100.0 & 156.7 & 56 & & Yes \\
J040151.09+545511.8 & 468403674405344256 & 3 & 23 & 9.91 & 100.0 & 188.1 & 0 & & No \\
J041750.14+532526.1 & 275370698611155584 & 4 & 21 & 9.68 & 100.0 & 222.4 & 0 & & No \\
J014620.23+611421.5 & 511217729481951616 & 1 & 27 & 9.49 & 98.9 & 158.6 & 0 & & No \\
J060806.15+130723.9 & 3344236736973246592 & 3 & 17 & 7.79 & 100.0 & 205.5 & 0 & & No \\
J035920.13+505620.1 & 250810048388652928 & 3 & 21 & 5.75 & 100.0 & 174.7 & 0 & & Yes \\
J054837.63+281710.2 & 3443128019662992128 & 5 & 17 & 5.51 & 100.0 & 46.8 & 28 & & No \\
J015645.74+635259.7 & 517984879952478592 & 1 & 30 & 5.34 & 100.0 & 128.2 & 0 & & No \\
J041238.36+463615.3 & 233962369156929920 & 4 & 17 & 5.17 & 100.0 & 84.5 & 20 & 123.19018 & No \\
J070724.42+035254.9 & 3116370751776866560 & 3 & 15 & 4.88 & 100.0 & 40.3 & 0 & 198.60501 & Yes \\
J050535.04+442648.4 & 205544627842417408 & 2 & 16 & 4.07 & 100.0 & 64.9 & 0 & & No \\
J040839.73+474244.5 & 246137776809763712 & 6 & 15 & 3.86 & 100.0 & 44.7 & 1 & & No \\
J064707.66+044232.0 & 3129237374282787584 & 10 & 16 & 3.79 & 100.0 & 60.1 & 3 & & No \\
J062108.40+044846.0 & 3317666596109523328 & 2 & 17 & 3.71 & 100.0 & 46.5 & 38 & & No \\
J044345.35+243908.8 & 147294464602944256 & 16 & 17 & 3.58 & 100.0 & 167.4 & 0 & & Yes \\
J025158.41+570147.2 & 460613256564267008 & 11 & 19 & 3.39 & 100.0 & 30.6 & 0 & & No \\
J204151.60+482727.9 & 2167790566919245568 & 1 & 29 & 2.90 & 100.0 & 41.2 & 0 & & No \\
J035622.12+484550.2 & 247082394733516032 & 6 & 21 & 2.89 & 100.0 & 52.5 & 48 & & No \\
J063752.74+083237.2 & 3326115827571757184 & 1 & 14 & 2.75 & 100.0 & 24.5 & 0 & & Yes \\
J065553.98+061214.4 & 3129869048013753344 & 2 & 14 & 2.52 & 100.0 & 25.9 & 0 & & No \\
J035101.91+493720.1 & 250284825427868800 & 5 & 20 & 2.38 & 100.0 & 24.0 & 45 & & No \\
J055546.16+305126.4 & 3444050814860576768 & 14 & 16 & 2.38 & 100.0 & 25.7 & 0 & & No \\
J063159.26-004955.7 & 3119142929834761600 & 1 & 16 & 2.34 & 100.0 & 16.1 & 38 & & Yes \\
J054154.33+360647.9 & 3455879949363891584 & 2 & 17 & 2.13 & 100.0 & 38.0 & 0 & & Yes \\
J035814.15+521024.2 & 251702194696320768 & 7 & 21 & 2.12 & 100.0 & 27.4 & 2 & & No \\
J034512.81+521437.7 & 443762965990465920 & 5 & 21 & 2.09 & 97.0 & 17.2 & 1 & & Yes \\
J040818.93+552455.0 & 276392351073731712 & 3 & 23 & 2.06 & 97.4 & 25.0 & 0 & & No \\
J043346.95+324528.0 & 172036564308761728 & 2 & 15 & 2.05 & 100.0 & 30.4 & 0 & & No \\ 
... & ... & ... & ... & ... & ... & ... & ... & ... & ... \\  
\enddata
     \begin{tablenotes}
        \item[1] \textbf{Notes.}
        \item[2] $^{\mathrm{a}}$: The percentile of CBe star's \texttt{RUWE} in the local \texttt{RUWE} distribution.  
        \item[3] $^{\mathrm{b}}$: Maximum RV separation published by \emph{Gaia} DR3 (km s$^{-1}$). 
        \item[4] (This table is available in its entirety in machine-readable form in the \href{https://iopscience.iop.org/article/10.3847/1538-4365/ae450d}{online article}.)         
     \end{tablenotes}
\end{deluxetable*}

\section{Spatial distribution, cluster membership and runaway stars}\label{sec5}
\begin{figure*}[ht!]
\plotone{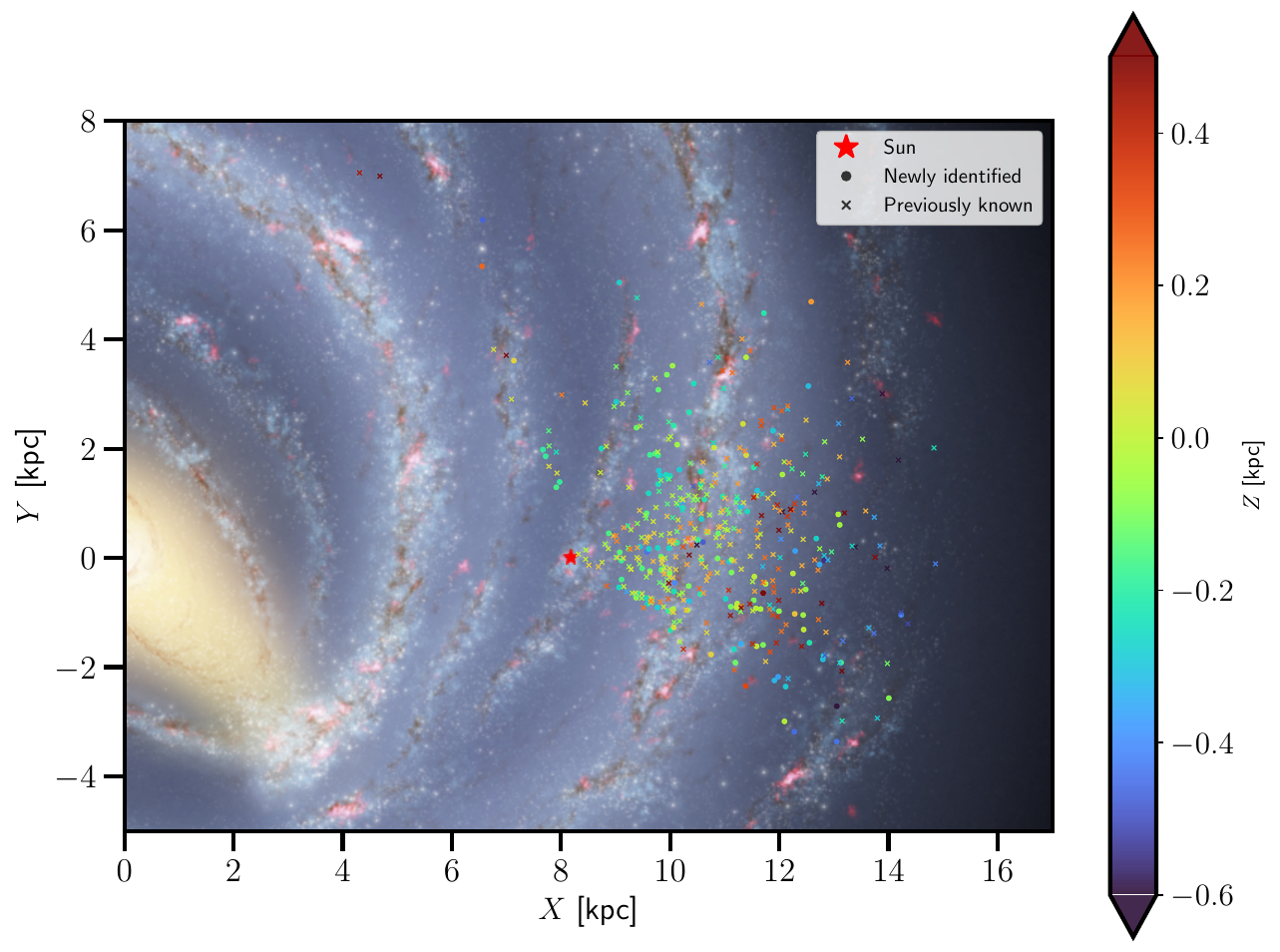}
\caption{Spatial distribution of CBe stars. The background image comes from NASA/JPL-Caltech, created by R. Hurt (SSC/Caltech) and loaded by \texttt{mw\_plot}. CBe stars newly identified by this work are marked by circles, while previously known CBe stars are marked by diagonal crosses. The heights of CBe stars above the Galactic plane are indicated by the colors of respective marks. Additionally, we mark the Sun with a red pentagram.  
\label{fig:distribution}}
\end{figure*}

The spatial distribution of these CBe stars in the Milky Way is shown in Figure \ref{fig:distribution}. The same as Section \ref{sec3.4}, sources analyzed here, and subsequent studies in this section are with relative distance uncertainty not exceeding 20\%. These CBe stars are primarily distributed at low Galactic latitudes toward the Galactic anticenter—a spatial trend observed in both newly identified and previously known sources. This trend may be attributable to the survey plan of LAMOST and the severe interstellar extinction toward the Galactic center. In particular, we notice that two sources, LAMOST J193507.42+262713.5 and LAMOST J193559.39+285042.8, located at ($X\sim$4.5 kpc, $Y\sim$7 kpc), appear spatially isolated from the main concentration of the sample. They do not have widely used cross-identifications in \texttt{SIMBAD} and were previously reported only as part of survey-level samples \citep{2021RAA....21..288S} rather than being targets of dedicated studies. As discussed in Section \ref{sec3.4}, LAMOST J193507.42+262713.5 is classified as a B3 II star and lies above the MS in the H-R diagram, with a dereddened (BP – RP) = -0.26 and an absolute $M_\mathrm{G}$ = -3.69. LAMOST J193559.39+285042.8 is classified as a B3 IV star and lies on or near the MS in the H-R diagram, with a dereddened (BP – RP) = -0.30 and an absolute $M_\mathrm{G}$ = -1.64. Their locations in the H-R diagram are broadly consistent with their spectroscopically derived luminosity classes and do not show any obvious contradictions, supporting the conclusion that estimates of their distances are (roughly) reliable. Their apparent isolation in the $XY$ plane also reflects distance-dependent selection effects: at such a large distance, only intrinsically bright stars remain detectable and classifiable.

\par We also attempt to identify potential cluster members in our sample. Clusters provide an advantageous context for astrophysical studies, as they consist of coeval stars with similar distances, ages, and chemical compositions. Confirmed membership would thus offer strong constraints on fundamental parameters of the CBe stars residing in such environments. We utilize the cluster catalog from \citet{2023A&A...673A.114H}, which was complied by the largest blind search for star clusters to date based on \emph{Gaia} data, containing 7167 clusters. We crossmatch our sample of CBe stars with this catalog and identify CBe stars as potential cluster members according to the following three criteria:
\begin{enumerate}
    \item The angular separations between CBe stars and the densest points of clusters are less than the total angular radii of corresponding clusters,
    \item The distances to the CBe stars and clusters are consistent within 1$\sigma$,
    \item The proper motions in both R.A. and decl. between CBe stars and clusters are consistent within 0.5 mas yr$^{-1}$.
\end{enumerate}
We identify 34 CBe stars as potential cluster members. Among these, seven are newly identified CBe stars in this work, and five are associated with two potential clusters each. These potential members and their host clusters are listed in Table \ref{tab:potential cluster members}.
\par Another interesting aspect of CBe stars is that how many of them are runaway stars. Runaway stars are OB-type stars with large peculiar velocities, first identified by \citet{1954ApJ...119..625B}. According to prevailing theories, such stars can be produced either by the supernova explosion of a binary companion \citep{1957moas.book.....Z, 1961BAN....15..265B} or through dynamical ejection from dense young clusters via gravitational interactions \citep{1967BOTT....4...86P}. The study of runaway stars, therefore, provides valuable insights into the kinematics in supernova explosions and interactions between stars. To identify runaway stars within our sample, we calculate their peculiar velocities for sources with measured RVs. For a star in binary, this RV is actually the systemic velocity, which needs to be determined by solving the orbital parameters. For simplicity, we exclude RV-variable sources from calculating peculiar velocities. We also omit sources for which all available spectra had a SNR below 10. For the remaining objects, we use the spectrum with the highest SNR per source to measure their RVs. Ultimately, a total of 203 sources with RV uncertainties below 10 km s$^{-1}$ are retained for the peculiar velocity calculation.    
\par To calculate peculiar velocities, we first calculate the 3D velocities in Galactocentric Cartesian frame ($V_{\rm x}$, $V_{\rm y}$ and $V_{\rm z}$) under the Milky way potential model \texttt{McMillan17} \citep{2017MNRAS.465...76M}, the most used Milky way potential model, by the Python package \texttt{galpy}\footnote{\url{http://github.com/jobovy/galpy}} \citep{2015ApJS..216...29B}. The Sun is placed at (\emph{X}, \emph{Z}) = (-8.178, 0.025) kpc \citep{2016ARA&A..54..529B, 2019A&A...625L..10G}, with the circular velocity at the solar position of 233.2 km s$^{-1}$ according to \texttt{McMillan17} at such place. The peculiar velocities of the Sun relative to the local standard of rest are set to be ($U_\odot$, $V_\odot$, $W_\odot$) = (7.01, 10.13, 4.95) km s$^{-1}$ \citep{2015MNRAS.449..162H}. To determine the local rotational velocities of the Milky Way at a given distance to the Galactic Center, $V_{\rm c}$, for subsequent calculations, the $V_{\rm x}$, $V_{\rm y}$ and $V_{\rm z}$ we calculate are at the moment of Galactic plane crossing ($Z=0$), conducted by backward orbit integration in time of 1 Gyr. $V_{\rm x}$, $V_{\rm y}$ and $V_{\rm z}$ are then converted into 3D Galactic space velocity ($U,~V,~W$) by the following matrix transformation:

$$
\begin{bmatrix}
    -\cos\alpha & -\sin\alpha & 0 \\
    -\sin\alpha & \cos\alpha & 0 \\ 0 & 0 & 1    
\end{bmatrix}
\begin{bmatrix}
    V_{\rm x} \\
    V_{\rm y} \\
    V_{\rm z}
\end{bmatrix}
=
\begin{bmatrix}
    U \\
    V \\
    W
\end{bmatrix}
,$$
where $\alpha$ is the angle between the Galactic center-to-source vector and the positive $x$-axis in Galactocentric Cartesian frame. Ultimately, 3D peculiar velocities are derived as $U_\mathrm{pec}=U$, $V_\mathrm{pec}=V-V_\mathrm{c}$, $W_\mathrm{pec}=W$ and $V_\mathrm{pec, tot}=\sqrt{U_\mathrm{pec}^{2} + V_\mathrm{pec}^{2} + W_\mathrm{pec}^{2}}$. Applying the classical criterion for runaway stars—a total peculiar velocity exceeding 40 km s$^{-1}$ \citep[e.g.,][]{1961BAN....15..265B, 2001ApJ...555..364B}—we identify 37 runaway stars (13 newly identified CBe stars) with total peculiar velocities $V_\mathrm{pec, tot}$ ranging from $\sim$40 to $\sim$101 km s$^{-1}$. In addition to \texttt{McMillan17}, we also calculate peculiar velocities utilizing other three Milky way potential models, including \texttt{MWPotential2014} \citep{2015ApJS..216...29B}, \texttt{Cautun20} \citep{2020MNRAS.494.4291C}, and \texttt{Irrgang13I} \citep{2013A&A...549A.137I}, for reference. Peculiar velocities calculated by all the four models are listed in Table \ref{tab:runaway stars}.
\\
\startlongtable
\begin{deluxetable*}{cccccccccc}
\renewcommand{\arraystretch}{1.1}
\tabletypesize{\footnotesize}
\setlength{\tabcolsep}{1.4pt}
\tablewidth{0pt}
\tablecaption{Potential cluster members of CBe stars from LAMOST DR12. \label{tab:potential cluster members}}
\tablehead{
\colhead{LAMOST}  &  \colhead{\emph{Gaia} DR3} & \colhead{pmra(s)$^{\mathrm{\textcolor{blue}{a}}}$}  &  \colhead{pmdec(s)$^{\mathrm{\textcolor{blue}{b}}}$}  & \colhead{Dis.(s)$^{\mathrm{\textcolor{blue}{c}}}$} &
\colhead{New} & \colhead{Cluster} & \colhead{pmra(c)$^{\mathrm{\textcolor{blue}{d}}}$} & \colhead{pmdec(c)$^{\mathrm{\textcolor{blue}{e}}}$} & \colhead{Dis.(c)$^{\mathrm{\textcolor{blue}{f}}}$} \\
\colhead{designation}  & \colhead{ID} & \colhead{(mas yr$^{-1}$)} & \colhead{(mas yr$^{-1}$)} & \colhead{(kpc)} & \colhead{CBe star?} & \colhead{name} & \colhead{(mas yr$^{-1}$)} & \colhead{(mas yr$^{-1}$)} & \colhead{(kpc)}}
\startdata
J000003.86+635429.6 & 431630851621447552 & -1.97$\pm$0.01 & -0.55$\pm$0.01 & 5.23$^{+0.33}_{-0.35}$ & No & HSC\_937 & -1.95$\pm$0.01 & -0.52$\pm$0.01 & 4.82$^{+0.09}_{-0.08}$ \\
J011038.68+613922.2 & 522583690414836736 & -1.82$\pm$0.01 & -0.54$\pm$0.01 & 3.91$^{+0.21}_{-0.18}$ & Yes & UBC\_1215 & -1.77$\pm$0.01 & -0.55$\pm$0.01 & 3.96$\pm$0.08 \\
J011352.27+590144.6 & 414110030746911488 & -0.99$\pm$0.01 & -0.64$\pm$0.01 & 3.08$^{+0.14}_{-0.11}$ & No & NGC\_436 & -0.98$\pm$0.00 & -0.66$\pm$0.00 & 3.19$\pm$0.02 \\
J013357.02+615318.3 & 510782151078382208 & -1.71$\pm$0.01 & -0.41$\pm$0.01 & 3.07$^{+0.19}_{-0.13}$ & Yes & HSC\_1025 & -1.67$\pm$0.01 & -0.25$\pm$0.01 & 2.92$\pm$0.06 \\
J030427.62+561337.7 & 459627063352742912 & 0.97$\pm$0.02 & -1.36$\pm$0.02 & 2.30$^{+0.08}_{-0.07}$ & Yes & UBC\_1245 & 0.66$\pm$0.01 & -1.03$\pm$0.01 & 2.35$\pm$0.03 \\
J033851.13+515341.2 & 442195161187958272 & 0.70$\pm$0.01 & -2.17$\pm$0.01 & 2.38$^{+0.08}_{-0.07}$ & No & UBC\_1252 & 0.75$\pm$0.02 & -2.13$\pm$0.02 & 2.43$^{+0.04}_{-0.03}$ \\
\multirow{2}{*}{J035221.20+522835.7} & \multirow{2}{*}{251943984175934720} & \multirow{2}{*}{0.77$\pm$0.01} & \multirow{2}{*}{-1.53$\pm$0.01} & \multirow{2}{*}{2.18$^{+0.07}_{-0.06}$} & \multirow{2}{*}{No} & CWNU\_1346 & 0.71$\pm$0.01 & -2.00$\pm$0.01 & 2.25$\pm$0.03 \\
 & & & & & & Teutsch\_194 & 0.89$\pm$0.02 & -1.53$\pm$0.02 & 2.13$\pm$0.03 \\
J035612.84+581824.4 & 470023942230011520 & -0.52$\pm$0.02 & -0.05$\pm$0.01 & 4.26$^{+0.20}_{-0.22}$ & No & Juchert\_9 & -0.30$\pm$0.01 & -0.11$\pm$0.01 & 4.43$\pm$0.11 \\
J035712.14+555244.5 & 468740674721437568 & -0.03$\pm$0.02 & 0.19$\pm$0.01 & 3.70$^{+0.27}_{-0.21}$ & No & HSC\_1181 & -0.01$\pm$0.00 & -0.04$\pm$0.00 & 3.77$\pm$0.02 \\
J044342.01+410424.7 & 203132441067857024 & 0.22$\pm$0.02 & -1.36$\pm$0.02 & 3.01$^{+0.22}_{-0.16}$ & No & FSR\_0723 & 0.00$\pm$0.03 & -1.23$\pm$0.02 & 2.97$\pm$0.03 \\
J052021.77+391925.9 & 187917223068585728 & 0.37$\pm$0.02 & -1.38$\pm$0.01 & 2.77$^{+0.09}_{-0.10}$ & No & NGC\_1857 & 0.47$\pm$0.00 & -1.38$\pm$0.00 & 2.71$\pm$0.02 \\
J052636.70+392052.4 & 193685540242156928 & 0.21$\pm$0.03 & -0.63$\pm$0.02 & 4.36$^{+0.80}_{-0.36}$ & No & HSC\_1322 & 0.10$\pm$0.01 & -0.70$\pm$0.01 & 4.45$^{+0.10}_{-0.09}$ \\
J052701.81+423622.7 & 195434554004220032 & -0.05$\pm$0.02 & -0.37$\pm$0.02 & 4.23$^{+0.28}_{-0.29}$ & No & HSC\_1300 & 0.10$\pm$0.02 & -0.29$\pm$0.01 & 4.09$\pm$0.08 \\
J052935.27+361306.0 & 183375007391969408 & -0.18$\pm$0.03 & -1.98$\pm$0.02 & 1.89$\pm$0.06 & No & UBC\_198 & -0.01$\pm$0.01 & -1.99$\pm$0.01 & 1.83$\pm$0.01 \\
J054153.93+323613.1 & 3448388529968087040 & -0.12$\pm$0.03 & -0.78$\pm$0.02 & 3.44$^{+0.53}_{-0.36}$ & No & HSC\_1381 & -0.10$\pm$0.01 & -0.78$\pm$0.02 & 3.65$\pm$0.08 \\
J055304.96+252606.6 & 3428324809286029056 & 0.36$\pm$0.03 & -1.67$\pm$0.02 & 2.48$^{+0.16}_{-0.14}$ & No & UBC\_1299 & 0.55$\pm$0.01 & -1.47$\pm$0.01 & 2.59$^{+0.04}_{-0.03}$ \\
J060017.47+234019.4 & 3424861897411791744 & 0.27$\pm$0.02 & -2.15$\pm$0.01 & 1.82$\pm$0.05 & No & NGC\_2129 & 0.22$\pm$0.01 & -2.19$\pm$0.01 & 1.86$\pm$0.01 \\
J060452.44+240331.3 & 3426200454038524928 & 0.93$\pm$0.02 & -3.07$\pm$0.02 & 1.94$\pm$0.07 & No & IC\_2157 & 0.98$\pm$0.01 & -3.02$\pm$0.01 & 1.94$\pm$0.01 \\
J061007.35+244201.1 & 3426316418155789696 & 0.23$\pm$0.02 & -0.45$\pm$0.01 & 3.73$^{+0.20}_{-0.21}$ & No & FSR\_0869 & 0.51$\pm$0.01 & -0.44$\pm$0.01 & 3.91$^{+0.10}_{-0.09}$ \\
\multirow{2}{*}{J061250.77+093035.0} & \multirow{2}{*}{3329082168206416000} & \multirow{2}{*}{-0.42$\pm$0.02} & \multirow{2}{*}{-1.01$\pm$0.02} & \multirow{2}{*}{2.32$^{+0.09}_{-0.10}$} & \multirow{2}{*}{Yes} & HSC\_1585 & -0.37$\pm$0.01 & -1.14$\pm$0.02 & 2.41$\pm$0.03 \\
 & & & & & & UBC\_1320 & 0.03$\pm$0.01 & -1.30$\pm$0.01 & 2.37$\pm$0.03 \\
J061931.75+140346.8 & 3344634107345773568 & 0.24$\pm$0.02 & -2.11$\pm$0.02 & 3.21$^{+0.14}_{-0.16}$ & Yes & UBC\_1316 & 0.53$\pm$0.01 & -1.83$\pm$0.02 & 3.09$\pm$0.05 \\
\multirow{2}{*}{J062104.61+221010.1} & \multirow{2}{*}{3376778299723392512} & \multirow{2}{*}{0.22$\pm$0.02} & \multirow{2}{*}{-2.36$\pm$0.02} & \multirow{2}{*}{1.74$^{+0.04}_{-0.05}$} & \multirow{2}{*}{No} & OC\_0315 & 0.62$\pm$0.02 & -2.50$\pm$0.02 & 1.69$\pm$0.01 \\
 & & & & & & OC\_0316 & 0.25$\pm$0.01 & -2.35$\pm$0.01 & 1.71$\pm$0.02 \\
J062806.24+174537.3 & 3369789906895933824 & 0.26$\pm$0.02 & -0.74$\pm$0.01 & 3.77$^{+0.13}_{-0.15}$ & No & HSC\_1540 & -0.07$\pm$0.01 & -0.41$\pm$0.02 & 3.90$^{+0.07}_{-0.06}$ \\
J062915.48+171630.7 & 3369689610820311552 & -0.15$\pm$0.02 & -0.21$\pm$0.02 & 3.61$^{+0.27}_{-0.21}$ & Yes & HSC\_1540 & -0.07$\pm$0.01 & -0.41$\pm$0.02 & 3.90$^{+0.07}_{-0.06}$ \\
J063500.45+171044.7 & 3358945453772420224 & -0.17$\pm$0.02 & -0.21$\pm$0.01 & 3.86$^{+0.23}_{-0.20}$ & No & HSC\_1552 & 0.00$\pm$0.02 & -0.51$\pm$0.02 & 3.79$\pm$0.07 \\
J063653.75+133219.2 & 3355441688172411648 & -0.03$\pm$0.02 & -0.17$\pm$0.02 & 4.09$^{+0.16}_{-0.18}$ & No & HSC\_1583 & 0.04$\pm$0.01 & -0.33$\pm$0.02 & 4.18$^{+0.08}_{-0.07}$ \\
\multirow{2}{*}{J063701.44+051307.1} & \multirow{2}{*}{3130981135303925120} & \multirow{2}{*}{-1.49$\pm$0.02} & \multirow{2}{*}{0.80$\pm$0.02} & \multirow{2}{*}{1.46$\pm$0.04} & \multirow{2}{*}{No} & Collinder\_104 & -1.39$\pm$0.01 & 0.65$\pm$0.01 & 1.51$\pm$0.01 \\
 & & & & & & HSC\_1629 & -1.21$\pm$0.02 & 0.58$\pm$0.02 & 1.43$\pm$0.01 \\
J063703.39+171837.0 & 3358961465412197888 & -0.37$\pm$0.02 & -0.30$\pm$0.02 & 3.95$^{+0.33}_{-0.28}$ & No & HSC\_1552 & 0.00$\pm$0.02 & -0.51$\pm$0.02 & 3.79$\pm$0.07 \\
J063809.24+034454.6 & 3130207762014534144 & -2.32$\pm$0.02 & -0.28$\pm$0.01 & 1.38$\pm$0.03 & Yes & Collinder\_107 & -2.43$\pm$0.01 & -0.20$\pm$0.02 & 1.41$\pm$0.01 \\
J063854.52+003418.5 & 3119809650489626880 & -0.35$\pm$0.03 & 0.61$\pm$0.02 & 5.81$^{+0.56}_{-0.49}$ & No & HSC\_1665 & -0.21$\pm$0.01 & 0.39$\pm$0.01 & 5.71$\pm$0.14 \\
J065025.97+113923.4 & 3351354734735237376 & 0.07$\pm$0.02 & -0.40$\pm$0.02 & 4.09$^{+0.37}_{-0.27}$ & No & Juchert\_J0649.4+1202 & 0.04$\pm$0.01 & -0.39$\pm$0.01 & 4.03$\pm$0.07 \\
\multirow{2}{*}{J202010.60+383732.1} & \multirow{2}{*}{2061287708824637440} & \multirow{2}{*}{-3.57$\pm$0.01} & \multirow{2}{*}{-5.73$\pm$0.02} & \multirow{2}{*}{1.73$\pm$0.04} & \multirow{2}{*}{No} & Berkeley\_86 & -3.45$\pm$0.01 & -5.47$\pm$0.01 & 1.68$\pm$0.01 \\
 & & & & & & FoF\_2179 & -3.25$\pm$0.01 & -5.99$\pm$0.01 & 1.70$\pm$0.01 \\
J213057.76+525020.6 & 2172650099079727488 & -2.52$\pm$0.02 & -2.77$\pm$0.02 & 2.85$\pm$0.13 & No & Theia\_5761 & -2.91$\pm$0.01 & -3.13$\pm$0.01 & 2.99$\pm$0.03 \\
J230456.23+573851.2 & 2010162896331826944 & -3.54$\pm$0.01 & -2.42$\pm$0.01 & 3.20$^{+0.15}_{-0.10}$ & No & Casado\_19 & -3.28$\pm$0.01 & -2.30$\pm$0.01 & 3.33$\pm$0.03 \\
\enddata
     \begin{tablenotes}
        \item[1] \textbf{Notes.}
        \item[2] $^{\mathrm{a}}$: Proper motion in the R.A. direction of the CBe star.
        \item[3] $^{\mathrm{b}}$: Proper motion in decl. direction of the CBe star.
        \item[4] $^{\mathrm{c}}$: Distance of the CBe star.
        \item[5] $^{\mathrm{d}}$: Proper motion in the R.A. direction of the cluster.
        \item[6] $^{\mathrm{e}}$: Proper motion in decl. direction of the cluster.
        \item[7] $^{\mathrm{f}}$: Distance of the cluster.
        \item[8] (This table is available in machine-readable form in the \href{https://iopscience.iop.org/article/10.3847/1538-4365/ae450d}{online article}.)
     \end{tablenotes}
\end{deluxetable*}

\startlongtable
\begin{deluxetable*}{ccccccccccc}
\renewcommand{\arraystretch}{1.1}
\tabletypesize{\footnotesize}
\setlength{\tabcolsep}{1.4pt}
\tablewidth{0pt}
\tablecaption{Peculiar velocities of partial CBe stars from LAMOST DR12. \label{tab:runaway stars}}
\tablehead{
\colhead{\emph{Gaia} DR3} & 
\colhead{RV} & \colhead{pmra(s)}  &  \colhead{pmdec(s)}  & \colhead{Dis.(s)} &
\colhead{New} & 
\colhead{N} & \colhead{$V_\mathrm{pec, McMillan17}$} & \colhead{$V_\mathrm{pec, MWPotential2014}$} & \colhead{$V_\mathrm{pec, Cautun20}$} & \colhead{$V_\mathrm{pec, Irrgang13I}$} \\
\colhead{ID} & \colhead{(km s$^{-1}$)} & \colhead{(mas yr$^{-1}$)} & \colhead{(mas yr$^{-1}$)} & \colhead{(kpc)} & \colhead{CBe star?} &
\colhead{} & \colhead{(km s$^{-1}$)} & \colhead{(km s$^{-1}$)} & \colhead{(km s$^{-1}$)} & \colhead{(km s$^{-1}$)}}
\startdata
943635960251342592 & -42.99$^{+4.32}_{-4.41}$ & 3.01$\pm$0.02 & -5.69$\pm$0.02 & 2.25$^{+0.11}_{-0.10}$ & No & 2 & 100.68 & 97.42 & 86.19 & 81.53 \\
2104951725038702336 & -64.86$^{+5.74}_{-5.62}$ & -4.49$\pm$0.03 & -10.17$\pm$0.03 & 4.08$^{+0.53}_{-0.48}$ & No & 6 & 99.69 & 106.05 & 112.61 & 138.61 \\
2024963284967616256 & 49.91$^{+5.97}_{-5.94}$ & -2.41$\pm$0.01 & -6.02$\pm$0.01 & 8.06$^{+0.80}_{-0.64}$ & No & 1 & 87.74 & 100.79 & 66.38 & 97.62 \\
3119809650489626880 & -33.57$^{+5.60}_{-5.29}$ & -0.35$\pm$0.03 & 0.61$\pm$0.02 & 5.81$^{+0.56}_{-0.49}$ & No & 1 & 78.87 & 95.32 & 84.09 & 100.91 \\
511217488963795584 & -112.65$^{+9.29}_{-9.81}$ & -1.07$\pm$0.01 & -0.23$\pm$0.01 & 2.88$^{+0.10}_{-0.11}$ & No & 1 & 72.03 & 67.09 & 77.81 & 90.27 \\
2013258876254822016 & 61.45$^{+6.74}_{-6.57}$ & -2.91$\pm$0.02 & -2.11$\pm$0.02 & 1.66$^{+0.07}_{-0.06}$ & No & 1 & 70.75 & 105.73 & 96.77 & 71.80 \\
3153828952571014528 & 87.84$^{+9.06}_{-8.94}$ & -0.63$\pm$0.02 & -1.52$\pm$0.02 & 1.68$\pm$0.07 & No & 4 & 70.17 & 47.87 & 49.04 & 63.85 \\
511257140102787584 & -109.95$^{+4.25}_{-4.23}$ & -1.50$\pm$0.01 & -0.19$\pm$0.01 & 3.44$^{+0.25}_{-0.22}$ & Yes & 2 & 67.36 & 68.10 & 49.81 & 70.44 \\
1871130372560308992 & 2.44$^{+4.41}_{-4.74}$ & 4.15$\pm$0.01 & -1.21$\pm$0.01 & 2.09$^{+0.06}_{-0.05}$ & No & 2 & 67.00 & 55.88 & 74.77 & 78.33 \\
273756340665255424 & -71.63$^{+5.34}_{-5.61}$ & 2.36$\pm$0.02 & -3.85$\pm$0.01 & 0.88$\pm$0.01 & No & 6 & 65.03 & 75.62 & 72.53 & 73.94 \\
442195161187958272 & -87.25$^{+8.11}_{-8.65}$ & 0.70$\pm$0.01 & -2.17$\pm$0.01 & 2.38$^{+0.08}_{-0.07}$ & No & 20 & 63.35 & 81.51 & 55.67 & 80.35 \\
459871498533427328 & -71.77$^{+9.58}_{-9.98}$ & 0.52$\pm$0.01 & -1.38$\pm$0.02 & 2.10$^{+0.06}_{-0.07}$ & Yes & 6 & 63.00 & 57.38 & 51.33 & 50.10 \\
2006919818073117952 & -76.13$^{+8.46}_{-8.55}$ & -2.63$\pm$0.02 & -3.57$\pm$0.02 & 2.08$^{+0.07}_{-0.06}$ & Yes & 2 & 62.57 & 46.14 & 52.92 & 64.31 \\
37694252194590976 & -33.68$^{+7.04}_{-7.29}$ & -2.02$\pm$0.04 & -1.65$\pm$0.03 & 2.76$^{+0.29}_{-0.24}$ & No & 5 & 61.60 & 71.78 & 71.87 & 73.60 \\
2003956939163275136 & -85.35$^{+3.99}_{-4.14}$ & -3.38$\pm$0.01 & -2.82$\pm$0.01 & 2.98$^{+0.09}_{-0.10}$ & Yes & 2 & 59.65 & 44.54 & 58.24 & 54.98 \\
1864444345433406592 & -82.45$^{+3.40}_{-3.29}$ & -3.93$\pm$0.01 & -5.60$\pm$0.01 & 6.41$^{+0.31}_{-0.38}$ & Yes & 1 & 56.74 & 73.75 & 55.77 & 52.32 \\
241341226047485312 & 18.05$^{+1.69}_{-1.65}$ & 1.94$\pm$0.03 & -3.54$\pm$0.03 & 2.65$^{+0.19}_{-0.16}$ & No & 3 & 51.74 & 47.97 & 64.27 & 56.57 \\
270346377148562816 & -99.84$^{+9.41}_{-9.44}$ & 0.26$\pm$0.02 & -0.71$\pm$0.02 & 3.74$^{+0.30}_{-0.27}$ & Yes & 1 & 51.67 & 76.04 & 48.87 & 75.07 \\
183440737570824448 & -54.18$^{+6.54}_{-6.77}$ & 0.80$\pm$0.02 & -1.72$\pm$0.02 & 2.81$^{+0.16}_{-0.12}$ & No & 16 & 51.67 & 37.81 & 54.77 & 45.92 \\
3088286686477378304 & -43.06$^{+8.13}_{-8.57}$ & -1.51$\pm$0.06 & -0.82$\pm$0.04 & 0.87$^{+0.04}_{-0.03}$ & No & 4 & 49.73 & 64.99 & 54.61 & 60.59 \\
3374580582137345408 & -46.61$^{+6.21}_{-6.25}$ & 0.55$\pm$0.03 & -3.86$\pm$0.02 & 1.47$\pm$0.06 & No & 1 & 49.66 & 40.62 & 54.12 & 51.78 \\
3326790721553428736 & -43.26$^{+6.05}_{-6.16}$ & -1.85$\pm$0.02 & -2.81$\pm$0.02 & 1.01$^{+0.02}_{-0.03}$ & Yes & 4 & 49.49 & 64.59 & 56.96 & 54.12 \\
448859820863017472 & -93.34$^{+8.27}_{-8.10}$ & 0.51$\pm$0.02 & -0.03$\pm$0.02 & 4.86$^{+0.40}_{-0.42}$ & No & 2 & 49.20 & 39.13 & 61.91 & 48.59 \\
2016382386934503168 & -74.41$^{+5.93}_{-5.88}$ & -5.27$\pm$0.01 & -0.46$\pm$0.01 & 2.67$^{+0.09}_{-0.07}$ & Yes & 1 & 49.04 & 48.66 & 48.10 & 45.97 \\
3425142311534938112 & -48.08$^{+6.21}_{-6.50}$ & 0.00$\pm$0.02 & -0.87$\pm$0.02 & 4.14$^{+0.28}_{-0.25}$ & No & 64 & 48.09 & 48.81 & 50.23 & 69.16 \\
189638096203389440 & -57.93$^{+6.70}_{-6.85}$ & -0.41$\pm$0.02 & -1.53$\pm$0.01 & 3.45$^{+0.17}_{-0.16}$ & No & 1 & 48.00 & 55.25 & 55.90 & 47.13 \\
184522519573582848 & -66.85$^{+6.19}_{-6.43}$ & 0.40$\pm$0.02 & -1.38$\pm$0.01 & 2.74$^{+0.14}_{-0.13}$ & Yes & 1 & 47.81 & 63.04 & 63.82 & 45.79 \\
247190868429680384 & 31.93$^{+6.85}_{-6.95}$ & 0.74$\pm$0.01 & -1.93$\pm$0.01 & 2.18$^{+0.09}_{-0.05}$ & No & 1 & 45.94 & 50.25 & 38.55 & 51.08 \\
3369789906895933824 & -16.13$^{+7.49}_{-7.26}$ & 0.26$\pm$0.02 & -0.74$\pm$0.01 & 3.77$^{+0.13}_{-0.15}$ & No & 18 & 44.17 & 44.77 & 33.98 & 46.88 \\
3107967257426335104 & 105.29$^{+7.86}_{-7.55}$ & -1.79$\pm$0.02 & 0.45$\pm$0.02 & 4.94$^{+0.82}_{-0.77}$ & Yes & 1 & 44.09 & 45.29 & 44.32 & 40.86 \\
458436120468296960 & -67.36$^{+3.31}_{-3.35}$ & -0.49$\pm$0.02 & -1.04$\pm$0.02 & 2.28$^{+0.11}_{-0.12}$ & Yes & 1 & 43.11 & 47.49 & 43.04 & 36.71 \\
3127578146818103040 & 69.44$^{+3.17}_{-3.19}$ & -1.07$\pm$0.03 & -0.40$\pm$0.02 & 2.04$\pm$0.09 & Yes & 6 & 42.05 & 40.55 & 40.80 & 35.26 \\
3120658331728894080 & 46.23$^{+1.69}_{-1.75}$ & -0.02$\pm$0.02 & -2.21$\pm$0.02 & 2.31$\pm$0.09 & No & 4 & 41.91 & 38.34 & 43.11 & 36.09 \\
2007475277604631040 & -54.98$\pm$7.07 & -1.74$\pm$0.02 & -2.37$\pm$0.02 & 2.14$\pm$0.06 & No & 2 & 41.74 & 29.29 & 42.80 & 38.09 \\
245140004363839488 & -60.31$^{+5.23}_{-5.15}$ & 0.91$\pm$0.02 & -2.16$\pm$0.01 & 1.57$^{+0.06}_{-0.05}$ & No & 4 & 41.67 & 51.96 & 51.17 & 52.77 \\
512033086070875136 & -68.39$^{+6.55}_{-6.73}$ & -1.26$\pm$0.01 & -0.28$\pm$0.01 & 2.46$\pm$0.06 & Yes & 3 & 41.66 & 29.97 & 42.65 & 40.37 \\
3455386161265334400 & -34.80$^{+8.78}_{-9.13}$ & 0.52$\pm$0.02 & -3.31$\pm$0.01 & 2.24$\pm$0.07 & No & 2 & 40.33 & 42.92 & 36.04 & 39.46 \\
... & ... & ... & ... & ... & ... & ... & ... & ... & ... & ...\\
\enddata
     \begin{tablenotes}
        \item[1] (This table is available in its entirety in machine-readable form in the \href{https://iopscience.iop.org/article/10.3847/1538-4365/ae450d}{online article}.) 
    \end{tablenotes}
\end{deluxetable*}

\section{Summary and discussion}\label{sec6}
CBe stars are rapidly rotating B-type stars whose high rotational speeds promote the formation of decretion disks—composed of material ejected from the stellar equator—that give rise to characteristic H$\alpha$ emission. Given that the majority of CBe stars are thought to form through binary interactions, identifying and characterizing these systems presents a valuable opportunity to study binary evolution processes. To this end, we conduct a systematic search for CBe stars and CBe star binaries using data from LAMOST. Our main results are as follows:
\begin{enumerate}
    \item \textit{CBe stars}: We construct the initial O/B star sample by utilizing \texttt{subclass} of LAMOST DR12. Among them, we identify 504 CBe stars including 141 newly identified by inspecting their spectra by eye checking with two criteria applied, crossmatching with CBe star catalogs, and IR test.
    \item \textit{CBe star binaries}: 14 CBe stars exhibit significant RV variability with maximum RV separations ranging from $\sim$20 to $\sim$130 km s$^{-1}$. One of them, ALS\,8814, has previously been reported as CBe star binaries, while the other two have been identified as (First Degree Trend) SB1 by \emph{Gaia} DR3. For other 11 newly identified CBe star binaries, we briefly discuss LAMOST J035933.84+555751.1.
    \item \textit{Potential CBe star binaries}: We also propose 60 CBe stars as potential CBe star binaries, which is with high \texttt{RUWE} but not confirmed by dynamics.
    \item \textit{Potential cluster members}: By comparing distances, angular separations, and proper motions between sources in our samples and clusters, we find 34 CBe stars are potential cluster members.
    \item \textit{Runaway stars}: We calculate peculiar velocities for 203 CBe stars and identify 37 runaway stars adopting the classical criterion for runaway stars (total peculiar velocity exceeding 40 km s$^{-1}$). 
\end{enumerate}
Of the CBe stars in our sample, only $\sim$3\% exhibit RV variability. This rate is undoubtedly sensitive to the number of observations of the sample to a certain extent. When considering only sources with at least five observations, the detection rate rises to 7.9\%. For comparison, \citet{2018ApJ...853..156W} performed a systematic search for Be+sdO binaries using far-ultraviolet (FUV) spectra from the International Ultraviolet Explorer (\emph{IUE}), reporting a detection rate of $\sim$4.5\%—somewhat lower than our detection rate. More recently, \citet{2025ApJ...993..192K} presented speckle observations of
76 Be stars taken using the Gemini North and South speckle imagers spanning angular separations of
20 mas–1.2$^{\prime \prime}$ and reported an observed binary detection rate of 14.5\%, the highest among the three studies.
\par Given that RV monitoring is less sensitive to widely separated and/or low-inclination systems, and considering that the spanning angular separations in \citet{2025ApJ...993..192K} remain below the spatial resolution of \emph{IUE}, the comparatively higher detection rates in this study and \citet{2025ApJ...993..192K} may suggest that companions with faint FUV radiations (compared to the primary CBe stars), such as compact objects or puffed-up helium stars, are more common than sdOs as companions to CBe stars. This inference is consistent with the theoretical predictions of \citet{2014ApJ...796...37S}. But overall, each of the three methods has its own advantages and disadvantages, and all three yield detection rates of the same order of magnitude, underscoring their complementary roles in identifying CBe star binaries.
\par We report a runaway rate of 18.2\% for CBe stars in our sample, which is significantly higher than the rates of $\sim$6.7\% and $\sim$5\% reported by \citet{2001ApJ...555..364B} and \citet{2022ApJS..260...35W}, respectively, both of which adopted the same threshold as us. We acknowledge that due to the
scarcity and/or concentration of corresponding phases of the available spectra, some CBe stars discussed here may be RV-variable but are mistaken for RV-constant, despite our filtering of objects with observed RV variations. There is no doubt that restricting our analysis to sources with multiple observations can significantly reduce potential binarity contamination. The runaway rates are 10.9\% ($\geq$ 3 observations), 12.1\% ($\geq$ 4 observations), 10.6\% ($\geq$ 5 observations), and 11.0\% ($\geq$ 6 observations), respectively. Therefore, the true runaway rate in our sample is likely not less than 10\%. The lower rate found by \citet{2001ApJ...555..364B} may be influenced by their sample selection, which included only stars brighter than $V=9$ and thus primarily probed the solar neighborhood. \citet{2022ApJS..260...35W} suggested that the absence of RV measurements in their study likely resulted in incomplete detection of runaway stars. Furthermore, \citet{2024ApJS..272...45G} reported a runaway rate of $\sim$5\% among 4432 early-type stars with measured RVs, a value much lower than our derived rate for CBe stars. If there are indeed differences between the runaway rates of early-type stars and CBe stars, it may be further evidence that CBe stars prefer to form through binary interactions, as a systematically higher runaway fraction of CBe stars would suggest that CBe stars are more likely to form in evolved systems that have experienced supernova explosions compared to the general population of early-type stars.

\begin{acknowledgments}
We thank Luqian Wang and Yang Huang for helpful discussion, and the anonymous referee for the constructive suggestions that improved the paper. This work was supported by the National Key R\&D Program of China under grants 2021YFA1600401 and 2023YFA1607901, the National Natural Science Foundation of China under grants 12433007 and 12221003. We acknowledge the science research grants from the China Manned Space Project with No. CMS-CSST-2025-A13. We acknowledge the Milky Way face-on background image provided by NASA/JPL-Caltech, created by R. Hurt (SSC/Caltech). This paper uses the data from the LAMOST survey. Guoshoujing Telescope (the Large Sky Area Multi-Object Fiber Spectroscopic Telescope LAMOST) is a National Major Scientific Project built by the Chinese Academy of Sciences. Funding for the project has been provided by the National Development and Reform Commission. LAMOST is operated and managed by the National Astronomical Observatories, Chinese Academy of Sciences. This work has made use of data from the European Space Agency (ESA) mission Gaia (\url{https://www.cosmos.esa.int/gaia}), processed by the Gaia Data Processing and Analysis Consortium (DPAC, \url{https://www.cosmos.esa.int/web/gaia/dpac/consortium}). Funding for the DPAC has been provided by national institutions, in particular the institutions participating in the Gaia Multilateral Agreement. This research has made use of the SIMBAD database, CDS, Strasbourg Astronomical Observatory, France. This research has made use of the VizieR catalog access tool, CDS,
Strasbourg, France \citep{1996BICDS..48...47O}. The original description 
of the VizieR service was published in \citet{2000A&AS..143...23O}.
\end{acknowledgments}

\facilities{LAMOST, \emph{Gaia}}
 
\software{
          Astropy \citep{2013A&A...558A..33A, 2018AJ....156..123A, 2022ApJ...935..167A},  
          Numpy \citep{harris2020array}, 
          Pandas \citep{reback2020pandas},
          laspec \citep{2020ApJS..246....9Z, 2021ApJS..256...14Z},
          matplotlib \citep{Hunter:2007},
          emcee \citep{2013PASP..125..306F}, 
          mw\_plot (\href{https://github.com/henrysky/milkyway_plot}{milkyway\_plot}), 
          Vizier \citep{2000A&AS..143...23O},
          Simbad \citep{2000A&AS..143....9W},       
          galpy \citep{2015ApJS..216...29B}.  
          }

\bibliography{ref}{}

@ARTICLE{2022ApJ...935..167A,
       author = {{Astropy Collaboration} and {Price-Whelan}, Adrian M. and {Lim}, Pey Lian and {Earl}, Nicholas and {Starkman}, Nathaniel and {Bradley}, Larry and {Shupe}, David L. and {Patil}, Aarya A. and {Corrales}, Lia and {Brasseur}, C.~E. and {N{\"o}the}, Maximilian and {Donath}, Axel and {Tollerud}, Erik and {Morris}, Brett M. and {Ginsburg}, Adam and {Vaher}, Eero and {Weaver}, Benjamin A. and {Tocknell}, James and {Jamieson}, William and {van Kerkwijk}, Marten H. and {Robitaille}, Thomas P. and {Merry}, Bruce and {Bachetti}, Matteo and {G{\"u}nther}, H. Moritz and {Aldcroft}, Thomas L. and {Alvarado-Montes}, Jaime A. and {Archibald}, Anne M. and {B{\'o}di}, Attila and {Bapat}, Shreyas and {Barentsen}, Geert and {Baz{\'a}n}, Juanjo and {Biswas}, Manish and {Boquien}, M{\'e}d{\'e}ric and {Burke}, D.~J. and {Cara}, Daria and {Cara}, Mihai and {Conroy}, Kyle E. and {Conseil}, Simon and {Craig}, Matthew W. and {Cross}, Robert M. and {Cruz}, Kelle L. and {D'Eugenio}, Francesco and {Dencheva}, Nadia and {Devillepoix}, Hadrien A.~R. and {Dietrich}, J{\"o}rg P. and {Eigenbrot}, Arthur Davis and {Erben}, Thomas and {Ferreira}, Leonardo and {Foreman-Mackey}, Daniel and {Fox}, Ryan and {Freij}, Nabil and {Garg}, Suyog and {Geda}, Robel and {Glattly}, Lauren and {Gondhalekar}, Yash and {Gordon}, Karl D. and {Grant}, David and {Greenfield}, Perry and {Groener}, Austen M. and {Guest}, Steve and {Gurovich}, Sebastian and {Handberg}, Rasmus and {Hart}, Akeem and {Hatfield-Dodds}, Zac and {Homeier}, Derek and {Hosseinzadeh}, Griffin and {Jenness}, Tim and {Jones}, Craig K. and {Joseph}, Prajwel and {Kalmbach}, J. Bryce and {Karamehmetoglu}, Emir and {Ka{\l}uszy{\'n}ski}, Miko{\l}aj and {Kelley}, Michael S.~P. and {Kern}, Nicholas and {Kerzendorf}, Wolfgang E. and {Koch}, Eric W. and {Kulumani}, Shankar and {Lee}, Antony and {Ly}, Chun and {Ma}, Zhiyuan and {MacBride}, Conor and {Maljaars}, Jakob M. and {Muna}, Demitri and {Murphy}, N.~A. and {Norman}, Henrik and {O'Steen}, Richard and {Oman}, Kyle A. and {Pacifici}, Camilla and {Pascual}, Sergio and {Pascual-Granado}, J. and {Patil}, Rohit R. and {Perren}, Gabriel I. and {Pickering}, Timothy E. and {Rastogi}, Tanuj and {Roulston}, Benjamin R. and {Ryan}, Daniel F. and {Rykoff}, Eli S. and {Sabater}, Jose and {Sakurikar}, Parikshit and {Salgado}, Jes{\'u}s and {Sanghi}, Aniket and {Saunders}, Nicholas and {Savchenko}, Volodymyr and {Schwardt}, Ludwig and {Seifert-Eckert}, Michael and {Shih}, Albert Y. and {Jain}, Anany Shrey and {Shukla}, Gyanendra and {Sick}, Jonathan and {Simpson}, Chris and {Singanamalla}, Sudheesh and {Singer}, Leo P. and {Singhal}, Jaladh and {Sinha}, Manodeep and {Sip{\H{o}}cz}, Brigitta M. and {Spitler}, Lee R. and {Stansby}, David and {Streicher}, Ole and {{\v{S}}umak}, Jani and {Swinbank}, John D. and {Taranu}, Dan S. and {Tewary}, Nikita and {Tremblay}, Grant R. and {de Val-Borro}, Miguel and {Van Kooten}, Samuel J. and {Vasovi{\'c}}, Zlatan and {Verma}, Shresth and {de Miranda Cardoso}, Jos{\'e} Vin{\'\i}cius and {Williams}, Peter K.~G. and {Wilson}, Tom J. and {Winkel}, Benjamin and {Wood-Vasey}, W.~M. and {Xue}, Rui and {Yoachim}, Peter and {Zhang}, Chen and {Zonca}, Andrea and {Astropy Project Contributors}},
        title = "{The Astropy Project: Sustaining and Growing a Community-oriented Open-source Project and the Latest Major Release (v5.0) of the Core Package}",
      journal = {\apj},
         year = 2022,
        month = aug,
       volume = {935},
       number = {2},
          eid = {167},
        pages = {167},
          doi = {10.3847/1538-4357/ac7c74},
archivePrefix = {arXiv},
       eprint = {2206.14220},
 primaryClass = {astro-ph.IM},
       adsurl = {https://ui.adsabs.harvard.edu/abs/2022ApJ...935..167A}
}

@ARTICLE{2018AJ....156..123A,
       author = {{Astropy Collaboration} and {Price-Whelan}, A.~M. and {Sip{\H{o}}cz}, B.~M. and {G{\"u}nther}, H.~M. and {Lim}, P.~L. and {Crawford}, S.~M. and {Conseil}, S. and {Shupe}, D.~L. and {Craig}, M.~W. and {Dencheva}, N. and {Ginsburg}, A. and {VanderPlas}, J.~T. and {Bradley}, L.~D. and {P{\'e}rez-Su{\'a}rez}, D. and {de Val-Borro}, M. and {Aldcroft}, T.~L. and {Cruz}, K.~L. and {Robitaille}, T.~P. and {Tollerud}, E.~J. and {Ardelean}, C. and {Babej}, T. and {Bach}, Y.~P. and {Bachetti}, M. and {Bakanov}, A.~V. and {Bamford}, S.~P. and {Barentsen}, G. and {Barmby}, P. and {Baumbach}, A. and {Berry}, K.~L. and {Biscani}, F. and {Boquien}, M. and {Bostroem}, K.~A. and {Bouma}, L.~G. and {Brammer}, G.~B. and {Bray}, E.~M. and {Breytenbach}, H. and {Buddelmeijer}, H. and {Burke}, D.~J. and {Calderone}, G. and {Cano Rodr{\'\i}guez}, J.~L. and {Cara}, M. and {Cardoso}, J.~V.~M. and {Cheedella}, S. and {Copin}, Y. and {Corrales}, L. and {Crichton}, D. and {D'Avella}, D. and {Deil}, C. and {Depagne}, {\'E}. and {Dietrich}, J.~P. and {Donath}, A. and {Droettboom}, M. and {Earl}, N. and {Erben}, T. and {Fabbro}, S. and {Ferreira}, L.~A. and {Finethy}, T. and {Fox}, R.~T. and {Garrison}, L.~H. and {Gibbons}, S.~L.~J. and {Goldstein}, D.~A. and {Gommers}, R. and {Greco}, J.~P. and {Greenfield}, P. and {Groener}, A.~M. and {Grollier}, F. and {Hagen}, A. and {Hirst}, P. and {Homeier}, D. and {Horton}, A.~J. and {Hosseinzadeh}, G. and {Hu}, L. and {Hunkeler}, J.~S. and {Ivezi{\'c}}, {\v{Z}}. and {Jain}, A. and {Jenness}, T. and {Kanarek}, G. and {Kendrew}, S. and {Kern}, N.~S. and {Kerzendorf}, W.~E. and {Khvalko}, A. and {King}, J. and {Kirkby}, D. and {Kulkarni}, A.~M. and {Kumar}, A. and {Lee}, A. and {Lenz}, D. and {Littlefair}, S.~P. and {Ma}, Z. and {Macleod}, D.~M. and {Mastropietro}, M. and {McCully}, C. and {Montagnac}, S. and {Morris}, B.~M. and {Mueller}, M. and {Mumford}, S.~J. and {Muna}, D. and {Murphy}, N.~A. and {Nelson}, S. and {Nguyen}, G.~H. and {Ninan}, J.~P. and {N{\"o}the}, M. and {Ogaz}, S. and {Oh}, S. and {Parejko}, J.~K. and {Parley}, N. and {Pascual}, S. and {Patil}, R. and {Patil}, A.~A. and {Plunkett}, A.~L. and {Prochaska}, J.~X. and {Rastogi}, T. and {Reddy Janga}, V. and {Sabater}, J. and {Sakurikar}, P. and {Seifert}, M. and {Sherbert}, L.~E. and {Sherwood-Taylor}, H. and {Shih}, A.~Y. and {Sick}, J. and {Silbiger}, M.~T. and {Singanamalla}, S. and {Singer}, L.~P. and {Sladen}, P.~H. and {Sooley}, K.~A. and {Sornarajah}, S. and {Streicher}, O. and {Teuben}, P. and {Thomas}, S.~W. and {Tremblay}, G.~R. and {Turner}, J.~E.~H. and {Terr{\'o}n}, V. and {van Kerkwijk}, M.~H. and {de la Vega}, A. and {Watkins}, L.~L. and {Weaver}, B.~A. and {Whitmore}, J.~B. and {Woillez}, J. and {Zabalza}, V. and {Astropy Contributors}},
        title = "{The Astropy Project: Building an Open-science Project and Status of the v2.0 Core Package}",
      journal = {\aj},
         year = 2018,
        month = sep,
       volume = {156},
       number = {3},
          eid = {123},
        pages = {123},
          doi = {10.3847/1538-3881/aabc4f},
archivePrefix = {arXiv},
       eprint = {1801.02634},
 primaryClass = {astro-ph.IM},
       adsurl = {https://ui.adsabs.harvard.edu/abs/2018AJ....156..123A}
}

@ARTICLE{2013A&A...558A..33A,
       author = {{Astropy Collaboration} and {Robitaille}, Thomas P. and
         {Tollerud}, Erik J. and {Greenfield}, Perry and {Droettboom}, Michael and
         {Bray}, Erik and {Aldcroft}, Tom and {Davis}, Matt and
         {Ginsburg}, Adam and {Price-Whelan}, Adrian M. and
         {Kerzendorf}, Wolfgang E. and {Conley}, Alexander and {Crighton}, Neil and
         {Barbary}, Kyle and {Muna}, Demitri and {Ferguson}, Henry and
         {Grollier}, Fr{\'e}d{\'e}ric and {Parikh}, Madhura M. and
         {Nair}, Prasanth H. and {Unther}, Hans M. and {Deil}, Christoph and
         {Woillez}, Julien and {Conseil}, Simon and {Kramer}, Roban and
         {Turner}, James E.~H. and {Singer}, Leo and {Fox}, Ryan and
         {Weaver}, Benjamin A. and {Zabalza}, Victor and {Edwards}, Zachary I. and
         {Azalee Bostroem}, K. and {Burke}, D.~J. and {Casey}, Andrew R. and
         {Crawford}, Steven M. and {Dencheva}, Nadia and {Ely}, Justin and
         {Jenness}, Tim and {Labrie}, Kathleen and {Lim}, Pey Lian and
         {Pierfederici}, Francesco and {Pontzen}, Andrew and {Ptak}, Andy and
         {Refsdal}, Brian and {Servillat}, Mathieu and {Streicher}, Ole},
        title = "{Astropy: A community Python package for astronomy}",
      journal = {\aap},
         year = "2013",
        month = "Oct",
       volume = {558},
          eid = {A33},
        pages = {A33},
          doi = {10.1051/0004-6361/201322068},
archivePrefix = {arXiv},
       eprint = {1307.6212},
 primaryClass = {astro-ph.IM},
       adsurl = {https://ui.adsabs.harvard.edu/abs/2013A&A...558A..33A}
}

@ARTICLE{2003PASP..115.1153P,
       author = {{Porter}, John M. and {Rivinius}, Thomas},
        title = "{Classical Be Stars}",
      journal = {\pasp},
         year = 2003,
        month = oct,
       volume = {115},
       number = {812},
        pages = {1153-1170},
          doi = {10.1086/378307},
       adsurl = {https://ui.adsabs.harvard.edu/abs/2003PASP..115.1153P}
}

@ARTICLE{2013A&ARv..21...69R,
       author = {{Rivinius}, Thomas and {Carciofi}, Alex C. and {Martayan}, Christophe},
        title = "{Classical Be stars. Rapidly rotating B stars with viscous Keplerian decretion disks}",
      journal = {\aapr},
         year = 2013,
        month = oct,
       volume = {21},
          eid = {69},
        pages = {69},
          doi = {10.1007/s00159-013-0069-0},
archivePrefix = {arXiv},
       eprint = {1310.3962},
 primaryClass = {astro-ph.SR},
       adsurl = {https://ui.adsabs.harvard.edu/abs/2013A&ARv..21...69R}
}

@ARTICLE{2016A&A...595A.132Z,
       author = {{Zorec}, J. and {Fr{\'e}mat}, Y. and {Domiciano de Souza}, A. and {Royer}, F. and {Cidale}, L. and {Hubert}, A. -M. and {Semaan}, T. and {Martayan}, C. and {Cochetti}, Y.~R. and {Arias}, M.~L. and {Aidelman}, Y. and {Stee}, P.},
        title = "{Critical study of the distribution of rotational velocities of Be stars. I. Deconvolution methods, effects due to gravity darkening, macroturbulence, and binarity}",
      journal = {\aap},
         year = 2016,
        month = nov,
       volume = {595},
          eid = {A132},
        pages = {A132},
          doi = {10.1051/0004-6361/201628760},
       adsurl = {https://ui.adsabs.harvard.edu/abs/2016A&A...595A.132Z}
}

@ARTICLE{1995ARA&A..33..199B,
       author = {{Bodenheimer}, Peter},
        title = "{Angular Momentum Evolution of Young Stars and Disks}",
      journal = {\araa},
         year = 1995,
        month = jan,
       volume = {33},
        pages = {199-238},
          doi = {10.1146/annurev.aa.33.090195.001215},
       adsurl = {https://ui.adsabs.harvard.edu/abs/1995ARA&A..33..199B}
}

@ARTICLE{2010ApJ...722..605H,
       author = {{Huang}, Wenjin and {Gies}, D.~R. and {McSwain}, M.~V.},
        title = "{A Stellar Rotation Census of B Stars: From ZAMS to TAMS}",
      journal = {\apj},
         year = 2010,
        month = oct,
       volume = {722},
       number = {1},
        pages = {605-619},
          doi = {10.1088/0004-637X/722/1/605},
archivePrefix = {arXiv},
       eprint = {1008.1761},
 primaryClass = {astro-ph.SR},
       adsurl = {https://ui.adsabs.harvard.edu/abs/2010ApJ...722..605H}
}

@ARTICLE{2004ApJ...601..979W,
       author = {{Wolff}, S.~C. and {Strom}, S.~E. and {Hillenbrand}, L.~A.},
        title = "{The Angular Momentum Evolution of 0.1-10 M$_{solar}$ Stars from the Birth Line to the Main Sequence}",
      journal = {\apj},
         year = 2004,
        month = feb,
       volume = {601},
       number = {2},
        pages = {979-999},
          doi = {10.1086/380503},
archivePrefix = {arXiv},
       eprint = {astro-ph/0310280},
 primaryClass = {astro-ph},
       adsurl = {https://ui.adsabs.harvard.edu/abs/2004ApJ...601..979W}
}

@ARTICLE{1979ApJ...230..485A,
       author = {{Abt}, H.~A.},
        title = "{The occurence of abnormal stars in open clusters.}",
      journal = {\apj},
         year = 1979,
        month = jun,
       volume = {230},
        pages = {485-496},
          doi = {10.1086/157104},
       adsurl = {https://ui.adsabs.harvard.edu/abs/1979ApJ...230..485A}
}

@ARTICLE{1982A&A...109...48M,
       author = {{Mermilliod}, J.~C.},
        title = "{Stellar content of young open clusters. II. Be stars.}",
      journal = {\aap},
         year = 1982,
        month = may,
       volume = {109},
        pages = {48-65},
       adsurl = {https://ui.adsabs.harvard.edu/abs/1982A&A...109...48M}
}

@ARTICLE{1985ApJS...59..769S,
       author = {{Slettebak}, A.},
        title = "{Be stars in open clusters.}",
      journal = {\apjs},
         year = 1985,
        month = dec,
       volume = {59},
        pages = {769-784},
          doi = {10.1086/191084},
       adsurl = {https://ui.adsabs.harvard.edu/abs/1985ApJS...59..769S}
}

@ARTICLE{2008A&A...478..467E,
       author = {{Ekstr{\"o}m}, S. and {Meynet}, G. and {Maeder}, A. and {Barblan}, F.},
        title = "{Evolution towards the critical limit and the origin of Be stars}",
      journal = {\aap},
         year = 2008,
        month = feb,
       volume = {478},
       number = {2},
        pages = {467-485},
          doi = {10.1051/0004-6361:20078095},
archivePrefix = {arXiv},
       eprint = {0711.1735},
 primaryClass = {astro-ph},
       adsurl = {https://ui.adsabs.harvard.edu/abs/2008A&A...478..467E}
}

@ARTICLE{2013A&A...553A..25G,
       author = {{Granada}, A. and {Ekstr{\"o}m}, S. and {Georgy}, C. and {Krti{\v{c}}ka}, J. and {Owocki}, S. and {Meynet}, G. and {Maeder}, A.},
        title = "{Populations of rotating stars. II. Rapid rotators and their link to Be-type stars}",
      journal = {\aap},
         year = 2013,
        month = may,
       volume = {553},
          eid = {A25},
        pages = {A25},
          doi = {10.1051/0004-6361/201220559},
archivePrefix = {arXiv},
       eprint = {1303.2393},
 primaryClass = {astro-ph.SR},
       adsurl = {https://ui.adsabs.harvard.edu/abs/2013A&A...553A..25G}
}

@ARTICLE{2020A&A...633A.165H,
       author = {{Hastings}, Ben and {Wang}, Chen and {Langer}, Norbert},
        title = "{The single star path to Be stars}",
      journal = {\aap},
         year = 2020,
        month = jan,
       volume = {633},
          eid = {A165},
        pages = {A165},
          doi = {10.1051/0004-6361/201937018},
archivePrefix = {arXiv},
       eprint = {1912.05290},
 primaryClass = {astro-ph.SR},
       adsurl = {https://ui.adsabs.harvard.edu/abs/2020A&A...633A.165H}
}

@ARTICLE{2011A&A...536A..65D,
       author = {{Dunstall}, P.~R. and {Brott}, I. and {Dufton}, P.~L. and {Lennon}, D.~J. and {Evans}, C.~J. and {Smartt}, S.~J. and {Hunter}, I.},
        title = "{The VLT-FLAMES survey of massive stars: Nitrogen abundances for Be-type stars in the Magellanic Clouds}",
      journal = {\aap},
         year = 2011,
        month = dec,
       volume = {536},
          eid = {A65},
        pages = {A65},
          doi = {10.1051/0004-6361/201117588},
archivePrefix = {arXiv},
       eprint = {1109.6661},
 primaryClass = {astro-ph.SR},
       adsurl = {https://ui.adsabs.harvard.edu/abs/2011A&A...536A..65D}
}

@ARTICLE{1866AN.....68...63S,
       author = {{Secchi}, Angelo},
        title = "{Schreiben des Herrn Prof. Secchi, Directors der Sternwarte des Collegio Romano, an den Herausgeber}",
      journal = {Astronomische Nachrichten},
         year = 1866,
        month = oct,
       volume = {68},
        pages = {63},
          doi = {10.1002/asna.18670680405},
       adsurl = {https://ui.adsabs.harvard.edu/abs/1866AN.....68...63S}
}

@ARTICLE{2015AJ....149....7C,
       author = {{Chojnowski}, S. Drew and {Whelan}, David G. and {Wisniewski}, John P. and {Majewski}, Steven R. and {Hall}, Matthew and {Shetrone}, Matthew and {Beaton}, Rachael and {Burton}, Adam and {Damke}, Guillermo and {Eikenberry}, Steve and {Hasselquist}, Sten and {Holtzman}, Jon A. and {M{\'e}sz{\'a}ros}, Szabolcs and {Nidever}, David and {Schneider}, Donald P. and {Wilson}, John and {Zasowski}, Gail and {Bizyaev}, Dmitry and {Brewington}, Howard and {Brinkmann}, J. and {Ebelke}, Garrett and {Frinchaboy}, Peter M. and {Kinemuchi}, Karen and {Malanushenko}, Elena and {Malanushenko}, Viktor and {Marchante}, Moses and {Oravetz}, Daniel and {Pan}, Kaike and {Simmons}, Audrey},
        title = "{High-Resolution H-Band Spectroscopy of Be Stars With SDSS-III/Apogee: I. New Be Stars, Line Identifications, and Line Profiles}",
      journal = {\aj},
         year = 2015,
        month = jan,
       volume = {149},
       number = {1},
          eid = {7},
        pages = {7},
          doi = {10.1088/0004-6256/149/1/7},
archivePrefix = {arXiv},
       eprint = {1409.4668},
 primaryClass = {astro-ph.SR},
       adsurl = {https://ui.adsabs.harvard.edu/abs/2015AJ....149....7C}
}

@ARTICLE{2015RAA....15.1325L,
       author = {{Lin}, Chien-Cheng and {Hou}, Jin-Liang and {Chen}, Li and {Shao}, Zheng-Yi and {Zhong}, Jing and {Yu}, Po-Chieh},
        title = "{Searching for classical Be stars in LAMOST DR1}",
      journal = {Research in Astronomy and Astrophysics},
         year = 2015,
        month = aug,
       volume = {15},
       number = {8},
          eid = {1325},
        pages = {1325},
          doi = {10.1088/1674-4527/15/8/015},
archivePrefix = {arXiv},
       eprint = {1505.07290},
 primaryClass = {astro-ph.SR},
       adsurl = {https://ui.adsabs.harvard.edu/abs/2015RAA....15.1325L}
}

@ARTICLE{2016RAA....16..138H,
       author = {{Hou}, Wen and {Luo}, A. -Li and {Hu}, Jing-Yao and {Yang}, Hai-Feng and {Du}, Chang-De and {Liu}, Chao and {Lee}, Chien-De and {Lin}, Chien-Cheng and {Wang}, Yue-Fei and {Zhang}, Yong and {Cao}, Zi-Huang and {Hou}, Yong-Hui},
        title = "{A catalog of early-type emission-line stars and H{\ensuremath{\alpha}} line profiles from LAMOST DR2}",
      journal = {Research in Astronomy and Astrophysics},
         year = 2016,
        month = sep,
       volume = {16},
       number = {9},
          eid = {138},
        pages = {138},
          doi = {10.1088/1674-4527/16/9/138},
archivePrefix = {arXiv},
       eprint = {1604.03704},
 primaryClass = {astro-ph.SR},
       adsurl = {https://ui.adsabs.harvard.edu/abs/2016RAA....16..138H}
}

@ARTICLE{1975BAICz..26...65K,
       author = {{Kriz}, S. and {Harmanec}, P.},
        title = "{A Hypothesis of the Binary Origin of Be Stars}",
      journal = {Bulletin of the Astronomical Institutes of Czechoslovakia},
         year = 1975,
        month = jan,
       volume = {26},
        pages = {65},
       adsurl = {https://ui.adsabs.harvard.edu/abs/1975BAICz..26...65K}
}

@ARTICLE{2011AJ....142..149N,
       author = {{Neiner}, C. and {de Batz}, B. and {Cochard}, F. and {Floquet}, M. and {Mekkas}, A. and {Desnoux}, V.},
        title = "{The Be Star Spectra (BeSS) Database}",
      journal = {\aj},
         year = 2011,
        month = nov,
       volume = {142},
       number = {5},
          eid = {149},
        pages = {149},
          doi = {10.1088/0004-6256/142/5/149},
       adsurl = {https://ui.adsabs.harvard.edu/abs/2011AJ....142..149N}
}

@ARTICLE{1991A&A...241..419P,
       author = {{Pols}, O.~R. and {Cote}, J. and {Waters}, L.~B.~F.~M. and {Heise}, J.},
        title = "{The formation of Be stars through close binary evolution.}",
      journal = {\aap},
         year = 1991,
        month = jan,
       volume = {241},
        pages = {419},
       adsurl = {https://ui.adsabs.harvard.edu/abs/1991A&A...241..419P}
}

@INPROCEEDINGS{1982IAUS...98..261J,
       author = {{Jaschek}, M. and {Egret}, D.},
        title = "{Catalog of Be stars.}",
    booktitle = {Be Stars},
         year = 1982,
       editor = {{Jaschek}, M. and {Groth}, H. -G.},
       series = {IAU Symposium},
       volume = {98},
        month = apr,
        pages = {261},
       adsurl = {https://ui.adsabs.harvard.edu/abs/1982IAUS...98..261J}
}

@ARTICLE{2005ApJS..161..118M,
       author = {{McSwain}, M. Virginia and {Gies}, Douglas R.},
        title = "{The Evolutionary Status of Be Stars: Results from a Photometric Study of Southern Open Clusters}",
      journal = {\apjs},
         year = 2005,
        month = nov,
       volume = {161},
       number = {1},
        pages = {118-146},
          doi = {10.1086/432757},
archivePrefix = {arXiv},
       eprint = {astro-ph/0505032},
 primaryClass = {astro-ph},
       adsurl = {https://ui.adsabs.harvard.edu/abs/2005ApJS..161..118M}
}

@ARTICLE{2014ApJ...796...37S,
       author = {{Shao}, Yong and {Li}, Xiang-Dong},
        title = "{On the Formation of Be Stars through Binary Interaction}",
      journal = {\apj},
         year = 2014,
        month = nov,
       volume = {796},
       number = {1},
          eid = {37},
        pages = {37},
          doi = {10.1088/0004-637X/796/1/37},
archivePrefix = {arXiv},
       eprint = {1410.0100},
 primaryClass = {astro-ph.HE},
       adsurl = {https://ui.adsabs.harvard.edu/abs/2014ApJ...796...37S}
}

@ARTICLE{2002MNRAS.329..897H,
       author = {{Hurley}, Jarrod R. and {Tout}, Christopher A. and {Pols}, Onno R.},
        title = "{Evolution of binary stars and the effect of tides on binary populations}",
      journal = {\mnras},
         year = 2002,
        month = feb,
       volume = {329},
       number = {4},
        pages = {897-928},
          doi = {10.1046/j.1365-8711.2002.05038.x},
archivePrefix = {arXiv},
       eprint = {astro-ph/0201220},
 primaryClass = {astro-ph},
       adsurl = {https://ui.adsabs.harvard.edu/abs/2002MNRAS.329..897H}
}

@ARTICLE{2011ApJ...728...86T,
       author = {{Tomsick}, John A. and {Heinke}, Craig and {Halpern}, Jules and {Kaaret}, Philip and {Chaty}, Sylvain and {Rodriguez}, Jerome and {Bodaghee}, Arash},
        title = "{Confirmation of IGR J01363+6610 as a Be X-ray Binary with Very Low Quiescent X-ray Luminosity}",
      journal = {\apj},
         year = 2011,
        month = feb,
       volume = {728},
       number = {2},
          eid = {86},
        pages = {86},
          doi = {10.1088/0004-637X/728/2/86},
archivePrefix = {arXiv},
       eprint = {1012.2817},
 primaryClass = {astro-ph.HE},
       adsurl = {https://ui.adsabs.harvard.edu/abs/2011ApJ...728...86T}
}

@ARTICLE{1999MNRAS.306..100R,
       author = {{Reig}, Pablo and {Roche}, Paul},
        title = "{Discovery of two new persistent Be/X-ray pulsar systems}",
      journal = {\mnras},
         year = 1999,
        month = jun,
       volume = {306},
       number = {1},
        pages = {100-106},
          doi = {10.1046/j.1365-8711.1999.02473.x},
archivePrefix = {arXiv},
       eprint = {astro-ph/9902221},
 primaryClass = {astro-ph},
       adsurl = {https://ui.adsabs.harvard.edu/abs/1999MNRAS.306..100R}
}

@ARTICLE{2018ApJ...853..156W,
       author = {{Wang}, Luqian and {Gies}, Douglas R. and {Peters}, Geraldine J.},
        title = "{Detection of Additional Be+sdO Systems from IUE Spectroscopy}",
      journal = {\apj},
         year = 2018,
        month = feb,
       volume = {853},
       number = {2},
          eid = {156},
        pages = {156},
          doi = {10.3847/1538-4357/aaa4b8},
archivePrefix = {arXiv},
       eprint = {1801.01066},
 primaryClass = {astro-ph.SR},
       adsurl = {https://ui.adsabs.harvard.edu/abs/2018ApJ...853..156W}
}

@ARTICLE{2021AJ....161..248W,
       author = {{Wang}, Luqian and {Gies}, Douglas R. and {Peters}, Geraldine J. and {G{\"o}tberg}, Ylva and {Chojnowski}, S. Drew and {Lester}, Kathryn V. and {Howell}, Steve B.},
        title = "{The Detection and Characterization of Be+sdO Binaries from HST/STIS FUV Spectroscopy}",
      journal = {\aj},
         year = 2021,
        month = may,
       volume = {161},
       number = {5},
          eid = {248},
        pages = {248},
          doi = {10.3847/1538-3881/abf144},
archivePrefix = {arXiv},
       eprint = {2103.13642},
 primaryClass = {astro-ph.SR},
       adsurl = {https://ui.adsabs.harvard.edu/abs/2021AJ....161..248W}
}

@ARTICLE{2025A&A...699L...1N,
       author = {{Nedhath}, Sneha and {Rani}, Sharmila and {Subramaniam}, Annapurni and {Pancino}, Elena},
        title = "{AstroSat/UVIT Study of NGC 663: First detection of Be+sdOB systems in a young star cluster}",
      journal = {\aap},
         year = 2025,
        month = jun,
       volume = {699},
          eid = {L1},
        pages = {L1},
          doi = {10.1051/0004-6361/202555495},
archivePrefix = {arXiv},
       eprint = {2506.08126},
 primaryClass = {astro-ph.SR},
       adsurl = {https://ui.adsabs.harvard.edu/abs/2025A&A...699L...1N}
}

@ARTICLE{2012RAA....12.1197C,
       author = {{Cui}, Xiang-Qun and {Zhao}, Yong-Heng and {Chu}, Yao-Quan and {Li}, Guo-Ping and {Li}, Qi and {Zhang}, Li-Ping and {Su}, Hong-Jun and {Yao}, Zheng-Qiu and {Wang}, Ya-Nan and {Xing}, Xiao-Zheng and {Li}, Xin-Nan and {Zhu}, Yong-Tian and {Wang}, Gang and {Gu}, Bo-Zhong and {Luo}, A. -Li and {Xu}, Xin-Qi and {Zhang}, Zhen-Chao and {Liu}, Gen-Rong and {Zhang}, Hao-Tong and {Yang}, De-Hua and {Cao}, Shu-Yun and {Chen}, Hai-Yuan and {Chen}, Jian-Jun and {Chen}, Kun-Xin and {Chen}, Ying and {Chu}, Jia-Ru and {Feng}, Lei and {Gong}, Xue-Fei and {Hou}, Yong-Hui and {Hu}, Hong-Zhuan and {Hu}, Ning-Sheng and {Hu}, Zhong-Wen and {Jia}, Lei and {Jiang}, Fang-Hua and {Jiang}, Xiang and {Jiang}, Zi-Bo and {Jin}, Ge and {Li}, Ai-Hua and {Li}, Yan and {Li}, Ye-Ping and {Liu}, Guan-Qun and {Liu}, Zhi-Gang and {Lu}, Wen-Zhi and {Mao}, Yin-Dun and {Men}, Li and {Qi}, Yong-Jun and {Qi}, Zhao-Xiang and {Shi}, Huo-Ming and {Tang}, Zheng-Hong and {Tao}, Qing-Sheng and {Wang}, Da-Qi and {Wang}, Dan and {Wang}, Guo-Min and {Wang}, Hai and {Wang}, Jia-Ning and {Wang}, Jian and {Wang}, Jian-Ling and {Wang}, Jian-Ping and {Wang}, Lei and {Wang}, Shu-Qing and {Wang}, You and {Wang}, Yue-Fei and {Xu}, Ling-Zhe and {Xu}, Yan and {Yang}, Shi-Hai and {Yu}, Yong and {Yuan}, Hui and {Yuan}, Xiang-Yan and {Zhai}, Chao and {Zhang}, Jing and {Zhang}, Yan-Xia and {Zhang}, Yong and {Zhao}, Ming and {Zhou}, Fang and {Zhou}, Guo-Hua and {Zhu}, Jie and {Zou}, Si-Cheng},
        title = "{The Large Sky Area Multi-Object Fiber Spectroscopic Telescope (LAMOST)}",
      journal = {Research in Astronomy and Astrophysics},
         year = 2012,
        month = sep,
       volume = {12},
       number = {9},
        pages = {1197-1242},
          doi = {10.1088/1674-4527/12/9/003},
       adsurl = {https://ui.adsabs.harvard.edu/abs/2012RAA....12.1197C}
}

@ARTICLE{2015RAA....15.1095L,
       author = {{Luo}, A. -Li and {Zhao}, Yong-Heng and {Zhao}, Gang and {Deng}, Li-Cai and {Liu}, Xiao-Wei and {Jing}, Yi-Peng and {Wang}, Gang and {Zhang}, Hao-Tong and {Shi}, Jian-Rong and {Cui}, Xiang-Qun and {Chu}, Yao-Quan and {Li}, Guo-Ping and {Bai}, Zhong-Rui and {Wu}, Yue and {Cai}, Yan and {Cao}, Shu-Yun and {Cao}, Zi-Huang and {Carlin}, Jeffrey L. and {Chen}, Hai-Yuan and {Chen}, Jian-Jun and {Chen}, Kun-Xin and {Chen}, Li and {Chen}, Xue-Lei and {Chen}, Xiao-Yan and {Chen}, Ying and {Christlieb}, Norbert and {Chu}, Jia-Ru and {Cui}, Chen-Zhou and {Dong}, Yi-Qiao and {Du}, Bing and {Fan}, Dong-Wei and {Feng}, Lei and {Fu}, Jian-Ning and {Gao}, Peng and {Gong}, Xue-Fei and {Gu}, Bo-Zhong and {Guo}, Yan-Xin and {Han}, Zhan-Wen and {He}, Bo-Liang and {Hou}, Jin-Liang and {Hou}, Yong-Hui and {Hou}, Wen and {Hu}, Hong-Zhuan and {Hu}, Ning-Sheng and {Hu}, Zhong-Wen and {Huo}, Zhi-Ying and {Jia}, Lei and {Jiang}, Fang-Hua and {Jiang}, Xiang and {Jiang}, Zhi-Bo and {Jin}, Ge and {Kong}, Xiao and {Kong}, Xu and {Lei}, Ya-Juan and {Li}, Ai-Hua and {Li}, Chang-Hua and {Li}, Guang-Wei and {Li}, Hai-Ning and {Li}, Jian and {Li}, Qi and {Li}, Shuang and {Li}, Sha-Sha and {Li}, Xin-Nan and {Li}, Yan and {Li}, Yin-Bi and {Li}, Ye-Ping and {Liang}, Yuan and {Lin}, Chien-Cheng and {Liu}, Chao and {Liu}, Gen-Rong and {Liu}, Guan-Qun and {Liu}, Zhi-Gang and {Lu}, Wen-Zhi and {Luo}, Yu and {Mao}, Yin-Dun and {Newberg}, Heidi and {Ni}, Ji-Jun and {Qi}, Zhao-Xiang and {Qi}, Yong-Jun and {Shen}, Shi-Yin and {Shi}, Huo-Ming and {Song}, Jing and {Song}, Yi-Han and {Su}, Ding-Qiang and {Su}, Hong-Jun and {Tang}, Zheng-Hong and {Tao}, Qing-Sheng and {Tian}, Yuan and {Wang}, Dan and {Wang}, Da-Qi and {Wang}, Feng-Fei and {Wang}, Guo-Min and {Wang}, Hai and {Wang}, Hong-Chi and {Wang}, Jian and {Wang}, Jia-Ning and {Wang}, Jian-Ling and {Wang}, Jian-Ping and {Wang}, Jun-Xian and {Wang}, Lei and {Wang}, Meng-Xin and {Wang}, Shou-Guan and {Wang}, Shu-Qing and {Wang}, Xia and {Wang}, Ya-Nan and {Wang}, You and {Wang}, Yue-Fei and {Wang}, You-Fen and {Wei}, Peng and {Wei}, Ming-Zhi and {Wu}, Hong and {Wu}, Ke-Fei and {Wu}, Xue-Bing and {Wu}, Yu-Zhong and {Xing}, Xiao-Zheng and {Xu}, Ling-Zhe and {Xu}, Xin-Qi and {Xu}, Yan and {Yan}, Tai-Sheng and {Yang}, De-Hua and {Yang}, Hai-Feng and {Yang}, Hui-Qin and {Yang}, Ming and {Yao}, Zheng-Qiu and {Yu}, Yong and {Yuan}, Hui and {Yuan}, Hai-Bo and {Yuan}, Hai-Long and {Yuan}, Wei-Min and {Zhai}, Chao and {Zhang}, En-Peng and {Zhang}, Hua-Wei and {Zhang}, Jian-Nan and {Zhang}, Li-Pin and {Zhang}, Wei and {Zhang}, Yong and {Zhang}, Yan-Xia and {Zhang}, Zheng-Chao and {Zhao}, Ming and {Zhou}, Fang and {Zhou}, Xu and {Zhu}, Jie and {Zhu}, Yong-Tian and {Zou}, Si-Cheng and {Zuo}, Fang},
        title = "{The first data release (DR1) of the LAMOST regular survey}",
      journal = {Research in Astronomy and Astrophysics},
         year = 2015,
        month = aug,
       volume = {15},
       number = {8},
          eid = {1095},
        pages = {1095},
          doi = {10.1088/1674-4527/15/8/002},
archivePrefix = {arXiv},
       eprint = {1505.01570},
 primaryClass = {astro-ph.GA},
       adsurl = {https://ui.adsabs.harvard.edu/abs/2015RAA....15.1095L}
}

@ARTICLE{2020arXiv200507210L,
       author = {{Liu}, Chao and {Fu}, Jianning and {Shi}, Jianrong and {Wu}, Hong and {Han}, Zhanwen and {Chen}, Li and {Dong}, Subo and {Zhao}, Yongheng and {Chen}, Jian-Jun and {Zhang}, Haotong and {Bai}, Zhong-Rui and {Chen}, Xuefei and {Cui}, Wenyuan and {Du}, Bing and {Hsia}, Chih-Hao and {Jiang}, Deng-Kai and {Hou}, Jinliang and {Hou}, Wen and {Li}, Haining and {Li}, Jiao and {Li}, Lifang and {Liu}, Jiaming and {Liu}, Jifeng and {Luo}, A-Li and {Ren}, Juan-Juan and {Tian}, Hai-Jun and {Tian}, Hao and {Wang}, Jia-Xin and {Wu}, Chao-Jian and {Xie}, Ji-Wei and {Yan}, Hong-Liang and {Yang}, Fan and {Yu}, Jincheng and {Zhang}, Bo and {Zhang}, Huawei and {Zhang}, Li-Yun and {Zhang}, Wei and {Zhao}, Gang and {Zhong}, Jing and {Zong}, Weikai and {Zuo}, Fang},
        title = "{LAMOST Medium-Resolution Spectroscopic Survey (LAMOST-MRS): Scientific goals and survey plan}",
      journal = {arXiv e-prints},
         year = 2020,
        month = may,
          eid = {arXiv:2005.07210},
        pages = {arXiv:2005.07210},
          doi = {10.48550/arXiv.2005.07210},
archivePrefix = {arXiv},
       eprint = {2005.07210},
 primaryClass = {astro-ph.SR},
       adsurl = {https://ui.adsabs.harvard.edu/abs/2020arXiv200507210L}
}

@INPROCEEDINGS{2018SPIE10707E..2BL,
       author = {{Luo}, A. -Li and {Chen}, Jia-Jun and {Hou}, Wen and {Zuo}, Fang and {Du}, Bing and {Song}, Yi-Han and {Kong}, Xiao and {Wang}, Rui and {Lu}, Yan and {Zhao}, Yong-Heng},
        title = "{LAMOST staller parameters pipeline for medium resolution spectra}",
    booktitle = {Software and Cyberinfrastructure for Astronomy V},
         year = 2018,
       editor = {{Guzman}, Juan C. and {Ibsen}, Jorge},
       series = {Society of Photo-Optical Instrumentation Engineers (SPIE) Conference Series},
       volume = {10707},
        month = jul,
          eid = {107072B},
        pages = {107072B},
          doi = {10.1117/12.2312433},
       adsurl = {https://ui.adsabs.harvard.edu/abs/2018SPIE10707E..2BL}
}

@ARTICLE{2021RAA....21..202D,
       author = {{Du}, Bing and {Luo}, A. -Li and {Zhang}, Shuo and {Kong}, Xiao and {Guo}, Yan-Xin and {Li}, Yin-Bi and {Zuo}, Fang and {Wang}, You-Fen and {Chen}, Jian-Jun and {Zhao}, Yong-Heng},
        title = "{LASPM: the LAMOST stellar parameter pipeline for M-type stars and application to the sixth and seventh data release (DR6 and DR7)}",
      journal = {Research in Astronomy and Astrophysics},
         year = 2021,
        month = oct,
       volume = {21},
       number = {8},
          eid = {202},
        pages = {202},
          doi = {10.1088/1674-4527/21/8/202},
       adsurl = {https://ui.adsabs.harvard.edu/abs/2021RAA....21..202D}
}

@ARTICLE{2021RAA....21..292W,
       author = {{Wang}, Song and {Zhang}, Hao-Tong and {Bai}, Zhong-Rui and {Yuan}, Hai-Long and {Xiang}, Mao-Sheng and {Zhang}, Bo and {Hou}, Wen and {Zuo}, Fang and {Du}, Bing and {Li}, Tan-Da and {Yang}, Fan and {Cui}, Kai-Ming and {Wang}, Yi-Lun and {Li}, Jiao and {Kovalev}, Mikhail and {Li}, Chun-Qian and {Tian}, Hao and {Zong}, Wei-Kai and {Han}, Heng-Geng and {Liu}, Chao and {Luo}, A. -Li and {Shi}, Jian-Rong and {Fu}, Jian-Ning and {Bi}, Shao-Lan and {Han}, Zhan-Wen and {Liu}, Ji-Feng},
        title = "{LAMOST Time-Domain survey: first results of four K2 plates}",
      journal = {Research in Astronomy and Astrophysics},
         year = 2021,
        month = dec,
       volume = {21},
       number = {11},
          eid = {292},
        pages = {292},
          doi = {10.1088/1674-4527/21/11/292},
archivePrefix = {arXiv},
       eprint = {2109.03149},
 primaryClass = {astro-ph.SR},
       adsurl = {https://ui.adsabs.harvard.edu/abs/2021RAA....21..292W}
}

@ARTICLE{2006A&A...458..285K,
       author = {{Kahabka}, P. and {Haberl}, F. and {Payne}, J.~L. and {Filipovi{\'c}}, M.~D.},
        title = "{The super-soft source XMMU J052016.0-692505 in the LMC. A likely white dwarf Be/X-ray binary}",
      journal = {\aap},
         year = 2006,
        month = oct,
       volume = {458},
       number = {1},
        pages = {285-292},
          doi = {10.1051/0004-6361:20065490},
       adsurl = {https://ui.adsabs.harvard.edu/abs/2006A&A...458..285K}
}

@ARTICLE{2021MNRAS.508..781K,
       author = {{Kennea}, J.~A. and {Coe}, M.~J. and {Evans}, P.~A. and {Townsend}, L.~J. and {Campbell}, Z.~A. and {Udalski}, A.},
        title = "{Swift J011511.0-725611: discovery of a rare Be star/white dwarf binary system in the SMC}",
      journal = {\mnras},
         year = 2021,
        month = nov,
       volume = {508},
       number = {1},
        pages = {781-788},
          doi = {10.1093/mnras/stab2632},
archivePrefix = {arXiv},
       eprint = {2109.05307},
 primaryClass = {astro-ph.HE},
       adsurl = {https://ui.adsabs.harvard.edu/abs/2021MNRAS.508..781K}
}

@ARTICLE{2025ApJ...980L..36M,
       author = {{Marino}, A. and {Yang}, H.~N. and {Coti Zelati}, F. and {Rea}, N. and {Guillot}, S. and {Jaisawal}, G.~K. and {Maitra}, C. and {Ness}, J. -U. and {Haberl}, F. and {Kuulkers}, E. and {Yuan}, W. and {Feng}, H. and {Tao}, L. and {Jin}, C. and {Sun}, H. and {Zhang}, W. and {Chen}, W. and {van den Heuvel}, E.~P.~J. and {Soria}, R. and {Zhang}, B. and {Weng}, S. -S. and {Ji}, L. and {Zhang}, G.~B. and {Pan}, X. and {Lv}, Z. and {Zhang}, C. and {Ling}, Z.~X. and {Chen}, Y. and {Jia}, S. and {Liu}, Y. and {Cheng}, H.~Q. and {Li}, D.~Y. and {Gendreau}, K. and {Ng}, M. and {Strohmayer}, T.},
        title = "{Einstein Probe Discovery of EP J005245.1‑722843: A Rare Be{\textendash}White Dwarf Binary in the Small Magellanic Cloud?}",
      journal = {\apjl},
         year = 2025,
        month = feb,
       volume = {980},
       number = {2},
          eid = {L36},
        pages = {L36},
          doi = {10.3847/2041-8213/ad9580},
archivePrefix = {arXiv},
       eprint = {2407.21371},
 primaryClass = {astro-ph.HE},
       adsurl = {https://ui.adsabs.harvard.edu/abs/2025ApJ...980L..36M}
}

@ARTICLE{2022ApJS..260...35W,
       author = {{Wang}, Luqian and {Li}, Jiao and {Wu}, You and {Gies}, Douglas R. and {Liu}, Jin Zhong and {Liu}, Chao and {Guo}, Yanjun and {Chen}, Xuefei and {Han}, Zhanwen},
        title = "{Identification of New Classical Be Stars from the LAMOST Medium Resolution Survey}",
      journal = {\apjs},
         year = 2022,
        month = jun,
       volume = {260},
       number = {2},
          eid = {35},
        pages = {35},
          doi = {10.3847/1538-4365/ac617a},
archivePrefix = {arXiv},
       eprint = {2203.15289},
 primaryClass = {astro-ph.SR},
       adsurl = {https://ui.adsabs.harvard.edu/abs/2022ApJS..260...35W}
}

@ARTICLE{2016MNRAS.463.1162C,
       author = {{Chen}, P.~S. and {Liu}, J.~Y. and {Shan}, H.~G.},
        title = "{A new approach to the infrared photometric study of Be stars}",
      journal = {\mnras},
         year = 2016,
        month = dec,
       volume = {463},
       number = {2},
        pages = {1162-1172},
          doi = {10.1093/mnras/stw1757},
       adsurl = {https://ui.adsabs.harvard.edu/abs/2016MNRAS.463.1162C}
}

@ARTICLE{Vioque2020,
       author = {{Vioque}, M. and {Oudmaijer}, R.~D. and {Schreiner}, M. and {Mendigut{\'\i}a}, I. and {Baines}, D. and {Mowlavi}, N. and {P{\'e}rez-Mart{\'\i}nez}, R.},
        title = "{Catalogue of new Herbig Ae/Be and classical Be stars. A machine learning approach to Gaia DR2}",
      journal = {\aap},
         year = 2020,
        month = jun,
       volume = {638},
          eid = {A21},
        pages = {A21},
          doi = {10.1051/0004-6361/202037731},
archivePrefix = {arXiv},
       eprint = {2005.01727},
 primaryClass = {astro-ph.SR},
       adsurl = {https://ui.adsabs.harvard.edu/abs/2020A&A...638A..21V}
}

@ARTICLE{2021RAA....21..288S,
       author = {{Shridharan}, Baskaran and {Mathew}, Blesson and {Nidhi}, Sabu and {Anusha}, Ravikumar and {Arun}, Roy and {Kartha}, Sreeja S. and {Kumar}, Yerra Bharat},
        title = "{Discovery of 2716 hot emission-line stars from LAMOST DR5}",
      journal = {Research in Astronomy and Astrophysics},
         year = 2021,
        month = dec,
       volume = {21},
       number = {11},
          eid = {288},
        pages = {288},
          doi = {10.1088/1674-4527/21/11/288},
archivePrefix = {arXiv},
       eprint = {2108.08025},
 primaryClass = {astro-ph.SR},
       adsurl = {https://ui.adsabs.harvard.edu/abs/2021RAA....21..288S}
}

@ARTICLE{2000A&AS..143....9W,
       author = {{Wenger}, M. and {Ochsenbein}, F. and {Egret}, D. and {Dubois}, P. and {Bonnarel}, F. and {Borde}, S. and {Genova}, F. and {Jasniewicz}, G. and {Lalo{\"e}}, S. and {Lesteven}, S. and {Monier}, R.},
        title = "{The SIMBAD astronomical database. The CDS reference database for astronomical objects}",
      journal = {\aaps},
         year = 2000,
        month = apr,
       volume = {143},
        pages = {9-22},
          doi = {10.1051/aas:2000332},
archivePrefix = {arXiv},
       eprint = {astro-ph/0002110},
 primaryClass = {astro-ph},
       adsurl = {https://ui.adsabs.harvard.edu/abs/2000A&AS..143....9W}
}

@ARTICLE{1984A&AS...55..109F,
       author = {{Finkenzeller}, U. and {Mundt}, R.},
        title = "{The Herbig Ae/Be stars associated with nebulosity.}",
      journal = {\aaps},
         year = 1984,
        month = jan,
       volume = {55},
        pages = {109-141},
       adsurl = {https://ui.adsabs.harvard.edu/abs/1984A&AS...55..109F}
}

@ARTICLE{2006AJ....131.1163S,
       author = {{Skrutskie}, M.~F. and {Cutri}, R.~M. and {Stiening}, R. and {Weinberg}, M.~D. and {Schneider}, S. and {Carpenter}, J.~M. and {Beichman}, C. and {Capps}, R. and {Chester}, T. and {Elias}, J. and {Huchra}, J. and {Liebert}, J. and {Lonsdale}, C. and {Monet}, D.~G. and {Price}, S. and {Seitzer}, P. and {Jarrett}, T. and {Kirkpatrick}, J.~D. and {Gizis}, J.~E. and {Howard}, E. and {Evans}, T. and {Fowler}, J. and {Fullmer}, L. and {Hurt}, R. and {Light}, R. and {Kopan}, E.~L. and {Marsh}, K.~A. and {McCallon}, H.~L. and {Tam}, R. and {Van Dyk}, S. and {Wheelock}, S.},
        title = "{The Two Micron All Sky Survey (2MASS)}",
      journal = {\aj},
         year = 2006,
        month = feb,
       volume = {131},
       number = {2},
        pages = {1163-1183},
          doi = {10.1086/498708},
       adsurl = {https://ui.adsabs.harvard.edu/abs/2006AJ....131.1163S}
}

@ARTICLE{2010AJ....140.1868W,
       author = {{Wright}, Edward L. and {Eisenhardt}, Peter R.~M. and {Mainzer}, Amy K. and {Ressler}, Michael E. and {Cutri}, Roc M. and {Jarrett}, Thomas and {Kirkpatrick}, J. Davy and {Padgett}, Deborah and {McMillan}, Robert S. and {Skrutskie}, Michael and {Stanford}, S.~A. and {Cohen}, Martin and {Walker}, Russell G. and {Mather}, John C. and {Leisawitz}, David and {Gautier}, III, Thomas N. and {McLean}, Ian and {Benford}, Dominic and {Lonsdale}, Carol J. and {Blain}, Andrew and {Mendez}, Bryan and {Irace}, William R. and {Duval}, Valerie and {Liu}, Fengchuan and {Royer}, Don and {Heinrichsen}, Ingolf and {Howard}, Joan and {Shannon}, Mark and {Kendall}, Martha and {Walsh}, Amy L. and {Larsen}, Mark and {Cardon}, Joel G. and {Schick}, Scott and {Schwalm}, Mark and {Abid}, Mohamed and {Fabinsky}, Beth and {Naes}, Larry and {Tsai}, Chao-Wei},
        title = "{The Wide-field Infrared Survey Explorer (WISE): Mission Description and Initial On-orbit Performance}",
      journal = {\aj},
         year = 2010,
        month = dec,
       volume = {140},
       number = {6},
        pages = {1868-1881},
          doi = {10.1088/0004-6256/140/6/1868},
archivePrefix = {arXiv},
       eprint = {1008.0031},
 primaryClass = {astro-ph.IM},
       adsurl = {https://ui.adsabs.harvard.edu/abs/2010AJ....140.1868W}
}

@ARTICLE{2000A&AS..143...23O,
       author = {{Ochsenbein}, F. and {Bauer}, P. and {Marcout}, J.},
        title = "{The VizieR database of astronomical catalogues}",
      journal = {\aaps},
         year = 2000,
        month = apr,
       volume = {143},
        pages = {23-32},
          doi = {10.1051/aas:2000169},
archivePrefix = {arXiv},
       eprint = {astro-ph/0002122},
 primaryClass = {astro-ph},
       adsurl = {https://ui.adsabs.harvard.edu/abs/2000A&AS..143...23O}
}

@ARTICLE{1996BICDS..48...47O,
       author = {{Ochsenbein}, F.},
        title = "{VizieR, the new catalogue interface at CDS}",
      journal = {Bulletin d'Information du Centre de Donnees Stellaires},
         year = 1996,
        month = mar,
       volume = {48},
        pages = {47},
       adsurl = {https://ui.adsabs.harvard.edu/abs/1996BICDS..48...47O}
}

@dataset{2003yCat.2246....0C,
       author = {{Cutri}, R.~M. and {Skrutskie}, M.~F. and {van Dyk}, S. and {Beichman}, C.~A. and {Carpenter}, J.~M. and {Chester}, T. and {Cambresy}, L. and {Evans}, T. and {Fowler}, J. and {Gizis}, J. and {Howard}, E. and {Huchra}, J. and {Jarrett}, T. and {Kopan}, E.~L. and {Kirkpatrick}, J.~D. and {Light}, R.~M. and {Marsh}, K.~A. and {McCallon}, H. and {Schneider}, S. and {Stiening}, R. and {Sykes}, M. and {Weinberg}, M. and {Wheaton}, W.~A. and {Wheelock}, S. and {Zacarias}, N.},
        title = "{VizieR Online Data Catalog: 2MASS All-Sky Catalog of Point Sources (Cutri+ 2003)}",
 howpublished = {VizieR On-line Data Catalog: II/246.  Originally published in: University of Massachusetts and Infrared Processing and Analysis Center, (IPAC/California Institute of Technology) (2003)},
         year = 2003,
        month = jun,
          eid = {II/246},
       adsurl = {https://ui.adsabs.harvard.edu/abs/2003yCat.2246....0C}
}

@dataset{2014yCat.2328....0C,
       author = {{Cutri}, R.~M. and {Wright}, E.~L. and {Conrow}, T. and {Fowler}, J.~W. and {Eisenhardt}, P.~R.~M. and {Grillmair}, C. and {Kirkpatrick}, J.~D. and {Masci}, F. and {McCallon}, H.~L. and {Wheelock}, S.~L. and {Fajardo-Acosta}, S. and {Yan}, L. and {Benford}, D. and {Harbut}, M. and {Jarrett}, T. and {Lake}, S. and {Leisawitz}, D. and {Ressler}, M.~E. and {Stanford}, S.~A. and {Tsai}, C. -W. and {Liu}, F. and {Helou}, G. and {Mainzer}, A. and {Gettngs}, D. and {Gonzalez}, A. and {Hoffman}, D. and {Marsh}, K.~A. and {Padgett}, D. and {Skrutskie}, M.~F. and {Beck}, R. and {Papin}, M. and {Wittman}, M.},
        title = "{VizieR Online Data Catalog: AllWISE Data Release (Cutri+ 2013)}",
 howpublished = {VizieR On-line Data Catalog: II/328.  Originally published in: IPAC/Caltech (2013)},
         year = 2021,
        month = feb,
          eid = {II/328},
       adsurl = {https://ui.adsabs.harvard.edu/abs/2014yCat.2328....0C}
}

@ARTICLE{2021MNRAS.508.1788A,
       author = {{Am{\^o}res}, Eduardo B. and {Jesus}, Ricardo M. and {Moitinho}, Andr{\'e} and {Arsenijevic}, Vladan and {Levenhagen}, Ronaldo S. and {Marshall}, Douglas J. and {Kerber}, Leandro O. and {K{\"u}nzel}, Roseli and {Moura}, Rodrigo A.},
        title = "{GALExtin: an alternative online tool to determine the interstellar extinction in the Milky Way}",
      journal = {\mnras},
         year = 2021,
        month = dec,
       volume = {508},
       number = {2},
        pages = {1788-1797},
          doi = {10.1093/mnras/stab2248},
archivePrefix = {arXiv},
       eprint = {2108.00561},
 primaryClass = {astro-ph.GA},
       adsurl = {https://ui.adsabs.harvard.edu/abs/2021MNRAS.508.1788A}
}

@ARTICLE{2019ApJ...887...93G,
       author = {{Green}, Gregory M. and {Schlafly}, Edward and {Zucker}, Catherine and {Speagle}, Joshua S. and {Finkbeiner}, Douglas},
        title = "{A 3D Dust Map Based on Gaia, Pan-STARRS 1, and 2MASS}",
      journal = {\apj},
         year = 2019,
        month = dec,
       volume = {887},
       number = {1},
          eid = {93},
        pages = {93},
          doi = {10.3847/1538-4357/ab5362},
archivePrefix = {arXiv},
       eprint = {1905.02734},
 primaryClass = {astro-ph.GA},
       adsurl = {https://ui.adsabs.harvard.edu/abs/2019ApJ...887...93G}
}

@ARTICLE{2021AJ....161..147B,
       author = {{Bailer-Jones}, C.~A.~L. and {Rybizki}, J. and {Fouesneau}, M. and {Demleitner}, M. and {Andrae}, R.},
        title = "{Estimating Distances from Parallaxes. V. Geometric and Photogeometric Distances to 1.47 Billion Stars in Gaia Early Data Release 3}",
      journal = {\aj},
         year = 2021,
        month = mar,
       volume = {161},
       number = {3},
          eid = {147},
        pages = {147},
          doi = {10.3847/1538-3881/abd806},
archivePrefix = {arXiv},
       eprint = {2012.05220},
 primaryClass = {astro-ph.SR},
       adsurl = {https://ui.adsabs.harvard.edu/abs/2021AJ....161..147B}
}

@ARTICLE{1989ApJ...345..245C,
       author = {{Cardelli}, Jason A. and {Clayton}, Geoffrey C. and {Mathis}, John S.},
        title = "{The Relationship between Infrared, Optical, and Ultraviolet Extinction}",
      journal = {\apj},
         year = 1989,
        month = oct,
       volume = {345},
        pages = {245},
          doi = {10.1086/167900},
       adsurl = {https://ui.adsabs.harvard.edu/abs/1989ApJ...345..245C}
}

@ARTICLE{2008A&A...492..277B,
       author = {{Bayo}, A. and {Rodrigo}, C. and {Barrado Y Navascu{\'e}s}, D. and {Solano}, E. and {Guti{\'e}rrez}, R. and {Morales-Calder{\'o}n}, M. and {Allard}, F.},
        title = "{VOSA: virtual observatory SED analyzer. An application to the Collinder 69 open cluster}",
      journal = {\aap},
         year = 2008,
        month = dec,
       volume = {492},
       number = {1},
        pages = {277-287},
          doi = {10.1051/0004-6361:200810395},
archivePrefix = {arXiv},
       eprint = {0808.0270},
 primaryClass = {astro-ph},
       adsurl = {https://ui.adsabs.harvard.edu/abs/2008A&A...492..277B}
}

@ARTICLE{2025arXiv250523151A,
       author = {{An}, Qian-Yu and {Huang}, Yang and {Gu}, Wei-Min and {Shao}, Yong and {Zhang}, Zhi-Xiang and {Yi}, Tuan and {Lailey}, B.~D. and {Sigut}, T.~A.~A. and {Akira Rocha}, Kyle and {Sun}, Meng and {Gossage}, Seth and {Gao}, Shi-Jie and {Weng}, Shan-Shan and {Wang}, Song and {Zhang}, Bowen and {Zhao}, Xinlin and {Qi}, Senyu and {Liao}, Shilong and {Ji}, Jianghui and {Wang}, Junfeng and {Wu}, Jianfeng and {Sun}, Mouyuan and {Li}, Xiang-Dong and {Liu}, Jifeng},
        title = "{A Be star-black hole binary with a wide orbit from LAMOST time-domain survey}",
      journal = {arXiv e-prints},
         year = 2025,
        month = may,
          eid = {arXiv:2505.23151},
        pages = {arXiv:2505.23151},
          doi = {10.48550/arXiv.2505.23151},
archivePrefix = {arXiv},
       eprint = {2505.23151},
 primaryClass = {astro-ph.SR},
       adsurl = {https://ui.adsabs.harvard.edu/abs/2025arXiv250523151A}
}

@ARTICLE{1979AJ.....84.1511T,
       author = {{Tonry}, J. and {Davis}, M.},
        title = "{A survey of galaxy redshifts. I. Data reduction techniques.}",
      journal = {\aj},
         year = 1979,
        month = oct,
       volume = {84},
        pages = {1511-1525},
          doi = {10.1086/112569},
       adsurl = {https://ui.adsabs.harvard.edu/abs/1979AJ.....84.1511T}
}

@ARTICLE{2024A&A...686L...2G,
       author = {{Gaia Collaboration} and {Panuzzo}, P. and {Mazeh}, T. and {Arenou}, F. and {Holl}, B. and {Caffau}, E. and {Jorissen}, A. and {Babusiaux}, C. and {Gavras}, P. and {Sahlmann}, J. and {Bastian}, U. and {Wyrzykowski}, {\L}. and {Eyer}, L. and {Leclerc}, N. and {Bauchet}, N. and {Bombrun}, A. and {Mowlavi}, N. and {Seabroke}, G.~M. and {Teyssier}, D. and {Balbinot}, E. and {Helmi}, A. and {Brown}, A.~G.~A. and {Vallenari}, A. and {Prusti}, T. and {de Bruijne}, J.~H.~J. and {Barbier}, A. and {Biermann}, M. and {Creevey}, O.~L. and {Ducourant}, C. and {Evans}, D.~W. and {Guerra}, R. and {Hutton}, A. and {Jordi}, C. and {Klioner}, S.~A. and {Lammers}, U. and {Lindegren}, L. and {Luri}, X. and {Mignard}, F. and {Nicolas}, C. and {Randich}, S. and {Sartoretti}, P. and {Smiljanic}, R. and {Tanga}, P. and {Walton}, N.~A. and {Aerts}, C. and {Bailer-Jones}, C.~A.~L. and {Cropper}, M. and {Drimmel}, R. and {Jansen}, F. and {Katz}, D. and {Lattanzi}, M.~G. and {Soubiran}, C. and {Th{\'e}venin}, F. and {van Leeuwen}, F. and {Andrae}, R. and {Audard}, M. and {Bakker}, J. and {Blomme}, R. and {Casta{\~n}eda}, J. and {De Angeli}, F. and {Fabricius}, C. and {Fouesneau}, M. and {Fr{\'e}mat}, Y. and {Galluccio}, L. and {Guerrier}, A. and {Heiter}, U. and {Masana}, E. and {Messineo}, R. and {Nienartowicz}, K. and {Pailler}, F. and {Riclet}, F. and {Roux}, W. and {Sordo}, R. and {Gracia-Abril}, G. and {Portell}, J. and {Altmann}, M. and {Benson}, K. and {Berthier}, J. and {Burgess}, P.~W. and {Busonero}, D. and {Busso}, G. and {Cacciari}, C. and {C{\'a}novas}, H. and {Carrasco}, J.~M. and {Carry}, B. and {Cellino}, A. and {Cheek}, N. and {Clementini}, G. and {Damerdji}, Y. and {Davidson}, M. and {de Teodoro}, P. and {Delchambre}, L. and {Dell'Oro}, A. and {Fraile Garcia}, E. and {Garabato}, D. and {Garc{\'\i}a-Lario}, P. and {Haigron}, R. and {Hambly}, N.~C. and {Harrison}, D.~L. and {Hatzidimitriou}, D. and {Hern{\'a}ndez}, J. and {Hestroffer}, D. and {Hodgkin}, S.~T. and {Jamal}, S. and {Jevardat de Fombelle}, G. and {Jordan}, S. and {Krone-Martins}, A. and {Lanzafame}, A.~C. and {L{\"o}ffler}, W. and {Lorca}, A. and {Marchal}, O. and {Marrese}, P.~M. and {Moitinho}, A. and {Muinonen}, K. and {Nu{\~n}ez Campos}, M. and {Oreshina-Slezak}, I. and {Osborne}, P. and {Pancino}, E. and {Pauwels}, T. and {Recio-Blanco}, A. and {Riello}, M. and {Rimoldini}, L. and {Robin}, A.~C. and {Roegiers}, T. and {Sarro}, L.~M. and {Schultheis}, M. and {Smith}, M. and {Sozzetti}, A. and {Utrilla}, E. and {van Leeuwen}, M. and {Weingrill}, K. and {Abbas}, U. and {{\'A}brah{\'a}m}, P. and {Abreu Aramburu}, A. and {Ahmed}, S. and {Altavilla}, G. and {{\'A}lvarez}, M.~A. and {Anders}, F. and {Anderson}, R.~I. and {Anglada Varela}, E. and {Antoja}, T. and {Baig}, S. and {Baines}, D. and {Baker}, S.~G. and {Balaguer-N{\'u}{\~n}ez}, L. and {Balog}, Z. and {Barache}, C. and {Barros}, M. and {Barstow}, M.~A. and {Bartolom{\'e}}, S. and {Bashi}, D. and {Bassilana}, J. -L. and {Baudeau}, N. and {Becciani}, U. and {Bedin}, L.~R. and {Bellas-Velidis}, I. and {Bellazzini}, M. and {Beordo}, W. and {Bernet}, M. and {Bertolotto}, C. and {Bertone}, S. and {Bianchi}, L. and {Binnenfeld}, A. and {Blanco-Cuaresma}, S. and {Bland-Hawthorn}, J. and {Blazere}, A. and {Boch}, T. and {Bossini}, D. and {Bouquillon}, S. and {Bragaglia}, A. and {Braine}, J. and {Bratsolis}, E. and {Breedt}, E. and {Bressan}, A. and {Brouillet}, N. and {Brugaletta}, E. and {Bucciarelli}, B. and {Butkevich}, A.~G. and {Buzzi}, R. and {Camut}, A. and {Cancelliere}, R. and {Cantat-Gaudin}, T. and {Capilla Guilarte}, D. and {Carballo}, R. and {Carlucci}, T. and {Carnerero}, M.~I. and {Carretero}, J. and {Carton}, S. and {Casamiquela}, L. and {Casey}, A. and {Castellani}, M. and {Castro-Ginard}, A. and {Ceraj}, L. and {Cesare}, V. and {Charlot}, P. and {Chaudet}, C. and {Chemin}, L. and {Chiavassa}, A. and {Chornay}, N. and {Chosson}, D.},
        title = "{Discovery of a dormant 33 solar-mass black hole in pre-release Gaia astrometry}",
      journal = {\aap},
         year = 2024,
        month = jun,
       volume = {686},
          eid = {L2},
        pages = {L2},
          doi = {10.1051/0004-6361/202449763},
archivePrefix = {arXiv},
       eprint = {2404.10486},
 primaryClass = {astro-ph.GA},
       adsurl = {https://ui.adsabs.harvard.edu/abs/2024A&A...686L...2G}
}

@ARTICLE{2025PASP..137i4202N,
       author = {{Nagarajan}, Pranav and {El-Badry}, Kareem and {Reggiani}, Henrique and {Lam}, Casey Y. and {Simon}, Joshua D. and {M{\"u}ller-Horn}, Johanna and {Seeburger}, Rhys and {Rix}, Hans-Walter and {Isaacson}, Howard and {Lu}, Jessica R. and {Chandra}, Vedant and {Andrae}, Rene},
        title = "{A Spectroscopic Search for Dormant Black Holes in Low-metallicity Binaries}",
      journal = {\pasp},
         year = 2025,
        month = sep,
       volume = {137},
       number = {9},
          eid = {094202},
        pages = {094202},
          doi = {10.1088/1538-3873/adffb7},
archivePrefix = {arXiv},
       eprint = {2507.12532},
 primaryClass = {astro-ph.SR},
       adsurl = {https://ui.adsabs.harvard.edu/abs/2025PASP..137i4202N}
}

@Article{harris2020array,
title = {Array programming with {NumPy}},
author = {Charles R. Harris and K. Jarrod Millman and St{\\'{e}}fan J
van der Walt and Ralf Gommers and Pauli Virtanen and David
Cournapeau and Eric Wieser and Julian Taylor and Sebastian
Berg and Nathaniel J. Smith and Robert Kern and Matti Picus
and Stephan Hoyer and Marten H. van Kerkwijk and Matthew
Brett and Allan Haldane and Jaime Fern{\\'{a}}ndez del
R{\\'{i}}o and Mark Wiebe and Pearu Peterson and Pierre
G{\\'{e}}rard-Marchant and Kevin Sheppard and Tyler Reddy and
Warren Weckesser and Hameer Abbasi and Christoph Gohlke and
Travis E. Oliphant},
year = {2020},
month = sep,
journal = {Nature},
volume = {585},
number = {7825},
pages = {357--362},
doi = {10.1038/s41586-020-2649-2},
publisher = {Springer Science and Business Media {LLC}},
url = {https://doi.org/10.1038/s41586-020-2649-2}
}

@software{reback2020pandas,
author = {The pandas development team},
title = {pandas-dev/pandas: Pandas},
month = feb,
year = 2020,
publisher = {Zenodo},
version = {latest},
doi = {10.5281/zenodo.3509134},
url = {https://doi.org/10.5281/zenodo.3509134}
}

@ARTICLE{2021ApJS..256...14Z,
       author = {{Zhang}, Bo and {Li}, Jiao and {Yang}, Fan and {Xiong}, Jian-Ping and {Fu}, Jian-Ning and {Liu}, Chao and {Tian}, Hao and {Li}, Yin-Bi and {Wang}, Jia-Xin and {Liang}, Cai-Xia and {Zhou}, Yu-Tao and {Zong}, Weikai and {Yang}, Cheng-Qun and {Liu}, Nian and {Hou}, Yong-Hui},
        title = "{Self-consistent Stellar Radial Velocities from LAMOST Medium-resolution Survey DR7}",
      journal = {\apjs},
         year = 2021,
        month = sep,
       volume = {256},
       number = {1},
          eid = {14},
        pages = {14},
          doi = {10.3847/1538-4365/ac0834},
archivePrefix = {arXiv},
       eprint = {2105.11624},
 primaryClass = {astro-ph.SR},
       adsurl = {https://ui.adsabs.harvard.edu/abs/2021ApJS..256...14Z}
}

@ARTICLE{2020ApJS..246....9Z,
       author = {{Zhang}, Bo and {Liu}, Chao and {Deng}, Li-Cai},
        title = "{Deriving the Stellar Labels of LAMOST Spectra with the Stellar LAbel Machine (SLAM)}",
      journal = {\apjs},
         year = 2020,
        month = jan,
       volume = {246},
       number = {1},
          eid = {9},
        pages = {9},
          doi = {10.3847/1538-4365/ab55ef},
archivePrefix = {arXiv},
       eprint = {1908.08677},
 primaryClass = {astro-ph.SR},
       adsurl = {https://ui.adsabs.harvard.edu/abs/2020ApJS..246....9Z}
}

@Article{Hunter:2007,
Author = {Hunter, J. D.},
Title = {Matplotlib: A 2D graphics environment},
Journal = {Computing in Science \& Engineering},
Volume = {9},
Number = {3},
Pages = {90--95},
publisher = {IEEE COMPUTER SOC},
doi = {10.1109/MCSE.2007.55},
year = 2007
}

@ARTICLE{2013PASP..125..306F,
       author = {{Foreman-Mackey}, Daniel and {Hogg}, David W. and {Lang}, Dustin and {Goodman}, Jonathan},
        title = "{emcee: The MCMC Hammer}",
      journal = {\pasp},
         year = 2013,
        month = mar,
       volume = {125},
       number = {925},
        pages = {306},
          doi = {10.1086/670067},
archivePrefix = {arXiv},
       eprint = {1202.3665},
 primaryClass = {astro-ph.IM},
       adsurl = {https://ui.adsabs.harvard.edu/abs/2013PASP..125..306F}
}

@ARTICLE{2023A&A...674A...1G,
       author = {{Gaia Collaboration} and {Vallenari}, A. and {Brown}, A.~G.~A. and {Prusti}, T. and {de Bruijne}, J.~H.~J. and {Arenou}, F. and {Babusiaux}, C. and {Biermann}, M. and {Creevey}, O.~L. and {Ducourant}, C. and {Evans}, D.~W. and {Eyer}, L. and {Guerra}, R. and {Hutton}, A. and {Jordi}, C. and {Klioner}, S.~A. and {Lammers}, U.~L. and {Lindegren}, L. and {Luri}, X. and {Mignard}, F. and {Panem}, C. and {Pourbaix}, D. and {Randich}, S. and {Sartoretti}, P. and {Soubiran}, C. and {Tanga}, P. and {Walton}, N.~A. and {Bailer-Jones}, C.~A.~L. and {Bastian}, U. and {Drimmel}, R. and {Jansen}, F. and {Katz}, D. and {Lattanzi}, M.~G. and {van Leeuwen}, F. and {Bakker}, J. and {Cacciari}, C. and {Casta{\~n}eda}, J. and {De Angeli}, F. and {Fabricius}, C. and {Fouesneau}, M. and {Fr{\'e}mat}, Y. and {Galluccio}, L. and {Guerrier}, A. and {Heiter}, U. and {Masana}, E. and {Messineo}, R. and {Mowlavi}, N. and {Nicolas}, C. and {Nienartowicz}, K. and {Pailler}, F. and {Panuzzo}, P. and {Riclet}, F. and {Roux}, W. and {Seabroke}, G.~M. and {Sordo}, R. and {Th{\'e}venin}, F. and {Gracia-Abril}, G. and {Portell}, J. and {Teyssier}, D. and {Altmann}, M. and {Andrae}, R. and {Audard}, M. and {Bellas-Velidis}, I. and {Benson}, K. and {Berthier}, J. and {Blomme}, R. and {Burgess}, P.~W. and {Busonero}, D. and {Busso}, G. and {C{\'a}novas}, H. and {Carry}, B. and {Cellino}, A. and {Cheek}, N. and {Clementini}, G. and {Damerdji}, Y. and {Davidson}, M. and {de Teodoro}, P. and {Nu{\~n}ez Campos}, M. and {Delchambre}, L. and {Dell'Oro}, A. and {Esquej}, P. and {Fern{\'a}ndez-Hern{\'a}ndez}, J. and {Fraile}, E. and {Garabato}, D. and {Garc{\'\i}a-Lario}, P. and {Gosset}, E. and {Haigron}, R. and {Halbwachs}, J. -L. and {Hambly}, N.~C. and {Harrison}, D.~L. and {Hern{\'a}ndez}, J. and {Hestroffer}, D. and {Hodgkin}, S.~T. and {Holl}, B. and {Jan{\ss}en}, K. and {Jevardat de Fombelle}, G. and {Jordan}, S. and {Krone-Martins}, A. and {Lanzafame}, A.~C. and {L{\"o}ffler}, W. and {Marchal}, O. and {Marrese}, P.~M. and {Moitinho}, A. and {Muinonen}, K. and {Osborne}, P. and {Pancino}, E. and {Pauwels}, T. and {Recio-Blanco}, A. and {Reyl{\'e}}, C. and {Riello}, M. and {Rimoldini}, L. and {Roegiers}, T. and {Rybizki}, J. and {Sarro}, L.~M. and {Siopis}, C. and {Smith}, M. and {Sozzetti}, A. and {Utrilla}, E. and {van Leeuwen}, M. and {Abbas}, U. and {{\'A}brah{\'a}m}, P. and {Abreu Aramburu}, A. and {Aerts}, C. and {Aguado}, J.~J. and {Ajaj}, M. and {Aldea-Montero}, F. and {Altavilla}, G. and {{\'A}lvarez}, M.~A. and {Alves}, J. and {Anders}, F. and {Anderson}, R.~I. and {Anglada Varela}, E. and {Antoja}, T. and {Baines}, D. and {Baker}, S.~G. and {Balaguer-N{\'u}{\~n}ez}, L. and {Balbinot}, E. and {Balog}, Z. and {Barache}, C. and {Barbato}, D. and {Barros}, M. and {Barstow}, M.~A. and {Bartolom{\'e}}, S. and {Bassilana}, J. -L. and {Bauchet}, N. and {Becciani}, U. and {Bellazzini}, M. and {Berihuete}, A. and {Bernet}, M. and {Bertone}, S. and {Bianchi}, L. and {Binnenfeld}, A. and {Blanco-Cuaresma}, S. and {Blazere}, A. and {Boch}, T. and {Bombrun}, A. and {Bossini}, D. and {Bouquillon}, S. and {Bragaglia}, A. and {Bramante}, L. and {Breedt}, E. and {Bressan}, A. and {Brouillet}, N. and {Brugaletta}, E. and {Bucciarelli}, B. and {Burlacu}, A. and {Butkevich}, A.~G. and {Buzzi}, R. and {Caffau}, E. and {Cancelliere}, R. and {Cantat-Gaudin}, T. and {Carballo}, R. and {Carlucci}, T. and {Carnerero}, M.~I. and {Carrasco}, J.~M. and {Casamiquela}, L. and {Castellani}, M. and {Castro-Ginard}, A. and {Chaoul}, L. and {Charlot}, P. and {Chemin}, L. and {Chiaramida}, V. and {Chiavassa}, A. and {Chornay}, N. and {Comoretto}, G. and {Contursi}, G. and {Cooper}, W.~J. and {Cornez}, T. and {Cowell}, S. and {Crifo}, F. and {Cropper}, M. and {Crosta}, M. and {Crowley}, C. and {Dafonte}, C. and {Dapergolas}, A. and {David}, M. and {David}, P. and {de Laverny}, P. and {De Luise}, F. and {De March}, R.},
        title = "{Gaia Data Release 3. Summary of the content and survey properties}",
      journal = {\aap},
         year = 2023,
        month = jun,
       volume = {674},
          eid = {A1},
        pages = {A1},
          doi = {10.1051/0004-6361/202243940},
archivePrefix = {arXiv},
       eprint = {2208.00211},
 primaryClass = {astro-ph.GA},
       adsurl = {https://ui.adsabs.harvard.edu/abs/2023A&A...674A...1G}
}

@ARTICLE{1983PASP...95..311P,
       author = {{Peters}, G.~J.},
        title = "{Orbital motion and mass flow in the interacting binary Be star HR 2142.}",
      journal = {\pasp},
         year = 1983,
        month = may,
       volume = {95},
        pages = {311-318},
          doi = {10.1086/131164},
       adsurl = {https://ui.adsabs.harvard.edu/abs/1983PASP...95..311P}
}

@ARTICLE{1991A&A...250..437W,
       author = {{Waters}, L.~B.~F.~M. and {Cote}, J. and {Pols}, O.~R.},
        title = "{The nature of the Be binary HR 2142.}",
      journal = {\aap},
         year = 1991,
        month = oct,
       volume = {250},
        pages = {437},
       adsurl = {https://ui.adsabs.harvard.edu/abs/1991A&A...250..437W}
}

@ARTICLE{2025ApJ...993..192K,
       author = {{Kalari}, V.~M. and {Salinas}, R. and {S{\'a}ez-Carvajal}, C. and {Oudmaijer}, R.~D. and {Howell}, S. and {Caballero-Nieves}, S. and {Kamp}, K. and {Matson}, R. and {Scott}, N. and {Cao}, T. and {Hartman}, Z. and {Kim}, H.},
        title = "{A Search for Be Stars in Multiple Systems within the Solar Neighborhood}",
      journal = {\apj},
         year = 2025,
        month = nov,
       volume = {993},
       number = {2},
          eid = {192},
        pages = {192},
          doi = {10.3847/1538-4357/ae0731},
archivePrefix = {arXiv},
       eprint = {2509.19286},
 primaryClass = {astro-ph.SR},
       adsurl = {https://ui.adsabs.harvard.edu/abs/2025ApJ...993..192K}
}

@ARTICLE{2023A&A...673A.114H,
       author = {{Hunt}, Emily L. and {Reffert}, Sabine},
        title = "{Improving the open cluster census. II. An all-sky cluster catalogue with Gaia DR3}",
      journal = {\aap},
         year = 2023,
        month = may,
       volume = {673},
          eid = {A114},
        pages = {A114},
          doi = {10.1051/0004-6361/202346285},
archivePrefix = {arXiv},
       eprint = {2303.13424},
 primaryClass = {astro-ph.GA},
       adsurl = {https://ui.adsabs.harvard.edu/abs/2023A&A...673A.114H}
}

@ARTICLE{1954ApJ...119..625B,
       author = {{Blaauw}, A. and {Morgan}, W.~W.},
        title = "{The Space Motions of AE Aurigae and {\ensuremath{\mu}} Columbae with Respect to the Orion Nebula.}",
      journal = {\apj},
         year = 1954,
        month = may,
       volume = {119},
        pages = {625},
          doi = {10.1086/145866},
       adsurl = {https://ui.adsabs.harvard.edu/abs/1954ApJ...119..625B}
}

@BOOK{1957moas.book.....Z,
       author = {{Zwicky}, Fritz},
        title = "{Morphological astronomy}",
         year = 1957,
       adsurl = {https://ui.adsabs.harvard.edu/abs/1957moas.book.....Z}
}

@ARTICLE{1961BAN....15..265B,
       author = {{Blaauw}, A.},
        title = "{On the origin of the O- and B-type stars with high velocities (the ``run-away'' stars), and some related problems}",
      journal = {\bain},
         year = 1961,
        month = may,
       volume = {15},
        pages = {265},
       adsurl = {https://ui.adsabs.harvard.edu/abs/1961BAN....15..265B}
}

@ARTICLE{1967BOTT....4...86P,
       author = {{Poveda}, A. and {Ruiz}, J. and {Allen}, C.},
        title = "{Run-away Stars as the Result of the Gravitational Collapse of Proto-stellar Clusters}",
      journal = {Boletin de los Observatorios Tonantzintla y Tacubaya},
         year = 1967,
        month = apr,
       volume = {4},
        pages = {86-90},
       adsurl = {https://ui.adsabs.harvard.edu/abs/1967BOTT....4...86P}
}

@ARTICLE{2015ApJS..216...29B,
       author = {{Bovy}, Jo},
        title = "{galpy: A python Library for Galactic Dynamics}",
      journal = {\apjs},
         year = 2015,
        month = feb,
       volume = {216},
       number = {2},
          eid = {29},
        pages = {29},
          doi = {10.1088/0067-0049/216/2/29},
archivePrefix = {arXiv},
       eprint = {1412.3451},
 primaryClass = {astro-ph.GA},
       adsurl = {https://ui.adsabs.harvard.edu/abs/2015ApJS..216...29B}
}

@ARTICLE{2019A&A...625L..10G,
       author = {{GRAVITY Collaboration} and {Abuter}, R. and {Amorim}, A. and {Baub{\"o}ck}, M. and {Berger}, J.~P. and {Bonnet}, H. and {Brandner}, W. and {Cl{\'e}net}, Y. and {Coud{\'e} Du Foresto}, V. and {de Zeeuw}, P.~T. and {Dexter}, J. and {Duvert}, G. and {Eckart}, A. and {Eisenhauer}, F. and {F{\"o}rster Schreiber}, N.~M. and {Garcia}, P. and {Gao}, F. and {Gendron}, E. and {Genzel}, R. and {Gerhard}, O. and {Gillessen}, S. and {Habibi}, M. and {Haubois}, X. and {Henning}, T. and {Hippler}, S. and {Horrobin}, M. and {Jim{\'e}nez-Rosales}, A. and {Jocou}, L. and {Kervella}, P. and {Lacour}, S. and {Lapeyr{\`e}re}, V. and {Le Bouquin}, J. -B. and {L{\'e}na}, P. and {Ott}, T. and {Paumard}, T. and {Perraut}, K. and {Perrin}, G. and {Pfuhl}, O. and {Rabien}, S. and {Rodriguez Coira}, G. and {Rousset}, G. and {Scheithauer}, S. and {Sternberg}, A. and {Straub}, O. and {Straubmeier}, C. and {Sturm}, E. and {Tacconi}, L.~J. and {Vincent}, F. and {von Fellenberg}, S. and {Waisberg}, I. and {Widmann}, F. and {Wieprecht}, E. and {Wiezorrek}, E. and {Woillez}, J. and {Yazici}, S.},
        title = "{A geometric distance measurement to the Galactic center black hole with 0.3\% uncertainty}",
      journal = {\aap},
         year = 2019,
        month = may,
       volume = {625},
          eid = {L10},
        pages = {L10},
          doi = {10.1051/0004-6361/201935656},
archivePrefix = {arXiv},
       eprint = {1904.05721},
 primaryClass = {astro-ph.GA},
       adsurl = {https://ui.adsabs.harvard.edu/abs/2019A&A...625L..10G}
}

@ARTICLE{2016ARA&A..54..529B,
       author = {{Bland-Hawthorn}, Joss and {Gerhard}, Ortwin},
        title = "{The Galaxy in Context: Structural, Kinematic, and Integrated Properties}",
      journal = {\araa},
         year = 2016,
        month = sep,
       volume = {54},
        pages = {529-596},
          doi = {10.1146/annurev-astro-081915-023441},
archivePrefix = {arXiv},
       eprint = {1602.07702},
 primaryClass = {astro-ph.GA},
       adsurl = {https://ui.adsabs.harvard.edu/abs/2016ARA&A..54..529B}
}

@ARTICLE{2015MNRAS.449..162H,
       author = {{Huang}, Y. and {Liu}, X. -W. and {Yuan}, H. -B. and {Xiang}, M. -S. and {Huo}, Z. -Y. and {Chen}, B. -Q. and {Zhang}, Y. and {Hou}, Y. -H.},
        title = "{Determination of the local standard of rest using the LSS-GAC DR1}",
      journal = {\mnras},
         year = 2015,
        month = may,
       volume = {449},
       number = {1},
        pages = {162-174},
          doi = {10.1093/mnras/stv204},
archivePrefix = {arXiv},
       eprint = {1501.07095},
 primaryClass = {astro-ph.GA},
       adsurl = {https://ui.adsabs.harvard.edu/abs/2015MNRAS.449..162H}
}

@ARTICLE{2001ApJ...555..364B,
       author = {{Berger}, D.~H. and {Gies}, D.~R.},
        title = "{A Search for High-Velocity Be Stars}",
      journal = {\apj},
         year = 2001,
        month = jul,
       volume = {555},
       number = {1},
        pages = {364-367},
          doi = {10.1086/321461},
       adsurl = {https://ui.adsabs.harvard.edu/abs/2001ApJ...555..364B}
}

@ARTICLE{2024ApJS..272...45G,
       author = {{Guo}, Yanjun and {Wang}, Luqian and {Liu}, Chao and {Wu}, You and {Han}, ZhanWen and {Chen}, XueFei},
        title = "{A Catalog of Early-type Runaway Stars from LAMOST DR8}",
      journal = {\apjs},
         year = 2024,
        month = jun,
       volume = {272},
       number = {2},
          eid = {45},
        pages = {45},
          doi = {10.3847/1538-4365/ad46f8},
archivePrefix = {arXiv},
       eprint = {2405.04750},
 primaryClass = {astro-ph.SR},
       adsurl = {https://ui.adsabs.harvard.edu/abs/2024ApJS..272...45G}
}

@ARTICLE{2013A&A...550A.107S,
       author = {{Sana}, H. and {de Koter}, A. and {de Mink}, S.~E. and {Dunstall}, P.~R. and {Evans}, C.~J. and {H{\'e}nault-Brunet}, V. and {Ma{\'\i}z Apell{\'a}niz}, J. and {Ram{\'\i}rez-Agudelo}, O.~H. and {Taylor}, W.~D. and {Walborn}, N.~R. and {Clark}, J.~S. and {Crowther}, P.~A. and {Herrero}, A. and {Gieles}, M. and {Langer}, N. and {Lennon}, D.~J. and {Vink}, J.~S.},
        title = "{The VLT-FLAMES Tarantula Survey. VIII. Multiplicity properties of the O-type star population}",
      journal = {\aap},
         year = 2013,
        month = feb,
       volume = {550},
          eid = {A107},
        pages = {A107},
          doi = {10.1051/0004-6361/201219621},
archivePrefix = {arXiv},
       eprint = {1209.4638},
 primaryClass = {astro-ph.SR},
       adsurl = {https://ui.adsabs.harvard.edu/abs/2013A&A...550A.107S}
}

@ARTICLE{2021A&A...652A..70B,
       author = {{Bodensteiner}, J. and {Sana}, H. and {Wang}, C. and {Langer}, N. and {Mahy}, L. and {Banyard}, G. and {de Koter}, A. and {de Mink}, S.~E. and {Evans}, C.~J. and {G{\"o}tberg}, Y. and {Patrick}, L.~R. and {Schneider}, F.~R.~N. and {Tramper}, F.},
        title = "{The young massive SMC cluster NGC 330 seen by MUSE. II. Multiplicity properties of the massive-star population}",
      journal = {\aap},
         year = 2021,
        month = aug,
       volume = {652},
          eid = {A70},
        pages = {A70},
          doi = {10.1051/0004-6361/202140507},
archivePrefix = {arXiv},
       eprint = {2104.13409},
 primaryClass = {astro-ph.SR},
       adsurl = {https://ui.adsabs.harvard.edu/abs/2021A&A...652A..70B}
}

@ARTICLE{2022A&A...658A..69B,
       author = {{Banyard}, G. and {Sana}, H. and {Mahy}, L. and {Bodensteiner}, J. and {Villase{\~n}or}, J.~I. and {Evans}, C.~J.},
        title = "{The observed multiplicity properties of B-type stars in the Galactic young open cluster NGC 6231}",
      journal = {\aap},
         year = 2022,
        month = feb,
       volume = {658},
          eid = {A69},
        pages = {A69},
          doi = {10.1051/0004-6361/202141037},
archivePrefix = {arXiv},
       eprint = {2108.07814},
 primaryClass = {astro-ph.SR},
       adsurl = {https://ui.adsabs.harvard.edu/abs/2022A&A...658A..69B}
}

@ARTICLE{2025arXiv251108675V,
       author = {{Villase{\~n}or}, J.~I. and {Sana}, H. and {Bodensteiner}, J. and {Britavskiy}, N. and {Patrick}, L.~R. and {Shenar}, T. and {the BLOeM Collaboration}},
        title = "{The binary landscape of massive stars at low $Z$: Insights from the BLOeM Campaign}",
      journal = {arXiv e-prints},
         year = 2025,
        month = nov,
          eid = {arXiv:2511.08675},
        pages = {arXiv:2511.08675},
          doi = {10.48550/arXiv.2511.08675},
archivePrefix = {arXiv},
       eprint = {2511.08675},
 primaryClass = {astro-ph.SR},
       adsurl = {https://ui.adsabs.harvard.edu/abs/2025arXiv251108675V}
}

@INPROCEEDINGS{2026enap....2...43M,
       author = {{Ma{\'\i}z Apell{\'a}niz}, Jes{\'u}s and {Negueruela}, Ignacio and {Caballero}, Jos{\'e} A.},
        title = "{Spectral classification}",
    booktitle = {Encyclopedia of Astrophysics},
         year = 2026,
       volume = {2},
        month = jan,
        pages = {43-84},
          doi = {10.1016/B978-0-443-21439-4.00057-2},
archivePrefix = {arXiv},
       eprint = {2410.07301},
 primaryClass = {astro-ph.SR},
       adsurl = {https://ui.adsabs.harvard.edu/abs/2026enap....2...43M}
}

@ARTICLE{2017MNRAS.465...76M,
       author = {{McMillan}, Paul J.},
        title = "{The mass distribution and gravitational potential of the Milky Way}",
      journal = {\mnras},
         year = 2017,
        month = feb,
       volume = {465},
       number = {1},
        pages = {76-94},
          doi = {10.1093/mnras/stw2759},
archivePrefix = {arXiv},
       eprint = {1608.00971},
 primaryClass = {astro-ph.GA},
       adsurl = {https://ui.adsabs.harvard.edu/abs/2017MNRAS.465...76M}
}

@ARTICLE{2020MNRAS.494.4291C,
       author = {{Cautun}, Marius and {Ben{\'\i}tez-Llambay}, Alejandro and {Deason}, Alis J. and {Frenk}, Carlos S. and {Fattahi}, Azadeh and {G{\'o}mez}, Facundo A. and {Grand}, Robert J.~J. and {Oman}, Kyle A. and {Navarro}, Julio F. and {Simpson}, Christine M.},
        title = "{The milky way total mass profile as inferred from Gaia DR2}",
      journal = {\mnras},
         year = 2020,
        month = may,
       volume = {494},
       number = {3},
        pages = {4291-4313},
          doi = {10.1093/mnras/staa1017},
archivePrefix = {arXiv},
       eprint = {1911.04557},
 primaryClass = {astro-ph.GA},
       adsurl = {https://ui.adsabs.harvard.edu/abs/2020MNRAS.494.4291C}
}

@ARTICLE{2013A&A...549A.137I,
       author = {{Irrgang}, A. and {Wilcox}, B. and {Tucker}, E. and {Schiefelbein}, L.},
        title = "{Milky Way mass models for orbit calculations}",
      journal = {\aap},
         year = 2013,
        month = jan,
       volume = {549},
          eid = {A137},
        pages = {A137},
          doi = {10.1051/0004-6361/201220540},
archivePrefix = {arXiv},
       eprint = {1211.4353},
 primaryClass = {astro-ph.GA},
       adsurl = {https://ui.adsabs.harvard.edu/abs/2013A&A...549A.137I}
}

@ARTICLE{2025A&A...694A.172R,
       author = {{Rivinius}, Th. and {Klement}, R. and {Chojnowski}, S.~D. and {Baade}, D. and {Abdul-Masih}, M. and {Przybilla}, N. and {Guarro Fl{\'o}}, J. and {Heathcote}, B. and {Hadrava}, P. and {Gies}, D. and {Shepard}, K. and {Buil}, C. and {Garde}, O. and {Thizy}, O. and {Monnier}, J.~D. and {Anugu}, N. and {Lanthermann}, C. and {Schaefer}, G. and {Davies}, C. and {Kraus}, S. and {Ennis}, J. and {Setterholm}, B.~R. and {Gardner}, T. and {Ibrahim}, N. and {Chhabra}, S. and {Gutierrez}, M. and {Codron}, I.},
        title = "{Newborn Be star systems observed shortly after mass transfer}",
      journal = {\aap},
         year = 2025,
        month = feb,
       volume = {694},
          eid = {A172},
        pages = {A172},
          doi = {10.1051/0004-6361/202347275},
archivePrefix = {arXiv},
       eprint = {2412.09720},
 primaryClass = {astro-ph.SR},
       adsurl = {https://ui.adsabs.harvard.edu/abs/2025A&A...694A.172R}
}

@ARTICLE{2026ApJ...996...61R,
       author = {{Rocha}, Danilo F. and {Emilio}, Marcelo and {Labadie-Bartz}, Jonathan and {Neiner}, Coralie and {Bodensteiner}, Julia and {Shenar}, Tomer and {Andrade}, Laerte and {Abdul-Masih}, Michael and {Navarete}, Felipe and {Melo}, Alessandro and {Janot-Pacheco}, Eduardo and {Eleut{\'e}rio}, Romualdo and {Pereira}, Alan W.},
        title = "{The Triple System V1371 Tau: An Eclipsing Binary with an Outer Be Star}",
      journal = {\apj},
         year = 2026,
        month = jan,
       volume = {996},
       number = {1},
          eid = {61},
        pages = {61},
          doi = {10.3847/1538-4357/ae1d57},
archivePrefix = {arXiv},
       eprint = {2511.05761},
 primaryClass = {astro-ph.SR},
       adsurl = {https://ui.adsabs.harvard.edu/abs/2026ApJ...996...61R}
}

@ARTICLE{2021ApJ...916...24K,
       author = {{Klement}, Robert and {Hadrava}, Petr and {Rivinius}, Thomas and {Baade}, Dietrich and {Cabezas}, Mauricio and {Heida}, Marianne and {Schaefer}, Gail H. and {Gardner}, Tyler and {Gies}, Douglas R. and {Anugu}, Narsireddy and {Lanthermann}, Cyprien and {Davies}, Claire L. and {Anderson}, Matthew D. and {Monnier}, John D. and {Ennis}, Jacob and {Labdon}, Aaron and {Setterholm}, Benjamin R. and {Kraus}, Stefan and {ten Brummelaar}, Theo A. and {Le Bouquin}, Jean-Baptiste},
        title = "{{\ensuremath{\nu}} Gem: A Hierarchical Triple System with an Outer Be Star}",
      journal = {\apj},
         year = 2021,
        month = jul,
       volume = {916},
       number = {1},
          eid = {24},
        pages = {24},
          doi = {10.3847/1538-4357/ac062c},
archivePrefix = {arXiv},
       eprint = {2105.13437},
 primaryClass = {astro-ph.SR},
       adsurl = {https://ui.adsabs.harvard.edu/abs/2021ApJ...916...24K}
}

@ARTICLE{2013ApJ...766..119M,
       author = {{Miroshnichenko}, A.~S. and {Pasechnik}, A.~V. and {Manset}, N. and {Carciofi}, A.~C. and {Rivinius}, Th. and {{\v{S}}tefl}, S. and {Gvaramadze}, V.~V. and {Ribeiro}, J. and {Fernando}, A. and {Garrel}, T. and {Knapen}, J.~H. and {Buil}, C. and {Heathcote}, B. and {Pollmann}, E. and {Mauclaire}, B. and {Thizy}, O. and {Martin}, J. and {Zharikov}, S.~V. and {Okazaki}, A.~T. and {Gandet}, T.~L. and {Eversberg}, T. and {Reinecke}, N.},
        title = "{The 2011 Periastron Passage of the Be Binary {\ensuremath{\delta}} Scorpii}",
      journal = {\apj},
         year = 2013,
        month = apr,
       volume = {766},
       number = {2},
          eid = {119},
        pages = {119},
          doi = {10.1088/0004-637X/766/2/119},
archivePrefix = {arXiv},
       eprint = {1302.4021},
 primaryClass = {astro-ph.SR},
       adsurl = {https://ui.adsabs.harvard.edu/abs/2013ApJ...766..119M}
}

@ARTICLE{2022ApJ...926..213K,
       author = {{Klement}, Robert and {Schaefer}, Gail H. and {Gies}, Douglas R. and {Wang}, Luqian and {Baade}, Dietrich and {Rivinius}, Thomas and {Gallenne}, Alexandre and {Carciofi}, Alex C. and {Monnier}, John D. and {M{\'e}rand}, Antoine and {Anugu}, Narsireddy and {Kraus}, Stefan and {Davies}, Claire L. and {Lanthermann}, Cyprien and {Gardner}, Tyler and {Wysocki}, Peter and {Ennis}, Jacob and {Labdon}, Aaron and {Setterholm}, Benjamin R. and {Le Bouquin}, Jean-Baptiste},
        title = "{Interferometric Detections of sdO Companions Orbiting Three Classical Be Stars}",
      journal = {\apj},
         year = 2022,
        month = feb,
       volume = {926},
       number = {2},
          eid = {213},
        pages = {213},
          doi = {10.3847/1538-4357/ac4266},
archivePrefix = {arXiv},
       eprint = {2112.05073},
 primaryClass = {astro-ph.SR},
       adsurl = {https://ui.adsabs.harvard.edu/abs/2022ApJ...926..213K}
}

@ARTICLE{2024A&A...690A.176N,
       author = {{Negueruela}, I. and {Sim{\'o}n-D{\'\i}az}, S. and {de Burgos}, A. and {Casasbuenas}, A. and {Beck}, P.~G.},
        title = "{The IACOB project: XII. New grid of northern standards for the spectral classification of B-type stars}",
      journal = {\aap},
         year = 2024,
        month = oct,
       volume = {690},
          eid = {A176},
        pages = {A176},
          doi = {10.1051/0004-6361/202449298},
archivePrefix = {arXiv},
       eprint = {2407.04163},
 primaryClass = {astro-ph.SR},
       adsurl = {https://ui.adsabs.harvard.edu/abs/2024A&A...690A.176N}
}

@ARTICLE{2025OJAp....8E.128E,
       author = {{El-Badry}, Kareem and {Fabry}, Matthias and {Sana}, Hugues and {Shenar}, Tomer and {Seeburger}, Rhys},
        title = "{Complex spectral variability and hints of a luminous companion in the Be star + black hole binary candidate ALS 8814}",
      journal = {The Open Journal of Astrophysics},
         year = 2025,
        month = sep,
       volume = {8},
          eid = {128},
        pages = {128},
          doi = {10.33232/001c.143907},
archivePrefix = {arXiv},
       eprint = {2509.01545},
 primaryClass = {astro-ph.SR},
       adsurl = {https://ui.adsabs.harvard.edu/abs/2025OJAp....8E.128E}
}
\bibliographystyle{aasjournalv7}

\end{document}